\documentclass[11pt]{article}
\usepackage[margin=1in]{geometry}
\usepackage[colorlinks = true,
            linkcolor = blue,
            urlcolor  = blue,
            citecolor = blue,
            anchorcolor = blue]{hyperref}
\usepackage[utf8]{inputenc}
\usepackage{authblk}
\usepackage{setspace,etoolbox}
\usepackage{amsfonts, amssymb, amsthm, amsmath, mathrsfs, mathtools}
\usepackage{natbib, enumitem, setspace, multicol}
\usepackage{graphicx, float, caption, subcaption}
\usepackage{bm}
\usepackage{verbatim}
\usepackage{pdfpages}
\usepackage{fancyvrb}
\usepackage{hyperref}
\usepackage[linesnumbered,ruled]{algorithm2e}
\usepackage{algorithmic}
\usepackage{listings}
\usepackage{bm}
\usepackage{multirow}

\usepackage{makecell}

\usepackage{natbib}
\newtheorem{corollary}{Corollary}[section]

\newtheorem{lemma}{Lemma}[section]
\theoremstyle{definition}
\newtheorem{example}{Example}[section]
\newtheorem{theorem}{Theorem}[section]

\newtheorem{ass}{Assumption}[section]

\newtheorem{defn}{Definition}[section]
\newtheorem{prop}{Proposition}[section]

\newtheorem*{result*}{Result}
\newtheorem*{remark*}{Remark}
\newtheorem*{theorem*}{Theorem}
\newtheorem*{claim*}{Claim}

\usepackage{bm}

\newcommand{\dto}{\stackrel{d}{\longrightarrow}}

\newcommand{\Pto}{\stackrel{\mathbb P}{\longrightarrow}}

\newcommand{\Var}{\mathrm{Var}}
\newcommand{\ind}{\mbox{$\perp\!\!\!\perp$}}

\newcommand{\iid}{\stackrel{i.i.d.}{\sim}}
\newcommand{\var}{\mathrm{Var}}

\allowdisplaybreaks
\numberwithin{equation}{section}

\title{\textbf{ Balancing Covariates for Continuous Treatments by Exploiting Continuity}}

\author{Souhardya Sengupta\thanks{PhD candidate, Department of Statistics, Harvard University. Email: \href{mailto:ssengupta@g.harvard.edu}{ssengupta@g.harvard.edu} } \qquad Kosuke Imai\thanks{Edith and Benjamin Geisinger Professor of Government and of Statistics, Harvard University.  1737 Cambridge Street, Institute for Quantitative Social Science, Cambridge MA 02138, U.S.A.  Email: \href{mailto:imai@harvard.edu}{imai@harvard.edu} URL:
\href{https://imai.fas.harvard.edu}{https://imai.fas.harvard.edu}}}

\date{}

\begin{document}
\maketitle

\begin{abstract}
    We consider causal effect estimation in observational studies with continuous treatment. We introduce a general class of causal estimands that, as a special case, includes the average causal effect of assigning treatment according to an arbitrary counterfactual distribution. Balancing estimators are particularly attractive tools for covariate adjustment in this setting because they avoid directly modeling the treatment assignment mechanism, thereby reducing the potential for instability and misspecification bias in subsequent causal effect estimation. In this paper, we develop a general framework of covariate balancing for continuous treatment with the proposed class of causal estimands, leveraging the assumption of continuity of treatment dose. As in the binary treatment setting, analysts specify the basis functions to be balanced, the allowable degree of covariate imbalance, and a criterion for selecting among admissible weights.  Under a minimum variance criterion for choosing weights, we establish consistency, characterize the effect of basis misspecification, and prove asymptotic normality. We develop consistent variance estimation, which yields asymptotically valid confidence intervals. Under homoskedasticity, our estimator is provably more efficient than inverse probability weighting based on the \emph{true} propensity score. Simulation studies support our theoretical results, and we illustrate the proposed method by estimating the effect of vocational training hours on expected future earnings using data from the U.S. Job Corps study.
\end{abstract}

\section{Introduction}
\label{sec:intro}
	
Many causal questions involve a treatment taking values in a continuous domain, such as drug dosage in medicine, wages in economics, and campaign spending in politics.  In observational studies, the treatment assignment mechanism is unknown and is likely to be confounded.  To address this problem, researchers typically estimate the generalized propensity score (GPS), which represents the conditional density of the treatment given the observed covariates \citep{hiranoandimbens2004,imai:vand:04}. There now exists a large literature based on this GPS-based approach \citep[e.g.,][]{zhuetal2015,galvao2015,kenedyetal2016,kallusandzhou2018,zhao2020propensity,colangeloandlee2025}. 

Estimation of conditional density, however, is a difficult task due to potential model misspecification, and poorly estimated GPS leads to unstable inverse probability weights and substantial bias. Covariate balancing is an attractive alternative because it sidesteps the modeling of the treatment assignment mechanism and directly weighs the sample to achieve covariate balance.  A large number of covariate balancing methods exist for a binary treatment, but they do not directly extend to a continuous treatment \citep[e.g,.][]{hainmueller2012,imaiandratkovic2013,zubuzaretta2015,chan2016globally,athey2018approximate,zhao2017entropy,wong2018kernel,zhao2019covariate,ning2020robust,wang2020minimal,benmichael2021balancingactcausalinference,fan2023optimal}.

In this paper, we develop a new covariate balancing method for a continuous treatment where the causal estimand of interest is the average outcome under an arbitrary stochastic intervention \citep{diaz2011}.  While there exist a relatively small number of covariate balancing methods for a continuous treatment, they either target the average outcome under a fixed treatment value that is common across individuals \citep[e.g.,][]{fongetal2018,vegetabile2020,tubbicke2021,bahadorietal2022,huling2023}, or the average outcome under the treatment policy that is directly modified from the observed treatment assignment in a possibly covariate-dependent manner \citep{jiangandhuling2025}. 

In contrast, we consider an arbitrary stochastic intervention with a continuous treatment that is substantively different from these settings of the existing methods.  In particular, our covariate balancing method is applicable for offline policy evaluation to assess the empirical performance of individualized treatment rules \citep[see e.g.,][for the binary treatment case]{LuedtkeVanderLaan2016,ImaiLi2023,liandbrunskill2026}. Another setting in which stochastic interventions arise naturally is when multiple versions of a treatment policy are possible \citep{vandeweeleandhernan2013}. For instance, consider a group of policymakers whose members may recommend different policies for the same individual. In this case, a stochastic intervention provides a more natural representation of the group's collective decision-making process than a deterministic rule, as it captures variation in policy choices across its members. Our balancing estimator can then be used to evaluate the effect of the group's collective policy on a population with a different covariate distribution.
    
We begin by introducing a general class of causal estimands, where the average causal effect of an arbitrarily specified counterfactual distribution on the continuous treatment space is a special case. We extend the traditional covariate balancing framework, in which researchers choose the set of basis functions to be balanced, specify the allowable degree of covariate imbalance, and choose the objective used to select among the admissible weights that satisfy balancing conditions. We show that one natural choice of such an objective function yields a closed-form estimator and derive the bias due to misspecification of basis functions. 

The key assumption we leverage is continuity of the outcome in the dose, which is commonly used in outcome modeling \citep[e.g.,][]{kenedyetal2016,bransonetal2023,huling2023,nagalapattietal2024}. We establish that the proposed estimator is asymptotically normal, and its asymptotic variance is substantially smaller than that of the IPW estimator.  This variance reduction is due to the fact that enforcing covariate balance removes the part of the variability of the weighting estimator explained by the basis functions. Extensive simulations demonstrate its practical performance. Finally, as our main application, we use the proposed method to estimate the effect of vocational training hours on the expected future earnings by analyzing the data from the U.S. Job Corps study.
	
\subsection*{Related literature}
\label{sec:literature}

Existing balancing methods for a continuous treatment fall into two groups, distinguished by the causal estimand they target. The first targets the average dose-response function (ADRF), the mean outcome when every unit receives a common dose. Methods in this group choose weights that render the treatment and the covariates uncorrelated in the weighted sample \citep[e.g.,][]{fongetal2018,vegetabile2020,tubbicke2021,bahadorietal2022,huling2023}. However, this target quantity is the {\it unconditional} dose-response curve, averaged over the covariates, rather than the \emph{heterogeneous} dose-response function $\mathbb{E}\left[Y(a)\mid \bm X = \bm x\right]$. In contrast, the proposed method evaluates the average outcome under an arbitrary intervention that is possibly a function of covariates.

The second group of balancing methods for a continuous treatment targets the average outcome under a modified treatment policy (MTP) \citep{haneuseandrotinzky2013}.  MTP alters the observed dose according to a prespecified rule, possibly as a function of the observed covariates. Applying an MTP to the observed data yields an explicit target sample whose average outcome is the object of inference, so estimation reduces to a two-sample balancing problem analogous to the binary-treatment setting. In particular, \cite{jiangandhuling2025} solves this problem by minimizing the energy distance between the two samples. MTP differs from an arbitrary intervention we consider in that the latter is not a function of observed dose and admits no exact target sample to balance against. 

The only work constructing weights for an arbitrary intervention like our method is \citet{sondhietal2020}, who convert the odds of a classifier separating observed from counterfactual covariate--treatment samples into a GPS-ratio estimate, but provide no method for inference. Indeed, valid statistical inference is rare in either group.  The exceptions are the augmented energy balancing estimators of \citet{huling2023} and \citet{jiangandhuling2025} which come with asymptotic guarantees.  But, as mentioned above, these methods target ADRF and MTP, respectively.

There are other approaches to observational studies with a continuous treatment. \citet{fongetal2018} and \citet{yiuandsu2018} model the treatment density parametrically. The incremental propensity score \citep{kennedy2017,diazhejazi2020,schindl2026incrementaleffectscontinuousexposures} takes a different route---the counterfactual is an exponential tilt, indexed by a single parameter, of the unknown treatment-generating distribution, where the estimand is the causal effect of this tilt  \citep[see][]{nabi2025semiparametricsensitivityanalysisunmeasured,rakshitetal2024,schindl2025causalgeodesycounterfactualestimation}. In this case, there is no need to achieve covariate balance since, by construction, this tilt can be identified without knowledge of the propensity score. 


\section{The Proposed methodology}

We begin by describing our setting and then define the class of causal quantities whose inference is the central goal of this paper.

\subsection{Set up}

Suppose that we have an independently and identically distributed (i.i.d.) sample of $n$ units from a super-population $\mathcal{G}$. For the $i^{\mathrm{th}}$ unit with $1\leq i \leq n$, we observe a $d$-dimensional covariate $\bm X_i\in \mathcal X\subseteq \mathbb R^p$, a continuous treatment exposure $A_i\in \mathcal A\subseteq \mathbb R$, and an observed outcome $Y_i\in \mathbb R$. Let $Y_i(a)$ denote the potential outcome we would observe if unit $i$ were to be treated with exposure level $a \in \mathcal{A}$. Then, the observed outcome relates to the potential outcomes via the equality $Y_i = Y_i(A_i)$.
\begin{ass}[Independent and identically distributed (i.i.d.) sample]
 \label{assm:iid}
     There exists a super-population $\mathcal G$ such that $\left(\bm X_i, A_i, \{Y_i(a)\}_{a\in \mathcal{A}}\right)\iid \mathcal G$ for $i=1,\dots, n$.
\end{ass}

Before we introduce a broad class of causal estimands, we first define the notion of a counterfactual weight \citep[see][for a similar estimand in the context of binary treatments with interference]{sengupta2025lowrankcovariatebalancingestimators}. 
\begin{defn}[Counterfactual weights]
    \label{defn:counterfactual_weights}
    Let $\mathscr A$ be a $\sigma$-algebra on $\mathcal A$. Then a counterfactual weight is a function
    \[
        f:\mathscr A\times \mathcal X\mapsto \mathbb R,
    \]
    such that for any given $\bm x\in \mathcal X$, $f(,\bm x)$ is a finite, signed measure on the space $(\mathcal A, \mathscr A)$. If $f(,\bm x)$ admits a density with respect to some base measure on $(\mathcal A, \mathscr A)$, we will abuse notations and use $f(,\bm x)$ to denote this density.
\end{defn}
As it is clear from this definition, a counterfactual weight represents a \textit{stochastic intervention} \citep{diaz2011,kennedy2017} if for any given $\bm x \in \mathcal X$, $f(,\bm x)$ is also a probability over the treatment space $\mathcal A$. Using this counterfactual weight, we can define a general class of causal estimand.
\begin{defn}[Causal estimand]
    \label{defn:causal_estimand}
    Consider a counterfactual weight $f$ that satisfies
    \begin{align}
    \label{eqn:existence}
        \mathbb E\left[\int |Y_i(a)||f|(da, \bm X_i)\right]<\infty.
    \end{align}
    Then, our estimand of interest is defined as,
    \[
        \mu_f := \mathbb E\left[\int Y_i(a)f(da, \bm X_i) \right].
    \]
\end{defn}
Throughout this paper, we assume that Equation~\eqref{eqn:existence}, which guarantees the existence of $\mu_f$, holds for the counterfactual weight $f$ under consideration. When $f$ is a stochastic intervention, the causal estimand $\mu_f$ equals its {\it policy value}, i.e., the average outcome under $f$. In the policy learning literature, estimating $\mu_f$ from previously collected data is referred to as off-policy evaluation, or OPE.

\begin{example}[Policy value of a deterministic intervention]
    \label{ex:useful_2}
    Consider an individualized deterministic treatment policy $\pi:\mathcal X\mapsto \mathcal A$. Let $f_{\pi}(,\bm x)$ denote the Dirac measure with its mass placed entirely at $\pi(\bm x)$. Then, the value of this deterministic policy $\pi$ is given by,
    \[
        \mu_{f_\pi} := \mathbb E[Y_i(\pi(\bm X_i))].
    \]
\end{example}

\begin{example}[Average treatment effect of a stochastic intervention]
    \label{ex:useful_3} The average treatment effect of a stochastic intervention $f_1$ over the baseline stochastic intervention $f_0$ is given by, 
    \[
        \mu_{f_1-f_0} = \mu_{f_1} - \mu_{f0}.
    \]
    If $f_1$ is an individualized treatment rule and $f_0$ is a uniform treatment rule with an identical budget constraint, $\mu_{f_1-f_0}$ represents the value of personalization \citep{ImaiLi2023,liandbrunskill2026}.
\end{example}

\subsection{Characterizing balancing equation for a continuous treatment}
\label{sec:weighting_estimators}

Like the binary case, our balancing estimator belongs to a broad class of weighting estimators. In particular, weighting estimators take the form
$$
T(\bm w) := \frac{\bm w^\top \bm Y}{n},
$$
where $\bm w=(w_1,\ldots,w_n)^\top$ is a vector of weights and $\bm Y=(Y_1,\ldots,Y_n)^\top$. Our choice of weights is aimed at an unbiased estimation of $\mu_f$. 

As a first step towards motivating the construction of these balancing weights, we begin with the choice of the basis functions to balance. In particular, let 
\begin{align}
\label{eqn:potential_outcome_regression}
    m(a, \bm x) = \mathbb E\left[Y_i(a)\mid \bm X_i = \bm x\right]
\end{align}
denote the potential outcome regression function of the covariates. In the binary case, i.e., $a\in \{0,1\}$, we choose basis functions $\{\phi_k(\cdot)\}_{k=1}^p$ of the covariates such that they span $m(0, \cdot)$ and $m(1,\cdot)$. Analogously, in the continuous case, we require that the basis functions satisfy $m(a,\cdot)\in \mathrm{span}\left(\{\phi_k(\cdot)\}_{k=1}^p\right)$ for all $a\in \mathcal A$. An appropriate choice of such basis functions has received considerable discussion in the literature; for example, see \cite[Section 5]{benmichael2021balancingactcausalinference} and the references therein. While we focus on balancing any given basis functions, we will later consider the effect of the misspecification of the basis function on the bias of the weighting estimator.

In the binary case, once the basis functions are specified, one usually obtains weights such that the mean of the weighted basis function of a source population matches the unweighted mean of a target. This task usually removes the conditional bias of the weighting estimator. In our continuous case, however, the notion of a target population is not well defined. Thus, we directly form the balancing equation with the goal of eliminating the conditional bias. To do so, first assume that we are in a setting where the basis functions have been correctly specified and hence, we can write $m(a,\bm x) = \bm \phi(\bm x)^\top\bm \beta(a)$ for some function $\bm \beta(\cdot)$ on $\mathcal A$, and $\bm \phi(\cdot):=\left(\phi_1(\cdot),\cdots, \phi_p(\cdot)\right)^\top$.

Next, we present our key assumption that a small change in the treatment dosage results in a small change in the outcome. Such an assumption is commonly used in the covariate balancing literature for continuous treatment, primarily in the nonparametric modelling of the outcome function \citep{vegetabile2020,huling2023}. However, to the best of our knowledge, the continuity assumption has not been used to derive the balancing weights themselves.
\begin{ass}[Continuity]
    \label{assm:continuity}
    For any $\bm x$, $a\mapsto m(a,\bm x)$ is a continuous function.
\end{ass}  

As we next state in Assumption~\ref{assm:bdd}, we will also assume that the treatment space is bounded. This is standard in the causal inference literature with continuous treatments \citep{hiranoandimbens2004}.  The latter implies that, without loss of generality, we can set $\mathcal A = [0,1]$, or equivalently $A_i\in [0,1]$, almost surely, for all $i=1,\dots, n$.

\begin{ass}[Bounded treatments]
\label{assm:bdd}
    We assume that $\mathcal{A}$ is a closed, bounded set.
\end{ass}

Next, given the decomposition $m(a,\bm x) = \bm \phi(\bm x)^\top\bm \beta(a)$, Assumption~\ref{assm:continuity} implies the continuity of $a\mapsto \bm \beta(a)$ as long as the domain of $\bm \phi(\cdot)$ is rich enough, i.e., there exist $\bm x_1,\cdots \bm x_d$ such that the matrix with columns $\bm \phi(\bm x_i)$ is non-singular. Next, we divide the treatment space, which is an interval $\mathcal{A}=[0,1]$, into $m_n$ equally spaced points, i.e., $1/m_n, 2/m_n,\cdots, 1$. For any $a\in [0,1]$, let $n_1(a)$ denote the nearest neighbor of $a$ from the set $\{i/m_n:1\leq i\leq m_n\}$, where, in case of ties, we choose a larger grid point. 

Define the \textit{$1$-nearest neighbor} estimator of $\bm \beta(a)$ as, 
\begin{align*}
    \tilde{\bm \beta}_n(a):=\bm \beta(n_1(a)).
\end{align*}
It is convenient to rewrite this estimator using matrix notation. Let us define $\bm v(a)$ as the $m_n$-dimensional binary indicator vector of 1-nearest neighbor with 1 in the $j^{\mathrm{th}}$ position if and only if $j/m_n = n_1(a)$. Furthermore, define $\bm B_n:=(\bm \beta(1/m_n)^\top,\cdots, \bm \beta(1)^\top)^\top$. Then, we can write, 
\begin{align}
\label{eqn:knn_estimation}
    \tilde{\bm \beta}_n(a) = \{\bm v(a)^\top\otimes \bm I_d\}\bm B_n,
\end{align}
where $\bm I_d$ denotes the $d\times d$ identity matrix. 

Under Assumption~\ref{assm:continuity}, $\tilde{\bm \beta}_n(\cdot)$ and  $\bm \beta(\cdot)$ becomes close as $m_n\to \infty$. Therefore, for sufficiently large $m_n$, $\mu_f$ and $\mathbb E [T(\bm w)]$ will be close to their proxies $\tilde{\mu}_f$ and $\tilde{\mathbb E}[T(\bm w)]$, respectively, which are defined as follows:
\begin{align}
\label{eqn:tilde_proxies}
    \tilde{\mu}_f := \mathbb E\left[\int_{\mathcal A}\bm \phi(\bm X_i)^\top\tilde{\bm \beta}_n(a)f(da, \bm X_i)\right],\quad \tilde{\mathbb E}\left[T(\bm w)\right] := \mathbb E\left[\frac{1}{n}\sum_{i=1}^n w_i \bm \phi(\bm X_i)^\top\tilde{\bm \beta}_n(A_i)\right].
\end{align}
We expect that the weights that make $\tilde{\mathbb E}[T(\bm w)]$ close to $\tilde{\mu}_f$, would asymptotically result in $T(\bm w)$ that approximates $\mu_f$, allowing us to estimate $\mu_f$ without bias. The following proposition formalizes this argument. 
\begin{prop}[Unbiased estimators]
\label{lem:beta_star_expressions} 
Suppose that $m(a, \bm x) = \bm \phi(\bm x)^\top\bm \beta(a)$ holds. Recall that $\tilde{\mu}_f$ and $\tilde{\mathbb E}[T(\bm w)]$ are defined in Equation~\eqref{eqn:tilde_proxies}. Then, unbiased estimators of $\tilde{\mu}_f$ and $\tilde{\mathbb E}[T(\bm w)]$ are given by:
\begin{align*}
    \hat{\tilde \mu}_f = \frac{1}{n}\left[\sum_{i=1}^n \bm \nu (f,\bm X_i)\otimes \bm \phi(\bm X_i)\right]^\top\bm B_n \quad \text{and} \quad \hat {\tilde T}(\bm w) = \frac{\bm w^\top\bm \Phi\bm B_n}{n},
\end{align*}
where 
\[
    \bm \Phi := \begin{bmatrix}
        \left(\bm v(A_1)\otimes \bm \phi(\bm X_1)\right)
        &\cdots
        &\left(\bm v(A_n)\otimes \bm \phi(\bm X_n)\right)
    \end{bmatrix}~{\rm and}~\bm \nu (f, \bm X_i):=\int\bm v(a)f(a, \bm X_i)da,\forall i.
\]
\end{prop}
The above proposition suggests that a choice of weights satisfying the following equation should result in a weighting estimator unbiased for ${\tilde \mu}_f$ uniformly across the choices of $\bm B_n$,
\begin{align}
\label{eqn:balancing}
    \bm \Phi^\top\bm w = \sum_{i=1}^n \bm \nu (f, \bm X_i)\otimes \bm \phi(\bm X_i).
\end{align}
This equation represents the balancing equation for continuous treatment. We refer to it as the \textit{approximate balancing equation}, because weights, which satisfy this equation, can only aim for asymptotic unbiasedness.

The balancing equation above can be rewritten as a system of $m_n$ simultaneous equations, each of which represents a balancing equation for a bin of continuous treatment. To see this, for any $a\in [0,1]$, the vector $\bm v(a)$ can take one of the $m_n$ possible different vectors $\{\bm v^{(1)}, \cdots, \bm v^{(m_n)}\}$ (see equation~\eqref{eqn:knn_estimation}), where $\bm v^{(j)}$ is a $m_n$-dimensional vector with entry 1 in the position $j$ and 0 everywhere else. In fact, we can partition $[0,1]$ into $m_n$ disjoint intervals, such that whenever $a$ belongs to an interval, $\bm v(a)$ equals the corresponding vector from $\left\{\bm v^{(j)}\right\}$. Proposition~\ref{lem:bj_intervals}(a) of the following proposition characterizes these intervals explicitly, while Proposition~\ref{lem:bj_intervals}(b) provides a simplification of the approximate balancing equation by breaking it into a simultaneous system of $m_n$ bin-wise equations.
\begin{prop}
    \label{lem:bj_intervals}
    Define the following partition of $[0,1]$:
    \begin{align}
    \label{eqn:Bj_sets}
    \begin{aligned}
    B^{(1)} & = \left[0,\frac{3}{2m_n}\right),\\
    B^{(j)} & = \left[\frac{2(j-1)+1}{2m_n}, \frac{2j+1}{2m_n}\right),\textrm{ for }j=2,\cdots, m_n - 1,\\
    B^{(m_n)} & = \left[1-\frac{1}{2m_n},1\right].
    \end{aligned}
    \end{align}
    Then, the following equations hold:
    \begin{enumerate}[label = (\alph*)]
        \item For any $j$, $a\in B^{(j)}\implies \bm v(a) = \bm v^{(j)}$.
        \item For any $j$, let $S_j = \{i: A_i \in B^{(j)}\}$ denote the set of observations that belong to $B^{(j)}$ where $n_j = |S_j|$. Furthermore, $\bm \Phi_{j}$ denotes the matrix formed with rows $\bm \phi(\bm X_i)$ corresponding to indices $i\in S_j$. Let $\hat{\bm r}_j = \frac{1}{n}\sum_{i=1}^n\bm \phi(\bm X_i)f(B^{(j)}, \bm X_i)$ where  $f(B,\bm x):=\int_{B}f(a,\bm x)da$, for any $\bm x$ and measurable set $B$,.
        Then, the approximate balancing equation~\eqref{eqn:balancing} is feasible if and only if the following binwise system of equations is feasible,
        \begin{align}
        \label{eqn:equivalent_balancing}
            \bm \Phi_j^\top\bm w_{S_j} = n\hat{\bm r}_j, \forall j=1,\dots, m_n,
        \end{align}
        where $\bm w_{S_j}$ represents the sub-vector of $\bm w$ corresponding to indices in $S_j$. Moreover, weights $\bm w$ satisfy equation~\eqref{eqn:balancing} if and only if $\bm w_{S_j}$ satisfies the balancing equation~\eqref{eqn:equivalent_balancing} indexed by $j$, for any $1\leq j \leq m_n$.
    \end{enumerate}
\end{prop}

The above proposition establishes that solving the approximate balancing equation is equivalent to solving the bin-wise system of equations given in equation~\eqref{eqn:equivalent_balancing}. While the approximate balancing equation is not always feasible, we can characterize a sufficient condition for its feasibility. For example, if $\bm \Phi_j$ is non-singular for all $j$, then all the $m_n$ equations above are feasible. This in turn holds if
for all $j$, $n_j$, the number of rows of $\bm \Phi_j$, is greater than $p$ and the columns of $\bm \Phi_j$ are in general position (see \citet[Section 2.2]{tibshirani2013lasso} for definition).

However, even when the approximate balancing equation is feasible, if $n_j$ is close to $p$, $\bm \Phi_j$ may be ill-conditioned, leading to an unstable solution for $\bm w_{S_j}$.
To address this issue, instead of solving the simultaneous equations exactly, we allow for slack terms $\{\bm \xi_j\}$,
\begin{align}
    \label{eqn:slack_balancing}
    \bm \Phi_j^\top\bm w_{S_j} = n\hat{\bm r}_j + \bm \xi_j, \forall j=1,\dots, m_n - k_n + 1.
\end{align}
In the binary treatment literature, researchers use a slack term to balance bias and variance (see \cite{benmichael2021balancingactcausalinference} for a review). In the next section, we show how to choose the values of slack parameters $\{\bm \xi_j\}$.


\subsection{Optimal choice of weights}
\label{sec:optimal_weight}

We choose optimal balancing weights by minimizing the norm,
\[
    \underset{\bm w \in \mathcal C_f}{\arg\min}~\|\bm w\|^2,
\]
where $\mathcal C_f$ denotes the class of weights that satisfy the approximate balancing equation~\eqref{eqn:balancing}. One can also optimize other objective functions and derive the properties of the resulting estimators, but we will leave that to future work. In our formulation, we further allow for the existence of slack term and so the optimal weights are given as a solution to the following optimization problem,
\begin{equation}
\label{eqn:final_optimization}
  \hat{\bm w}_f :=  \underset{\bm w_{S_j}}{\arg\min}~\left(\frac{1}{2}\|\bm w_{S_j}\|^2 + \frac{1}{2n\lambda_{nj}}\|\bm \xi_j\|^2\right) \quad
    \textrm{subject to } \bm \Phi_j^\top\bm w_{S_j} = n\hat{\bm r}_j + \bm \xi_j,
\end{equation}
where $\lambda_{nj}> 0$ is the regularization term and $\hat{\bm w}_{f,S_j}$ is the sub-vector of $\hat{\bm w}_f$ for each $j$.  This is equivalent to
\begin{equation}
    \hat{\bm w}_{f,S_j} = \underset{\bm w_{S_j}}{\arg\min}~\left(\frac{1}{2}\|\bm w_{S_j}\|^2 + \frac{1}{2n\lambda_{nj}}\|\bm \Phi_j^\top\bm w_{S_j} - n\hat{\bm r}_j\|^2\right).
\end{equation}

We note that this optimization problem is always feasible.
Thus, for each $j$, instead of controlling the slack terms $\{\bm \xi_j\}$ within a pre-specified range, we directly penalize the norm of weights in the objective function (along with the norm of $\bm w_j$), with the ridge parameter $\lambda_{nj}$. This is a common technique used in the literature to obtain stable weights (see \cite[Section~3.2]{benmichael2021balancingactcausalinference}). In the next section, we show that a rate-aware choice of $\lambda_{nj}$ does not incur any asymptotic bias of the resulting weighting estimator.

As is common in the covariate balancing literature \citep{zubuzaretta2015,benmichael2021balancingactcausalinference}, one can allow for a prespecified degree of covariate imbalance rather than requiring exact balance. For example, in our setting, this can correspond to imposing bin-wise imbalance constraints of the form
\[
    \|\bm \Phi_j^\top\bm w_{S_j} - n\hat{\bm r}_j\|\leq \delta_j, \quad j=1,\ldots,m_n.
\]
Allowing such imbalance can improve the stability of the weighting estimator at the cost of introducing some bias, whose magnitude is related to the degree of imbalance. However, the resulting constrained optimization problem need not be feasible for arbitrary choices of $\delta_j$. In Appendix~\ref{sec:ridge_imbalance_euqivalence}, we show that the class of minimum-norm solutions to the feasible imbalance-constrained problems coincides exactly with the class of weights obtained from Equation~\eqref{eqn:final_optimization} by varying $\lambda_{nj}$, including its limiting cases.



Finally, we derive the closed-form expression of the weights that solve the above optimization problem \eqref{eqn:final_optimization} in the following proposition.
\begin{prop}[Closed-form expression of weights]
\label{lem:weights_characterization}
        For any $1\leq j \leq m_n$, let $\lambda_{nj}>0$ and define,
        \begin{align*}
            \hat{\bm M}_j = \frac{1}{n}\bm \Phi_j^\top\bm \Phi_j \textrm{ and } \hat{\bm c}^{(n)}_j = \left(\hat{\bm M}_j + \lambda_{nj} \bm I_p\right)^{-1}\hat{\bm r}_j.
        \end{align*}
        Then, for any $i\in S_j$, we have
        \[
            \hat w_{f,i} = \bm \phi(\bm X_i)^\top\hat{\bm c}^{(n)}_{j}.
        \]
\end{prop}

\section{Asymptotic inference}
\label{sec:asymp}

In this section, we establish the asymptotic properties of our proposed balancing estimator. In particular, we derive the conditions for consistency and asymptotic normality of our balancing estimator, and construct asymptotically valid confidence intervals.  We also analyze the efficiency gain of the balancing estimator over the standard IPW estimator. Finally, based on our asymptotic theory, we propose a choice for the hyperparameters, $m_n$ and $\lambda_{nj}$, and discuss a practical way to assess finite-sample imbalance.

\subsection{Consistency}
\label{sec:asymp_consistency}

We consider a general setting with the set of basis functions $\bm \phi(\cdot)$.  We assume the following signal-plus-noise model, which allows for the possibility that the outcome regression function $m(\cdot,\cdot)$ does not lie in the span of $\bm \phi(\cdot)$. 
\begin{ass}[Signal-plus-noise model]
\label{assm:signal_plus_noise_general}
    The potential outcome can be decomposed into the following signal and noise components:
    \begin{align}
        \label{eqn:signal_plus_noise}
        Y_i(a) = m(a, \bm X_i) + \sigma(\bm X_i)\epsilon_i(a),\forall a \in \mathcal{A},
    \end{align}
    where, $\{\epsilon_i(a)\}_{a \in \mathcal{A}}$ is an independent process of error terms that is independent of $\bm X_i$, such that for any $a$, $\mathbb E [\epsilon_i(a) \mid \bm X_i] = 0$ and $\Var[\epsilon_i(a) \mid \bm X_i] = 1$. The entire noise term is also scaled by $\sigma^2(\bm x)$ to allow for heteroskedasticity. We use $\epsilon_i:=\epsilon_i(A_i)$ to denote the realized error term. 
\end{ass}

Next, we make some standard assumptions required for identification of our causal effects.
\begin{ass}[Unconfoundedness]
    \label{assm:unconfound}
    We have $\{Y_i(a)\}_{a\in\mathcal{A}}\ind A_i \mid \bm X_i$ for any $i$.
\end{ass}
\begin{ass}[Strong positivity]
    \label{assm:pos}
    Let $e(a, \bm x):=p(A_i = a \mid \bm X_i = \bm x)$ denote the conditional density of $A_i$ given $\bm X_i = \bm x$ for $a \in \mathcal{A}$. Then, there exists some $0<\alpha\leq \beta$, such that for all $a \in \mathcal{A}$, $\alpha\leq e(a,\bm x)\leq \beta$.
\end{ass}

Assumption~\ref{assm:unconfound} is commonly required for the identification of causal effects \citep{rosenbaumetal1983,vegetabile2020,huling2023}. Assumption~\ref{assm:pos} is also a standard positivity assumption in causal inference with observational data and ensures that the treatment assignment density is positive every where, so that for large enough sample sizes, every part of the treatment space is well-represented in the treatment sample. This assumption has been used in the literature on balancing for continuous treatment, including \cite{huling2023} (see Section~2.1). We next state some regularity conditions that will be used is establishing the consistency of our causal effects.

\begin{ass}[Regularity conditions]
    \label{assm:regularity}
    The following conditions hold:
    \begin{enumerate}[label = (\alph*)]
        \item (Bounded bases) There exists a $C>0$, such that $\|\bm\phi(\bm x)\|\leq C, \forall \bm x$.
        \item (Bounded counterfactual weights) There is some $C_f<\infty$, such that $|f|(a,\bm x)\leq C_f, \forall a, \bm x$.
        \item (Local regularity of counterfactual weights) There is some $C_f^\ast<\infty$ such that,
        \[
            \mathbb E\left[|f|(B, \bm X_1)\right]\leq C_f^\ast|B|,
        \]
        for any measurable set $B\subseteq \mathcal A(=[0,1])$, with non-zero Lebesgue measure, that is, $|B|>0$.
        \item (Full rank basis functions) $\lambda_{\min}\left(\mathbb E(\bm \phi(\bm X_1)\bm \phi(\bm X_1)^\top)\right)>0$.
    \end{enumerate}
\end{ass}

Assumption~\ref{assm:regularity}(a) states that the basis functions must be bounded.  This requirement is met if $\bm \phi$ is continuous with a compact domain. Assumptions~\ref{assm:regularity}(b)~and~(c) ensure that the counterfactual weight and its expectation remain bounded. These requirements are satisfied, for example, if $f$ is a bounded density for an absolutely continuous counterfactual stochastic intervention. Moreover, if we are interested in estimating the value of a deterministic policy $\pi(\bm x)\in \mathcal A$ so that $f(a, \bm x) = \mathbb I(a = \pi(\bm x))$, then, Assumption~\ref{assm:regularity}(c) is satisfied if $\pi(\bm X_1)$ has a bounded density because we have $\mathbb E [f(B, \bm X_1)] = \mathbb P(\pi(\bm X_1)\in B)$. Assumption~\ref{assm:regularity}(d) is also a mild condition, requiring that the components of the vector $\mathbb \phi(\bm X_1)$ must be of full rank.

Finally, we assume some rate conditions on the choice of number of bins $m_n$ and regularizes $\{\lambda_{nj}\}$.
\begin{ass}[Choice of $m_n$ and $\{\lambda_{nj}\}$]
    \label{assm:rate_choice_consistency}
    \begin{enumerate}[label = (\alph*)]
    \item (Choice of $m_n$) $m_n = n^\gamma$, for $\gamma\in (0,1)$ such that $m_n/n\to 0$.
    \item (Choice of $\lambda_{nj}$). $\lambda_{nj}$ is a function of $\{(\bm X_i, A_i)\}_{1\leq i \leq n}$ alone and $m_n\max_{1\leq j \leq m_n}\lambda_{nj} = o_{\mathbb P}(1)$, as $n\to \infty$.
    \end{enumerate}
\end{ass}

Assumption~\ref{assm:rate_choice_consistency}(a) requires that $\{m_n\}$ should not grow too fast as the sample size increases, while item (b) specifies the choice of $\lambda_{nj}$, which is allowed to be data-dependent though it cannot depend on the response $Y_i$.

Theorem~\ref{thm:consistency} establishes the consistency of the proposed estimator.
\begin{theorem}[Consistency]
\label{thm:consistency}
Suppose that Assumptions~\ref{assm:iid}--\ref{assm:signal_plus_noise} hold. Also assume that $m$ is bounded, $\mathbb E[\sigma^2(\bm X_1)]<\infty$ and that for every $\bm x$ the map $a\mapsto e(a,\bm x)$ is continuous on $\mathcal A$. Then, as $n\to \infty$, where
$$
T(\hat{\bm w}_f)\Pto \mu_f^* \ := \ \mathbb E\left[\int_{\mathcal A}\bm \phi(\bm X_1)^\top\bm \beta^*(a)f(da, \bm X_1)\right],
$$
where $\bm \beta^*(a) := \mathbb E[\bm \phi(\bm X_1)\bm \phi(\bm X_1)^\top e(a,\bm X_1)]^{-1}\mathbb E[\bm \phi(\bm X_1)m(a, \bm X_1)e(a, \bm X_1)]$.
\end{theorem}

The above theorem shows that our balancing estimator is consistent for a causal quantity that is similar to $\mu_f$, but replaces the true outcome regression function $m(a,\bm x)$ with $m^*(a,\bm x) = \bm \phi(\bm x)^\top\bm \beta^*(a)$. This quantity can be interpreted as a projection of $m$ onto the span of $\{\bm \phi_k(\cdot)\}_{1\leq k \leq p}$. Thus, the closer the original outcome regression function $m(\cdot)$ lies to the span of our specified basis function, the smaller we would expect the resulting bias to be. Finally, as formally stated below, $T(\hat{\bm w}_f)$ is consistent for $\mu_f$ if the basis functions are correctly specified.

\begin{corollary}[Correct specification]
\label{cor:correct_specification}
    Under the setting of Theorem~\ref{thm:consistency}, if the relation $m(a,\bm x) = \bm \phi(\bm x)^\top\bm \beta(a)$ holds, then $\mu_f = \mu_f^*$.
\end{corollary}

\subsection{Asymptotic normality}
\label{sec:asymp_normality}

We now establish a central limit theorem for our balancing estimator, which yields an asymptotically valid confidence interval. For this result, we assume correct specification of the basis functions under the same signal and noise model as before. 
\begin{ass}[Signal-plus-noise model with correct specification]
\label{assm:signal_plus_noise}
    There exists a vector set of basis functions $\bm \phi(\bm x)$, such that the outcome regression function has the form: $m(\bm a, \bm x) = \bm \phi(\bm x)^\top\bm \beta(a)$, for some function $\bm \beta(\cdot)$.
\end{ass}
We state the additional assumptions required for the asymptotic normality result.
\begin{ass}[Additional regularity conditions]
    \label{assm:asymp_regularity} The following conditions hold:
    \begin{enumerate}[label = (\alph*)]
        \item There exists some $\delta>0$, such that $\sup_{a\in \mathcal A}\mathbb E\left[| \epsilon_1(a)|^{2+\delta}\right]<\infty$.
        \item There exist some $\sigma^*\geq \sigma_*>0$ such that ${\sigma^*}^2\geq \sigma^2(\bm x)\geq \sigma_*^2$, for all $\bm x$.
        \item 
        \[
            \underset{n\to\infty}{\lim\inf}\sum_{j=1}^{m_n}\left\|\mathbb E\left[\bm \phi(\bm X_1)f(B^{(j)}, \bm X_1)\right] \right\|>0.
        \]
    \end{enumerate}
\end{ass}
Assumption~\ref{assm:asymp_regularity}(a) places a moment restriction on $\epsilon_i$.  This moment restriction can be made more interpretable by assuming instead that $\epsilon_i(a)$ is identically distributed across $a\in \mathcal A$. Then, $\sup_{a\in \mathcal A}\mathbb E\left[| \epsilon_1(a)|^{2+\delta}\right]$ is equal to a quantity that does not depend on $a$, e.g., $\mathbb E\left[|\epsilon_i(0)|^{2+\delta}\right]$. Assumption~\ref{assm:asymp_regularity}(b) places upper and lower bounds on the heteroskedastic noise variability. Assumption~\ref{assm:asymp_regularity}(c) ensures that the asymptotic variance of the balancing estimator is bounded away from zero. Appendix~\ref{sec:more_on_item_e} shows, by writing $\bm Q(B) = \mathbb E[\bm\phi(\bm X_1) f(B, \bm X_1)]$, that a necessary and sufficient condition for Assumption~\ref{assm:asymp_regularity}~(c) is $\bm Q$ is not the zero measure. This is a mild condition and can be satisfied, for example, as long as $\bm Q$ has an intercept term and $\mathbb E[f(\cdot, \bm X_1)]$ is not trivially the zero measure (the latter condition is likely true for any interesting counterfactual weight). In addition, because correct specification of basis functions implies $\mu_f = \mathbb E\left[\int \bm \beta(a)^\top d\bm Q(a)\right]$, when $\bm Q$ is zero measure, we have a trivial case where our causal estimand is identically 0, regardless of $\bm \beta$. Finally, we impose some refined conditions on the choice of $m_n$ and $\{\lambda_{nj}\}$.

\begin{ass}
\label{assm:rate_choice_normality}
\begin{enumerate}[label = (\alph*)]
     \item $\sqrt{n}M(1/m_n)\to 0$, where,
        \begin{align}
        \label{eqn:modulus of continuity}
            M(\eta):=\sup_{x,  y: |x - y|\leq \eta}\|\bm \beta(x) - \bm \beta(y)\|,
        \end{align}
        is the modulus of continuity of the function $\bm \beta$.
     \item $\sqrt{n}m_n\max_{1\leq j \leq m_n}\lambda_{nj} = o_{\mathbb P}(1)$, as $n\to \infty$.
\end{enumerate}
\end{ass}

Assumption~\ref{assm:rate_choice_normality}(a) further refines Assumption~\ref{assm:rate_choice_consistency}(a) by requiring that to achieve $\sqrt{n}$-consistency, $m_n$ needs to grow at a sufficiently faster rate governed by the quantity $M(\cdot)$. In particular, if $\bm \beta$ is Lipschitz (e.g., $\bm \beta$ has a continuous derivative on $\mathcal A$), then this condition is satisfied so long as $\sqrt{n}/m_n\to 0$ holds. We emphasize that one can always choose $m_n$ such that this condition holds. Assumption~\ref{assm:rate_choice_normality}(b) imposes a refined condition on the growth rate of $\lambda_{nj}$.

We now state the asymptotic normality result.
\begin{theorem}[Asymptotic normality]
    \label{thm:clt} Define $H_n(\bm x) := \sum_{j=1}^{m_n}\bm \beta(j/m_n)^\top\bm \phi(\bm x)f(B^{(j)}, \bm x)$. 
    Under Assumptions~\ref{assm:iid}--\ref{assm:rate_choice_normality}, we have that,
    \begin{align}
        \label{eqn:clt}
        \frac{\sqrt{n}(T(\hat{\bm w}_f) - \mu_f)}{\eta_n} \to \mathcal N\left(0, 1\right),
    \end{align}
    where
    \begin{align}
        \label{eqn:var_exprn}
        \eta_n^2 := {\rm Var}\left(H_n(\bm X_1)\right) + \mathbb E\left[\sigma^2(\bm X_1)\left\{\bm\phi(\bm X_1)^\top\bm c^{(n)}_{j(A_1)}\right\}^2\right],
    \end{align}
    with $j(a)$ denoting the index such that $a\in B^{(j(a))}$ for any $a$, and
    \[
        \bm c_j^{(n)} := \mathbb E\left[\bm \phi(\bm X_1)\bm \phi(\bm X_1)^\top e(B^{(j)}, \bm X_1)\right]^{-1}\mathbb E\left[\bm \phi(\bm X_1)f(B^{(j)}, \bm X_1)\right],
    \]
    is the population counterpart of $\hat{\bm c}^{(n)}_j$, which is defined in Proposition~\ref{lem:weights_characterization}.
\end{theorem}

Under Assumption~\ref{assm:signal_plus_noise}, $H_n(\bm x)$ is the conditional causal estimand $\mathbb E\left[\int Y_1(a)f(da, \bm X_1)\mid \bm X_1 = \bm x\right]$ with $\bm \beta(\cdot)$ replaced by its nearest-neighbor approximation $\tilde{\bm \beta}_n(\cdot)$. The proof of Theorem~\ref{thm:clt} hinges on the following representation (see Lemma~\ref{lem:clt_linear_representation} in the appendix):
\[
    \sqrt{n}\left(T(\hat{\bm w}_f) - \mu_f\right) = \frac{1}{\sqrt{n}}\sum_{i=1}^n \left\{H_n(\bm X_i) - \mathbb E[H_n(\bm X_1)]\right\} + \frac{1}{\sqrt{n}}\sum_{i=1}^n \sigma(\bm X_i) \epsilon_i\, \bm \phi(\bm X_i)^\top\bm c_{j(A_i)}^{(n)} + o_P(1),
\]
where $\bm \phi(\bm X_i)^\top\bm c_{j(A_i)}^{(n)}$ is the population analogue of the balancing weight $\hat w_{f,i} = \bm \phi(\bm X_i)^\top\hat{\bm c}_{j(A_i)}^{(n)}$. Since $\mathbb E[\epsilon_i \mid A_i, \bm X_i] = 0$, the two terms are uncorrelated, and the two components of $\eta_n^2$ in Equation~\eqref{eqn:var_exprn} are their respective variances. Under an additional restriction, $\eta_n^2$ has an asymptotic limit, which enables us to derive the following asymptotic variance expression.

\begin{prop}[Asymptotic variance]
    \label{lem:limiting_variance}
    Consider the setting and assumptions of Theorem~\ref{thm:clt}. Define a $p$-dimensional, finite, signed measure $\bm Q$ on $\mathcal A$ as follows,
    \[
        \bm Q(B) = \mathbb E\left[\bm \phi(\bm X_1)f(B, \bm X_1)\right],
    \]
    for any measurable set $B$ such that each coordinate of $\bm Q$ is a finite, real valued signed measure on $\mathcal A$. Then, $\bm Q$ is absolutely continuous. Now, suppose $\bm Q$ admits a continuous density $\bm q$, i.e., we have $\bm Q(B) = \int_B \bm q(a)da,\forall$ measurable sets $B$ and that the map $a\mapsto e(a,\bm x)$ is continuous on $\mathcal A$. Then, we have,
    \begin{equation}
    \label{eqn:var_exprn_limit}
    \begin{aligned}
         \sigma^2_{\rm bal} \ := \ \lim_{n\to \infty}\eta_n^2 \ = \ & {\rm Var}\left(\mathbb E\left[\int_{\mathcal A}Y_1(a)f(da,\bm X_1) \ \middle| \ \bm X_1\right]\right) \\
         & \quad + \int_{\mathcal A}\mathbb E\left[\sigma^2(\bm X_1)e(a,\bm X_1)\left\{\bm \phi(\bm X_1)^\top\bm M(a)^{-1}\bm q(a)\right\}^2\right]da.
    \end{aligned}
    \end{equation}
\end{prop}

\subsection{Efficiency gain over the IPW estimator with known propensity score}
\label{sec:ipw_estimator_comparison}

As a benchmark, we compare the limiting variance of our balancing estimator against that of the IPW estimator with the known propensity score. For analytical tractability, we consider a case where our counterfactual weight $f$ represents a stochastic intervention so that $f(da, \bm x) = f(a,\bm x)da$. In this case, the IPW estimator has the following simple expression,
\[
    T_{\rm IPW} = \frac{1}{n}\sum_{i=1}^n Y_i \frac{f(A_i, \bm X_i)}{e(A_i, \bm X_i)},
\]
where $w_{{\rm IPW},i} = f(A_i, \bm X_i)/e(A_i, \bm X_i)$. One can invoke the classic CLT to establish the asymptotic normality of $T_{\rm IPW}$, with asymptotic variance $\sigma^2_{\rm IPW}$. The following theorem compares $\sigma^2_{\rm bal}$ against $\sigma^2_{\rm IPW}$ under a homoskedastic error setting, and demonstrates the efficiency gain of the balancing estimator over the IPW estimator.
\begin{theorem}[Efficiency gain of balancing over IPW estimation]
    \label{thm:bal_eff_gain}
    Assume that $f$ is a density and there exists some $\sigma>0$ such that $\sigma^2(\bm x) = \sigma^2$, for all $\bm x$. Suppose that Assumptions~\ref{assm:iid}--\ref{assm:rate_choice_normality} hold, and that the map $a\mapsto e(a,\bm x)$ is continuous on $\mathcal A$ for every $\bm x$. Then, we have,
    \begin{align*}
        &\sigma^2_{\rm IPW} = \Var\left(m(A_1, \bm X_1)\frac{f(A_1, \bm X_1)}{e(A_1, \bm X_1)}\right) + \sigma^2\int_{\mathcal A}\mathbb E\left[\frac{f^2(a, \bm X_1)}{e(a, \bm X_1)}\right]da\\
        &\sigma^2_{\rm bal} = \Var\left(\mathbb E\left[m(A_1, \bm X_1)\frac{f(A_1, \bm X_1)}{e(A_1, \bm X_1)}\middle| \bm X_1\right]\right) + \sigma^2 \int_{\mathcal A}\mathbb E\left[\left\{\bm \Pi_a\left(\frac{f(a, \bm X_1)}{\sqrt{e(a, \bm X_1)}}\right)\right\}^2\right]da,
    \end{align*}
    where, 
    $$   \bm \Pi_a(l):= \bm \phi(\bm X_1)^\top\sqrt{e(a, \bm X_1)}\,\mathbb E\left[\bm \phi(\bm X_1)\bm \phi(\bm X_1)^\top e(a, \bm X_1)\right]^{-1}\mathbb E\left[\bm \phi(\bm X_1)\sqrt{e(a, \bm X_1)}\, l\right],
    $$ 
    is the theoretical projection operator onto $\bm \phi(\bm X_1)\sqrt{e(a, \bm X_1)}$. 
    Consequently, $\sigma^2_{\rm bal}\leq \sigma^2_{\rm IPW}$.
\end{theorem}

Theorem~\ref{thm:bal_eff_gain} shows that both of these variances can be decomposed into the sum of two components, where the signal strength (through the outcome regression function $m$) only affects the first component, while the noise variance only affects the second. The first component of $\sigma^2_{\rm bal}$ is provably smaller than that of $\sigma^2_{\rm IPW}$, due to the law of total variance and the fact that the first component of $\sigma^2_{\rm bal}$ in this case is given by $\Var\left(\mathbb E\left[m(A_1, \bm X_1)\frac{f(A_1, \bm X_1)}{e(A_1, \bm X_1)}\middle| \bm X_1\right]\right)$. This efficiency gain is due to the fact that the balancing estimator satisfies the balancing equation, while the IPW estimator satisfies it only in expectation. 

The second component of $\sigma^2_{\rm bal}$ is also smaller than the corresponding component of $\sigma^2_{\rm IPW}$. In fact, this difference becomes greater as the rank of the projection operator $\bm \Pi_a$ decreases. That is, if the (correct) specification of our bases is more compact, the reduction of the second component is greater for the balancing estimator. The result highlights the fact that even though the balancing estimator does not require the knowledge of the propensity score, it is more efficient than the IPW estimator with known propensity score due to its balancing property and knowledge of the basis functions.


\subsection{Consistent Variance Estimation}
\label{sec:var_estimation}

We now turn to the consistent estimation of the variance, which is necessary for constructing an asymptotically valid confidence interval. We first state the required regularity conditions.
\begin{ass}[Regularity conditions]
    \label{assm:variance_estimation}
    The following conditions hold:
    \begin{enumerate}[label = (\alph*)]
        \item There exist some $\zeta>2$ and some $\delta'>0$ such that $\sup_{a\in \mathcal A}\mathbb E\left[\left\lvert \epsilon_1(a)\right\rvert^{\zeta+\delta'}\right]<\infty$.
        \item $\frac{m_n^{1+2/\zeta}\log n}{n}\to 0$, as $n\to \infty$.
    \end{enumerate}
\end{ass}
Assumption~\ref{assm:variance_estimation}(a) imposes an additional moment restriction on the error term, while Condition~(b), together with Assumption~\ref{assm:rate_choice_consistency}(a)~and~\ref{assm:rate_choice_normality}(a) for a Lipschitz $\bm \beta(\cdot)$, implies that $m_n$ takes the form $n^\gamma$, where $0.5<\gamma < \frac{\zeta}{2+\zeta}$. More generally, the admissible values of $\gamma$ depend on the modulus of continuity $M$ of $\bm \beta(\cdot)$. Although Lipschitz continuity is stronger than the continuity assumed elsewhere, it is not particularly restrictive (see the discussion following Assumption~\ref{assm:rate_choice_normality}), and provides a simple, interpretable lower bound $\gamma>0.5$. Thus, for simplicity, we assume that $\bm \beta(\cdot)$ is Lipschitz whenever discussing the choice of $\gamma$. Moreover, by imposing stronger moment restrictions on $|\epsilon_i(a)|$ through a larger value of $\zeta$ in Assumption~\ref{assm:variance_estimation}(a), we can expand the admissible range of $\gamma$. In particular, if we assume that all moments of the error terms are finite, this range expands to $(0.5, 1)$. 

Finally, the following theorem establishes consistent variance estimation and, consequently, asymptotically valid confidence intervals for $\mu_f$. For a more compact presentation, we defer the expression of the variance estimator to Equation~\eqref{eqn:var_estimator} in Appendix~\ref{sec:proof_variance_estimation}.

\begin{theorem}[Consistent variance estimation]
\label{thm:variance_estimation}
    Suppose that Assumption~\ref{assm:variance_estimation} holds in addition to the assumptions of Theorem~\ref{thm:clt}. Let $\hat{\eta}^2_n$ be the variance estimator, as defined in Equation~\eqref{eqn:var_estimator}. Then, we have $\hat\eta_n^2-\eta_n^2=o_{\mathbb P}(1)$, as $n\to\infty$. Consequently, for any $\alpha\in (0,1)$, an asymptotic $100(1-\alpha)\%$ confidence interval for $\mu_f$ is given by
    \begin{align*}
        C_f^{(1-\alpha)} = \left(T(\hat{\bm w}_f)\pm z_{1-\alpha/2}\frac{\hat\eta_n}{\sqrt n}\right).
    \end{align*}
\end{theorem}

\subsection{Choice of $m_n$ and $\lambda_{nj}$}
\label{sec:choice_m_lambda}

As discussed above, Assumptions~\ref{assm:rate_choice_normality}(a)~and~\ref{assm:variance_estimation}(b) imply the choice of hyperparameter $m_n = n^\gamma$, $\gamma \in (0.5, 0.5 + \zeta)$. Even though a larger $\gamma$ implies a faster approximation of $\bm \beta(\cdot)$ through $\tilde{\bm \beta}_n(\cdot)$, we also face a trade-off; with a large $\gamma$, the interval is divided into a larger number of bins, leading to a smaller number of points within each bin and hence exacerbating finite sample bias. Our experiments presented in Appendix~\ref{sec:mn_choice} suggest that this latter effect is substantial. 
Indeed, we have found that the value of $\gamma$ closer to 0.5 attains the desired coverage faster. Our recommendation uses $\gamma = 0.51$, and hence, $m_n = n^{0.51}$ (this is plausible with the mildly stricter restriction $\zeta>2.09$ in Assumption~\ref{assm:variance_estimation}(a)).

For the choice of $\lambda_{nj}$, we first parameterize it as $\lambda_{nj} = h_j\lambda_n$ where $\lambda_n$ controls the rate of $\lambda_{nj}$ and $h_j$ scales it appropriately for the bin-wise ridge regression such that $\max_{1\leq j\leq m_n}h_j = O_{p}(1)$.  Under this parameterization, $\lambda_n$ does not depend on $j$, and so the rate of shrinkage of all the $\lambda_{nj}$ is the same. Then, Assumption~\ref{assm:rate_choice_normality}(b) implies the condition $\sqrt{n}m_n\lambda_n\to 0$, which leads to $\lambda_n = n^{-(1+c)(\gamma + 0.5)}$. Again, the choice of $c$ has a trade-off: a greater value of $c$ results in a smaller value of $\lambda_n$ with little regularization, leading to potentially unstable weights. On the other hand, a smaller value of $c$ can exacerbate finite sample biases. The results of our experiments, presented in Appendix~\ref{sec:lambda_choice}, suggest that $c = 1.5$ is a plausible choice. Thus, our recommendation is $m_n = n^{0.51}$ and $\lambda_{n} = n^{-2.525}$. Lastly, we choose
$$
\begin{aligned}
    h_j 
    = & \max\left\{\frac{m_n}{p}{\rm trace}(\hat{\bm M}_j),\ \frac{1}{np}\sum_{i=1}^n \|\bm \phi(\bm X_i)\|^2 \right\},
\end{aligned}
$$
which allows $\lambda_{nj}$ to scale for each individual bin-wise ridge regression problem.

\subsection{Assessing imbalance}
\label{sec:imbalance}

Although our regularized weights do not incur asymptotic bias, it is useful to understand their impact on finite-sample bias. In the balancing literature, weighting estimators are commonly evaluated by how well the weighted covariate distribution matches that of a target population \citep{zubuzaretta2015}. In this section, we develop a similar practical criterion for assessing covariate imbalance under our balancing weights.

Recall from Equation~\eqref{eqn:equivalent_balancing} that the balancing equation is a system of $m_n$ equations, with $\bm \Phi_j^\top\bm w_{S_j} = n\hat{\bm r}_j$, for each bin $j$, where $S_j$ denotes the set of indices $i$ such that $A_i \in B^{(j)}$. The imbalance incurred by the sub-vector $\bm w_{S_j}$ is therefore determined entirely by the $j^{\rm th}$ balancing equation. We first define covariate imbalance for each bin-wise equation and then discuss ways to aggregate it across bins. The imbalance for the $l^{\rm th}$ covariate in the $j^{\rm th}$ bin-wise balancing equation is
\[
    \nu_{l j}
    =
    \left\lvert\frac{1}{n}\sum_{i\in S_j} \phi_l(\bm X_i)\hat w_{f,i}
    - \left(\hat{\bm r}_j\right)_l \right\rvert
    =
    \left\lvert\frac{1}{n}\sum_{i=1}^n
    \left[
        \phi_l(\bm X_i)\hat w_{f,i}\mathbb I(i \in S_j)
        -
        \phi_l(\bm X_i)f(B^{(j)}, \bm X_i)
    \right]\right\rvert.
\]
This definition extends directly to any balancing estimator by replacing $\hat w_{f,i}$ with the corresponding weights.

To place this imbalance on a relative scale, note that the target for the $l^{\rm th}$ covariate in the $j^{\rm th}$ bin-wise balancing equation is
\[
    \left(\hat{\bm r}_j\right)_l
    =
    \frac{1}{n}\sum_{i=1}^n \hat r_{ij\cdot l},
    \qquad
    \hat r_{ij\cdot l}
    =
    \phi_l(\bm X_i)f(B^{(j)}, \bm X_i).
\]
We therefore compare $\nu_{lj}$ with the sample standard deviation of $\{\hat r_{ij\cdot l}\}_{i=1}^n$, denoted by $\sigma_{lj}$, and define the \textit{relative imbalance} of the $l^{\rm th}$ covariate in the $j^{\rm th}$ bin as
\[
    \nu^*_{l j} = \frac{\nu_{lj}}{\sigma_{lj}}.
\]

The quantity $\{\nu^*_{l j}\}_{l,j}$ describes the imbalance of each covariate across bins. A more compact summary with one relative imbalance measure per covariate may also be useful and is closer to the convention in the balancing literature. For the $l^{\rm th}$ covariate, we therefore recommend reporting a summary statistic of $\nu^*_{lj}$ across $j$. For example, one may report $\bar \nu^*_{l \cdot}= \frac{1}{m_n}\sum_{j=1}^{m_n}\nu^*_{lj}$. Following standard practice in causal inference, a practitioner would like the values of $\{\nu^*_{l j}\}_{l,j}$, or their per-covariate summaries such as $\bar \nu^*_{l \cdot}$, to be small. As a rule of thumb, we recommend a threshold of 10$\%$ for these quantities.

\section{Doubly-robust estimation}
\label{sec:outcome_modelling}

So far, we have required the knowledge of basis functions that span the outcome regressions. However, a balancing estimator can also be viewed as an estimator based on an outcome model \citep{chattopdhyayetal2023}.  We can adopt this perspective and write our balancing estimator as:
\begin{align}
\label{eqn:or_model}
    T(\hat{\bm w}_f) = \frac{1}{n}\sum_{i=1}^n Y_i\,\bm\phi(\bm X_i)^\top\hat{\bm c}^{(n)}_{j(A_i)} = \frac{1}{n}\sum_{i=1}^n\int_{\mathcal A}\bm\phi(\bm X_i)^\top\,\hat{\bm\beta}^{(n)}(a)\,f(a,\bm X_i)\,da,
\end{align}
where the fitted regression function is given by $\hat m(a, \bm x) = \bm \phi(\bm x)^\top\hat{\bm \beta}^{(n)}(a)$ with,
\[
    \hat{\bm\beta}^{(n)}(a):=(\hat{\bm M}_{j(a)}+\lambda_{n,j(a)}\bm I_p)^{-1}\hat{\bm q}_{j(a)},\quad \text{and} \quad 
    \hat{\bm q}_{j} = \frac{1}{n}\sum_{i=1}^n Y_i\,\bm\phi(\bm X_i)\,\mathbb{I}(A_i\in B^{(j)}).
\]
See Appendix~\ref{sec:or_model_proof} for a justification. Thus, one can interpret $\hat{\bm \beta}^{(n)}(a)$ as the ridge coefficient estimate with the regularizer $\lambda_{nj}$.  

Let $\hat{m}_{\rm bal}$ denote the estimator of $m$ based on the balancing equation. Then, because $\hat{m}_{\rm bal}$ is $\sqrt{n}$-consistent (see Section~\ref{sec:asymp_normality}), we can also construct a doubly-robust AIPW estimator based on the fitted outcome model $\hat{m}_{\rm bal}$  and the estimated propensity score $\hat{e}$ \citep{diaz2011}. The resulting AIPW estimator has the rate-robustness property. That is, the estimator converges at a product rate of the outcome and propensity score models. The following theorem states the asymptotic normality of this AIPW estimator and derives the rate condition.
\begin{theorem}[Doubly robust AIPW estimator based on the balancing estimator]
\label{thm:aipw_clt}
Fix an integer $K\ge 2$ and randomly partition the index set $\{1,\dots,n\}$ into $K$ folds $I_1,\dots,I_K$, and write
$I_k^c:=\{1,\dots,n\}\setminus I_k$. For each $k$, let $\widehat m_{(k)}$ and
$\widehat e_{(k)}$ be estimators of $m$ and $e$ constructed using only the
observations $\mathcal D_k = \{(Y_i,\bm X_i,A_i):i\in I_k^c\}$ such that $\hat e_{(k)}\geq c_e$, almost surely, for some constant $c_e>0$ that does not depend on $n$. Assume that the counterfactual weight $f$ is a probability density and consider the cross-fitted AIPW estimator
$$
\begin{aligned}
  \hat T^{\hat m_{\rm bal}, \hat e}_{\rm AIPW}
  \ = \ & \frac{1}{n}\sum_{k=1}^{K}\sum_{i\in I_k}
       \varphi\!\left(A_i,\bm X_i;\,\widehat m_{(k)},\widehat e_{(k)}\right)\\
    \ = \ & \frac{1}{n}\sum_{k=1}^{K}\sum_{i\in I_k}\left[\,\int_{\mathcal A}\widehat m_{(k)}(a,\bm X_i)\,f(a,\bm X_i)\,da
  \;+\;\frac{f(A_i,\bm X_i)}{\widehat e_{(k)}(A_i,\bm X_i)}
       \bigl(Y_i-\widehat m_{(k)}(A_i,\bm X_i)\bigr)\right],
\end{aligned}
$$
where, for each $k=1,\dots,K$, the fold-wise nuisance estimators satisfy
$\lVert\widehat m_{(k)}-m\rVert=o_{\mathbb P}(1)$,
$\lVert\widehat e_{(k)}-e\rVert=o_{\mathbb P}(1)$, and
$\lVert\widehat m_{(k)}-m\rVert\cdot\lVert\widehat e_{(k)}-e\rVert=o_{\mathbb P}(n^{-1/2})$, where note that $\lVert\widehat m_{(k)}-m\rVert:= \mathbb E\left[\left(\widehat m_{(k)}(A_i,\bm X_i)-m(A_i, \bm X_i)\right)^2\mid \mathcal D_k\right]^{1/2}$ for $i\in I_k$, and  $\lVert\widehat e_{(k)}-e\rVert$ is also defined similarly.

Then, under Assumptions~\ref{assm:iid}--\ref{assm:unconfound}, \ref{assm:asymp_regularity}(b), if $\mathbb{E}[\varphi^2(A_1,\bm X_1;m,e)]<\infty$, we have
\begin{equation*}
  \sqrt{n}\left(\hat T^{\hat m_{\rm bal}, \hat e}_{\rm AIPW}-\mu_f\right)
  \;\xrightarrow{\;d\;}\;\mathcal N\!\left(0,\ \sigma^2_{\mathrm{eff}}\right),
\end{equation*}
where $\sigma^2_{\mathrm{eff}}=\operatorname{Var}\!\big(\varphi(A_1,\bm X_1;m,e)\big)=\mathbb{E}[\varphi^2(A_1,\bm X_1;m,e)]-\mu_f^2$.
\end{theorem}

Because $T(\hat{\bm w}_f)$ is $\sqrt{n}$-consistent, using it for the outcome regression model yields $\hat T^{\hat m_{\rm bal}, \hat e}_{\rm AIPW}$ asymptotically normal as long as $\hat e$ is consistent, that is, $\|\hat e - e\| = o_{\mathbb P}(1)$. Being completely agnostic to the propensity score model, one can, for example, use kernel-density estimation to estimate $e$.

When the choice of basis functions is correct, however, the AIPW estimator can be less efficient than the balancing estimator, as the following theorem shows.
\begin{theorem}[Efficiency gain over the AIPW estimator]
\label{thm:eff_dominance}
    Under the settings and assumptions of Theorem~\ref{thm:clt} and homoskedasticity, i.e., $\sigma^2(\bm x) = \sigma^2,\forall \bm x$, the asymptotic variance of the AIPW estimator is given by,
    \[
        \sigma^2_{\rm eff} =  {\rm Var}\left(\mathbb E\left[\int_{\mathcal A}Y_1(a)f(a,\bm X_1)da\ \middle| \ \bm X_1\right]\right)  + \sigma^2\int_{\mathcal A}\mathbb E\left[\frac{f^2(a, \bm X_1)}{e(a, \bm X_1)}\right]da.
    \]
    This implies $\sigma^2_{\rm bal}\leq \sigma^2_{\rm eff}$.
\end{theorem}
This result, however, comes as no surprise. The AIPW estimator is designed to offer protection against possible model misspecification. In particular, the augmentation term $f(A_i,\bm X_i)/\hat e(A_i, \bm X_i)\cdot (Y_i - \hat m(A_i, \bm X_i))$  protects against the mis-specification of the outcome model. Thus, even if the AIPW estimator is supplied with a correct outcome model derived from the correct specification of the bases, it does not utilize this information about the correct outcome model. The augmentation only reintroduces back the IPW-weighted residual variance, which was removed by the balancing estimator. This is also demonstrated by the fact that the second term of $\sigma^2_{\rm eff}$ exactly matches the second term of $\sigma^2_{\rm IPW}$.

\section{Simulation studies}
\label{sec:expt_density}

We consider the following simulation setup. Assume we have $n$ i.i.d. observations $\{(Y_i, \bm X_i, A_i)\}_{i=1}^n$, where $Y_i = Y_i(A_i)$. Here, $\bm X_i$ is a $d$-dimensional covariate comprised of $d$ independent $U(0,1)$ draws, 
and $Y_i(a)$ follows the model:
\begin{align*}
    Y_i(a)\sim \mathcal N\left(\bm X_i^\top\bm \beta(a),1\right),
\end{align*}
where $\bm \beta(a)\in \mathbb R^d$ is defined as follows:
\begin{align*}
    \beta_1(a) = \sin a~\textrm{and}~\beta_{1+j}(a) = \alpha_{\rm amp}\cdot \frac{a^j}{\sqrt{\sum_{j=1}^{d-1}a^{2j}}}, 1\leq j \leq d-1,
\end{align*}
where the amplitude parameter $\alpha_{\rm amp}$ controls the signal size. The above choice of $\bm \beta$ represents a highly non-linear dose-response relation. 

To specify the treatment assignment mechanism, we first consider the following class of densities. For any $c$, define $\gamma_c:=\gamma_c(\bm x) = 0.1 + 0.8\cdot{\rm sig}(2\cdot \bm 1^\top\bm x - d+c)$, where ${\rm sig}()$ denotes the sigmoidal transformation and consider the distribution
\[
    \mu_c = \alpha_{\rm elv}U(0,1) + (1-\alpha_{\rm elv})\Delta(\gamma_c, 0,1),
\]
where $\Delta(\gamma_c,0,1)$ represents the triangle distribution over $[0,1]$ with mode at $\gamma_c$. We use $\alpha_{\rm elv} = 0.2$.  Then, $\mu_c$ represents the triangular distribution on $[0,1]$, but whose density function has been \textit{elevated} by $\alpha_{\rm elv}$ everywhere. 
This ensures the strong positivity condition in Assumption~\ref{assm:pos}. The treatment is generated as $A_i\sim \mu_0\mid \bm X_i$. 

In our experiments, we consider four counterfactual stochastic interventions based on the degree to which each intervention deviates from the propensity score: (low) $\mu_{0.05}$, (moderate) $\mu_{1.5}$, and (high) $\mu_{3}$. In this section, we present the results for $d=3$, while in Appendix~\ref{sec:further_simul_dens} contains additional results corresponding to $d=5$ and $d=10$. In each of these settings, we calibrate $\alpha_{\rm amp}$ so that the two components of the expression of $\sigma^2_{\rm IPW}$ discussed in Theorem~\ref{thm:bal_eff_gain} are the same; this ensures that the two components affected by the signal and the noise variance are equal placing us in a medium signal-to-noise ratio regime. The specific values of $\alpha_{\rm amp}$ are mentioned below for each of the settings.

We compare the 95\% confidence intervals (CIs) based on the balancing estimator against the following: (1) The CI based on the IPW estimator with the knowledge of the true propensity score, and (2) The AIPW estimator, which uses our balancing estimator for outcome regression as discussed in Section~\ref{sec:outcome_modelling} (with $k=5$ folds for cross-fitting), and the correctly specified propensity score. We also evaluate the performance of the balancing estimator by comparing its mean squared error (MSE) against the following methods: (1) the IPW, (2) the AIPW estimators discussed above, (3) the B-OPE method of \cite{sondhietal2020} (referred to hereon as `B-OPE') discussed in Section~\ref{sec:literature}. Further details of implementation are available in Section~\ref{sec:implementation_details}. 

\begin{figure}[!t]
    \centering\includegraphics{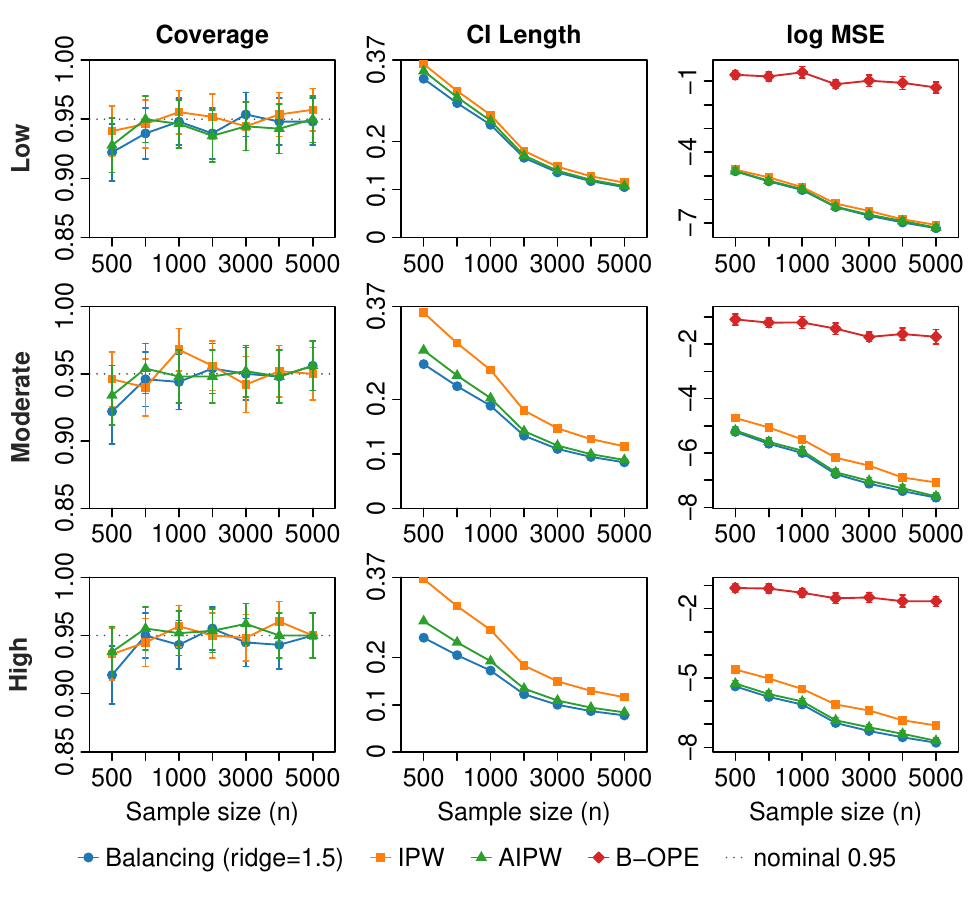}
    \caption{Empirical performance of the estimators across different sample sizes. The leftmost and center columns summarize the coverage and the length of the various confidence intervals while the rightmost column summarizes the MSE of the various point estimators. For `low', `moderate' and `high' settings, $\alpha_{\rm amp}$ take values 6.9, 3.7 and 2.7, respectively. All error bars represent plus or minus two standard errors.}
    \label{fig:main_density}
\end{figure}

Figure~\ref{fig:main_density} presents the results. 
The left panel shows that as the the sample size ($x$-axis) increases, all confidence intervals attain the $95\%$ nominal coverage.  The middle column shows that the CI of our balancing estimator has a shorter length than both its AIPW-based counterpart (with the true propensity score) and the CI of the IPW estimator. In fact, when the sample size is large and hence asymptotic approximation is better, the CI based on the balancing estimator is about 7.5$\%$, 20.9$\%$, and 26.6$\%$ shorter than the IPW-CI under the `low', `moderate' and `high' settings, respectively. In comparison with the AIPW-CI, the CI based on the balancing estimator is about $4.0\%$, $7.2\%$ and $10.4\%$ shorter. As discussed in Sections~\ref{sec:ipw_estimator_comparison}~and~\ref{sec:outcome_modelling}, these empirical results are consistent with the fact that the balancing estimator is the most efficient, followed by the AIPW and the IPW estimators, respectively.

Finally, the rightmost panels compare the performance of these estimators measured by their mean-squared error. We find that the B-OPE method has a significantly higher MSE than the other methods. This is at least in part because, unlike the other methods, the B-OPE method neither uses the knowledge about the true propensity score nor the set of correct basis functions. 

Among the other methods, averaged across all the sample sizes $n$, the balancing estimator has 9.6$\%$, 33.4$\%$ and 41.8$\%$ smaller MSE than the IPW estimator in the `low', `moderate' and `high' settings, respectively. The degree of improvement is less when compared against the AIPW estimator; $5.4\%$, 11.3$\%$ and 16.5$\%$, respectively. This confirms our theoretical result that the balancing estimator outperforms the other methods.

\section{Application to the Job Corps Study}
\label{sec:empirical}

\subsection{Data}
\label{sec:data}

The Job Corps program was the largest residential career training and education program in the U.S. for low-income youths aged 16 through 24 years. Our analysis in this section concerns a study conducted between November 1994 and February 1996 to evaluate the effectiveness of this program. Many scholars have analyzed related datasets, using the length of time exposed to training during the first year, which was measured in hours, as a continuous treatment to estimate its causal effects on various outcomes \citep{floresetal2012,lee2018,colangeloandlee2025,hsuetal2026}. 

For the individuals who enrolled in the Job Corps program, there may be selection bias.  In particular, the enrollees self-selected the hours of training by carrying out training at their own pace \citep[][Section D]{floresetal2012}.  Consequently, causal effect estimation for continuous dose must rely on tools for confounder adjustment as our balancing method does.
Following \cite{hsuetal2026}, we estimate the average causal effect of vocational training hours on weekly earnings in the fourth year.

Our dataset consists of 4024 observations and 65 covariates. In the full dataset, the treatment variable $A$ ranges from 0 to over 5000 hours. However, many authors have restricted attention to observations with $A\geq 40$ hours \citep{floresetal2012,colangeloandlee2025,hsuetal2026}, while \cite{hsuetal2026} notes that the number of observations with $A>3000$ is small. Hence, we consider only observations for which $A$ lies between 40 and 3000, reducing the dataset size to 3669. Finally, we transform the treatment to $(A - 40)/(3000-40)$ so that its range is $[0,1]$.

\subsection{Analysis setup}
\label{sec:empirical_analysis}

We estimate the causal effect of increasing vocational training hours on expected earnings. Specifically, for a given threshold $g$, we consider the difference in expected potential earnings under two counterfactual training-hour distributions: one uniformly assigning fewer than $g$ hours and the other uniformly assigning more than $g$ hours. This corresponds to a counterfactual weight given by the difference between the densities of the $U(g,2000)$ and $U(40,g)$ distributions. 
We cap the former distribution at 2000 rather than 3000 hours because the number of individuals receiving more than 2000 hours of training declines sharply, with only 13$\%$ of the sample exceeding this threshold. 

An additional motivation for this choice of counterfactual distributions is that, because they do not depend on individual covariates, they eliminate the dependence of assigned training hours on covariates arising from self-selection. In contrast, prior analyses have focused on estimating dose-response curves \citep{colangeloandlee2025,hsuetal2026}.

Rather than expressing $g$ on the original scale, we standardize it using the same transformation that maps the treatment range to $[0,1]$, namely, $(g - 40)/(3000-40)$. Thus, the counterfactual weight of interest is
\begin{align*}
    f_g(a) =  \frac{\mathbb I(g\leq a \leq \frac{1960}{2960})}{\frac{1960}{2960} - g} - \frac{\mathbb I(0\leq a \leq g)}{g},~\forall a\in [0,1].
\end{align*}
For the choice of $g$, we consider the following equally spaced grid of training hours on the original scale: $\mathcal G = \{400, 800, 1200, 1600\}$.

\subsection{Findings}
\label{sec:empirical_findings}

\begin{figure}[t!]
    \centering \includegraphics{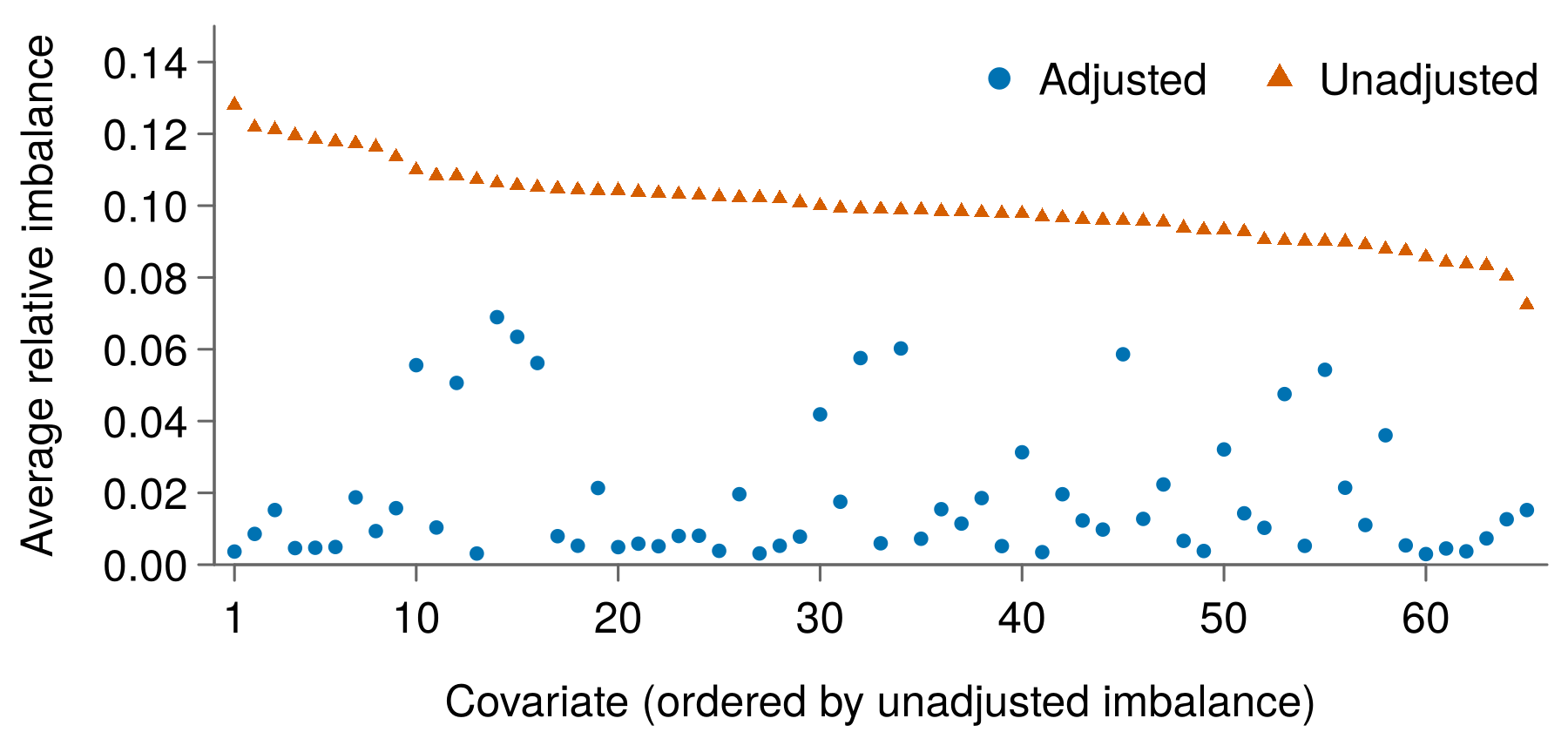}
    \caption{Box plots of the average relative imbalance across the adjusted (using balancing weights with full covariates) and unadjusted (using intercept-only model balancing weights) covariates.}
    \label{fig:imbalance_job_corps}
\end{figure}

Figure~\ref{fig:imbalance_job_corps} presents the average relative imbalance across bins for each covariate, as discussed in Section~\ref{sec:imbalance}. For comparison, we also report the corresponding imbalance without covariate adjustment. Specifically, for a given $g$, we use the following weights while evaluating the imbalance without adjustment:
\[
    w_i^{\rm(unadj)} = \frac{n\cdot f_g(B^{(j(A_i))})}{n_{j(A_i)}},
\]
where $j(A_i)$ denotes the bin containing $A_i$, $n_{j(A_i)}$ is the number of points in the bin, and $f_g(B^{(j(A_i))})$ represents the mass assigned to the bin by the counterfactual weight, which in this application does not depend on individual covariates. Equivalently, $\left\{w_i^{\rm(unadj)}\right\}_{i=1}^n$ are the balancing weights obtained under an intercept-only model without covariates.

Note that because $f_g$ does not depend on the covariates and its support does not depend on $g$, changing $g$ only multiplies each $\hat{\bm r}_j$ (defined in Proposition~\ref{lem:bj_intervals}(b)) by a bin-specific factor. Since $\lambda_{nj}$ does not depend on $f_g$ (Section~\ref{sec:choice_m_lambda}), Proposition~\ref{lem:weights_characterization} implies that the binwise balancing weights $\hat{\bm w}_{f_g,S_j}$ are scaled by the same factor. Consequently, the binwise and averaged relative imbalances of every covariate do not change across $g$. The same holds for the unadjusted weights $w^{\rm (unadj)}_i$, which are proportional to $f_g(B^{(j(A_i))})$ for all $i$. Figure~\ref{fig:imbalance_job_corps} therefore presents a single imbalance plot that is common to all $g \in \mathcal G$.

\begin{figure}[!t]
    \centering\includegraphics{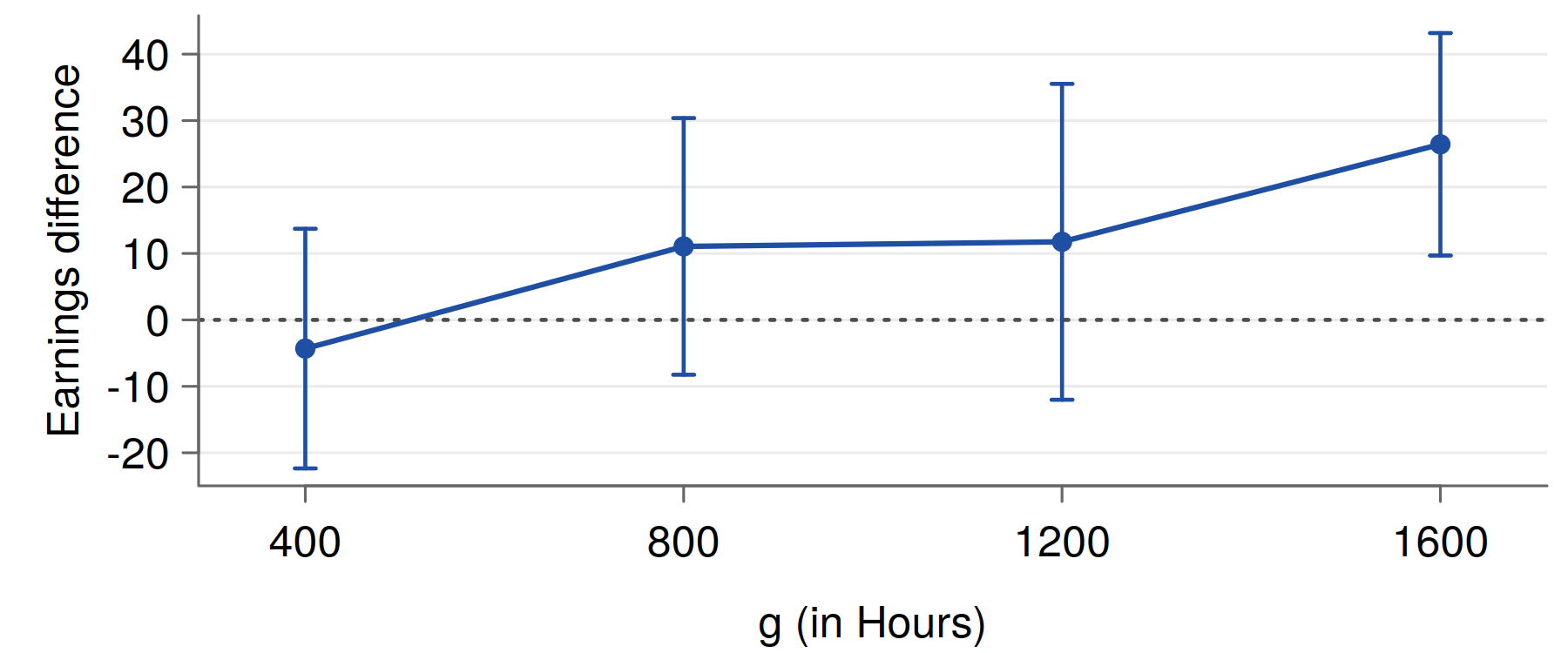}
    \caption{Estimated difference in expected potential earnings between the counterfactual distributions $U(g,\frac{1960}{2960})$ and $U(0,g)$, for various values of $g$ shown on the x-axis. The error bars represent plus or minus two standard errors.}
    \label{fig:job_corps_earnings}
\end{figure}

Figure~\ref{fig:job_corps_earnings} presents the estimated causal effects for different values of $g$, representing the difference in expected potential earnings between two counterfactual distributions assigning more or fewer than $g$ hours of training. We find that the estimated difference in earnings is generally non-decreasing in $g$. In particular, for the largest threshold considered, which corresponds to 1600 hours, the entire error bar lies above zero, indicating towards a statistically significant positive effect. Thus, our results provide evidence that assigning individuals more than 1600 hours of training, rather than fewer than 1600 hours, leads to higher expected earnings.

\section{Conclusion}
\label{sec:conclusions}

In this work, we developed a covariate balancing estimator for a general class of causal estimands under continuous treatment. Because it does not require estimating the propensity score, the proposed estimator is well suited to observational studies and is more efficient than the IPW estimator with the known propensity score. Our theory also establishes asymptotic inference for this estimator, and we demonstrate their applicability through simulations and a real-data application.

The key idea behind the proposed methodology is to exploit the continuity of causal effect in the treatment dose. This continuity assumption enables us to replace the causal effect with a discretized proxy defined on a grid of $m_n$ points in the treatment range. Conceptually, this work relates to the balancing estimator of \cite{sengupta2025lowrankcovariatebalancingestimators} that exploits a low-rank structure among the outcome regression functions in the presence of interference. Here, the low-dimensional structure is the grid approximation enabled by the continuity assumption.

The choice of $m_n$ involves a tradeoff. A smaller $m_n$ relative to $n$ gives more replicates for each of the $m_n$ distinct values of the proxy, but a coarser grid incurs a larger finite-sample approximation bias. Conversely, as $m_n\to \infty$ at a sufficient rate, the proxy approaches the true causal effect and the asymptotic bias vanishes. Our theory shows that $m_n$ can be chosen to grow slowly enough to retain sufficient replicates per grid point, yet fast enough that the discretization introduces no asymptotic bias. 

Several directions remain open for future work. Like covariate balancing methods in general, our method requires the practitioner to correctly specify a set of basis functions for the resulting estimator to be unbiased. Recent works relax this requirement substantially by assuming only that the outcome regression functions are smooth enough to lie in a reproducing kernel Hilbert space \citep{hulingandmak2024,huling2023,santraetal2026}. It would be interesting to incorporate such an approach under our framework. 

Next, although we consider one-dimensional continuous treatment, a similar methodology applies to multidimensional continuous treatments. Such extensions, however, would require a finer discretization with a larger $m_n$, and hence a much larger sample size $n$ for the asymptotic approximations to be accurate. This motivates studying whether a grid other than an equispaced one could speed up convergence or improve other aspects of the estimator, such as its efficiency.

\bibliography{refs}

\appendix

\section{Equivalence between the ridge penalty and the imbalance}
\label{sec:ridge_imbalance_euqivalence}

In Section~\ref{sec:optimal_weight}, we propose a ridge-regularized optimization problem for constructing balancing weights. An alternative, commonly used approach in the covariate balancing literature is to minimize the dispersion of the weights subject to a prespecified tolerance for covariate imbalance, rather than penalizing imbalance directly in the objective function. Since our balancing equations decompose into a system of binwise equations, this approach yields the following optimization problem for each bin $j$:
\begin{align}
\label{eqn:with_imbalance}
\begin{aligned}
    &\arg\min_{\bm w_{S_j}}~\frac{1}{2}\|\bm w_{S_j}\|^2\\
    &\text{subject to}~\|\bm \Phi_j^T\bm w_{S_j} - n\hat{\bm r}_j\|^2\leq \delta_j^2,
\end{aligned}
\qquad j=1,\ldots,m_n,
\end{align}
where $\{\delta_j\}_{j=1}^{m_n}$ are prespecified nonnegative thresholds controlling the allowable imbalance within each bin. We next establish that, for an appropriate choice of these thresholds, the solutions to~\eqref{eqn:with_imbalance} coincide with our proposed ridge-regularized balancing weights in Equation~\eqref{eqn:final_optimization}.

\begin{lemma}
\label{lem:ridge_imbalance_equiv}
Fix any bin $j$ and let $\lambda_{nj}>0$. Denote by $\hat{\bm w}^{\rm ridge}_{S_j}(\lambda_{nj})$ the unique solution to Equation~\eqref{eqn:final_optimization}, and by $\hat{\bm w}^{\rm imbal}_{S_j}(\delta_j)$ the unique solution to Equation~\eqref{eqn:with_imbalance}, whenever the latter is feasible. Define
\[
    \delta_j(\lambda_{nj})
    =
    n\lambda_{nj}
    \left\|
    (\hat{\bm M}_j+\lambda_{nj}\bm I_p)^{-1}
    \hat{\bm r}_j
    \right\|.
\]
Then, for every $\lambda_{nj}>0$, \eqref{eqn:with_imbalance} is feasible with $\delta_j = \delta_j(\lambda_{nj})$, and
\[
    \hat{\bm w}^{\rm ridge}_{S_j}(\lambda_{nj})
    =
    \hat{\bm w}^{\rm imbal}_{S_j}(\delta_j(\lambda_{nj})).
\]

Furthermore, let $\hat{\bm r}_j^{\perp}$ denote the projection of $\hat{\bm r}_j$ onto the null space of $\bm\Phi_j$. Then, the constrained formulation in \eqref{eqn:with_imbalance} is feasible if and only if
\[
    \delta_j\geq\delta_{\min}:=n\|\hat{\bm r}_j^{\perp}\|.
\]
For every feasible $\delta_j$, the constrained solution coincides with a ridge solution for some $\lambda_{nj}>0$ if $\delta_{\min}<\delta_j<n\|\hat{\bm r}_j\|$, or with its limit as $\lambda_{nj}\to0^+$ if $\delta_j=\delta_{\min}$, or as $\lambda_{nj}\to\infty$ if $\delta_j\geq n\|\hat{\bm r}_j\|$. In the latter case, $\hat{\bm w}_{S_j}^{\rm imbal}(\delta_j) = \bm 0$.
\end{lemma}

Lemma~\ref{lem:ridge_imbalance_equiv} establishes a correspondence between the ridge parameter $\lambda_{nj}$ and the allowable degree of covariate imbalance $\delta_j$, under which the two formulations yield identical balancing weights. In particular, the ridge parameter implicitly determines the imbalance threshold, and conversely, every feasible imbalance threshold yields a solution obtainable from the ridge formulation, including its limiting cases.

We adopt the ridge formulation throughout this paper because, unlike the constrained formulation with arbitrarily prespecified imbalance thresholds, it is always feasible and admits a unique solution. Moreover, as established above, it yields the same balancing weights as the constrained formulation for an appropriate choice of the imbalance threshold which results in a feasible Equation~\eqref{eqn:with_imbalance}.

Finally, the equivalence in Lemma~\ref{lem:ridge_imbalance_equiv} relies on measuring imbalance using the Euclidean norm. Alternative measures of imbalance lead to different penalized formulations through Lagrangian duality. For instance, constraining the $\ell_1$ norm of the imbalance vector yields a formulation with an $\ell_1$ penalty on imbalance. Similarly, imposing separate imbalance thresholds for individual covariates leads, under suitable constraint qualifications, to a formulation with coordinate-specific penalties. These alternatives offer additional flexibility in constructing balancing weights, although we focus on the squared Euclidean penalty in Section~\ref{sec:optimal_weight}.
\section{Proofs}
\subsection{Proof of Proposition~\ref{lem:beta_star_expressions}}
\label{sec:app_proof_lem_beta_star_expressions}
\begin{proof}
Note that we have,
\begin{align*}
    \mu_f &= \mathbb E\left[\int Y_i(a)f(da,\bm X_i)\right]\\
        &= \mathbb E\left[\int \bm \phi(\bm X_i)^T\bm \beta(a)f(da, \bm X_i)\right] + \mathbb E\left[\int \sigma(\bm X_i)\epsilon_i(a)f(da, \bm X_i)\right]~[\textrm{Assumption~\ref{assm:signal_plus_noise}}]\\
\end{align*}
Let $f=f^+-f^-$ be the Jordan decomposition of the signed measure and $|f|=f^++f^-$. By Assumption~\ref{assm:signal_plus_noise}, $\{\epsilon_i(a)\}_{a\in\mathcal A}$ is independent of $\bm X_i$ with $\mathbb E[|\epsilon_i(a)|]\le\{\mathbb E[\epsilon_i^2(a)]\}^{1/2}=1$. Tonelli's theorem then gives
\[
\mathbb E\left[\int\sigma(\bm X_i)|\epsilon_i(a)|\,|f|(da,\bm X_i)\right]
=\mathbb E\left[\sigma(\bm X_i)\int\mathbb E|\epsilon_i(a)|\,|f|(da,\bm X_i)\right]
\le\mathbb E\left[\sigma(\bm X_i)\,|f|(\mathcal A,\bm X_i)\right]<\infty .
\]
Hence Fubini's theorem applies to $f^+$ and $f^-$ separately, and
\[
\mathbb E\left[\int\sigma(\bm X_i)\epsilon_i(a)f(da,\bm X_i)\right]
=\mathbb E\left[\sigma(\bm X_i)\int\mathbb E[\epsilon_i(a)]\,f(da,\bm X_i)\right]=0 .
\]
Thus, it follows that,
\begin{align*}
        \mu_f=\frac{1}{n}\mathbb E\left[\int \sum_{i=1}^n\bm \phi(\bm X_i)^T\bm \beta(a)f(da, \bm X_i)\right]~\textrm{[Assumption~\ref{assm:iid}]}
\end{align*}
Thus, replacing $\bm \beta(a)$ with $\bm\beta^*_n(a)$ yields,
\begin{align*}
    \mu_f^* &=\frac{1}{n}\mathbb E\left[\int \sum_{i=1}^n\bm \phi(\bm X_i)^T\bm \beta^*_n(a)f(da, \bm X_i)\right]\\
    &= \frac{1}{n}\mathbb E\left[\int \sum_{i=1}^n\bm \phi(\bm X_i)^T\bm v(a)^T\otimes\bm I_df(da, \bm X_i)\right]\bm B_n\\
    &=\frac{1}{n}\mathbb E\left[\int \sum_{i=1}^n\bm v(a)\otimes \bm \phi(\bm X_i)f(da, \bm X_i)\right]^T\bm B_n\\
    &=\frac{1}{n}\mathbb E\left[\sum_{i=1}^n\left(\int \bm v(a)f(da, \bm X_i)\right)\otimes \bm \phi(\bm X_i)\right]^T\bm B_n\\
    &=\frac{1}{n}\mathbb E\left[\sum_{i=1}^n\bm \nu (f, \bm X_i)\otimes \bm \phi(\bm X_i)\right]^T\bm B_n
\end{align*}
This shows that $\tilde\mu_f=\mathbb E\big[\hat{\tilde\mu}_f\big]$.

We now turn to $\tilde{\mathbb E}[T(\bm w)]$. Throughout, the weights $\bm w=(w_1,\dots,w_n)^\top$ may depend on the data only through $\{(A_k,\bm X_k)\}_{k=1}^n$. We assume $\mathbb E[w_i^2\sigma^2(\bm X_i)]<\infty$ and $\mathbb E\big[|w_i|\max_{j}|\bm\phi(\bm X_i)^T\bm\beta(j/m_n)|\big]<\infty$ for every $i$. Since $Y_i=Y_i(A_i)$, Assumption~\ref{assm:signal_plus_noise} gives
\begin{align*}
\mathbb E[T(\bm w)]
=\frac1n\sum_{i=1}^n\mathbb E[w_iY_i]
=\frac1n\sum_{i=1}^n\mathbb E\left[w_i\bm\phi(\bm X_i)^T\bm\beta(A_i)\right]
+\frac1n\sum_{i=1}^n\mathbb E\left[w_i\eta_i(A_i)\right],
\end{align*}
where $\eta_i(a):=\sigma(\bm X_i)\epsilon_i(a)=Y_i(a)-m(a,\bm X_i)$. We show that every term of the second sum is zero. Let $\mathcal D:=\sigma\big(\{(A_k,\bm X_k)\}_{k=1}^n\big)$. By Assumption~\ref{assm:iid}, $\big(\{\eta_i(a)\}_{a\in\mathcal A},A_i,\bm X_i\big)$ is independent of $\{(A_k,\bm X_k)\}_{k\neq i}$. Each $\eta_i(a)$ is a function of $(Y_i(a),\bm X_i)$, so Assumption~\ref{assm:unconfound} gives $\{\eta_i(a)\}_{a\in\mathcal A}\ind A_i\mid\bm X_i$. Therefore
\begin{align*}
\mathbb E\left[\eta_i(A_i)\mid\mathcal D\right]
=\mathbb E\left[\eta_i(a)\mid\bm X_i\right]\Big|_{a=A_i}
=\sigma(\bm X_i)\,\mathbb E\left[\epsilon_i(a)\mid\bm X_i\right]\Big|_{a=A_i}=0,
\qquad
\mathbb E\left[\eta_i^2(A_i)\mid\mathcal D\right]=\sigma^2(\bm X_i).
\end{align*}
Since $w_i$ is $\mathcal D$-measurable, conditional Jensen and Cauchy--Schwarz give
$\mathbb E|w_i\eta_i(A_i)|=\mathbb E\big[|w_i|\,\mathbb E[|\eta_i(A_i)|\mid\mathcal D]\big]\le\mathbb E[|w_i|\sigma(\bm X_i)]\le\sqrt{\mathbb E[w_i^2\sigma^2(\bm X_i)]}<\infty$. The tower property then yields $\mathbb E[w_i\eta_i(A_i)]=\mathbb E\big[w_i\,\mathbb E[\eta_i(A_i)\mid\mathcal D]\big]=0$. Hence
\begin{align*}
\mathbb E[T(\bm w)]=\mathbb E\left[\frac1n\sum_{i=1}^n w_i\bm\phi(\bm X_i)^T\bm\beta(A_i)\right].
\end{align*}
Replacing $\bm\beta(\cdot)$ by $\tilde{\bm\beta}_n(\cdot)$ and using \eqref{eqn:knn_estimation},
\begin{align*}
\tilde{\mathbb E}[T(\bm w)]
&=\mathbb E\left[\frac1n\sum_{i=1}^n w_i\bm\phi(\bm X_i)^T\tilde{\bm\beta}_n(A_i)\right]
=\mathbb E\left[\frac1n\sum_{i=1}^n w_i\bm\phi(\bm X_i)^T\big(\bm v(A_i)^T\otimes\bm I_p\big)\bm B_n\right]\\
&=\mathbb E\left[\frac1n\sum_{i=1}^n w_i\big(\bm v(A_i)\otimes\bm\phi(\bm X_i)\big)^T\bm B_n\right]
=\mathbb E\left[\frac{\bm w^\top\bm\Phi\bm B_n}{n}\right]
=\mathbb E\big[\hat{\tilde T}(\bm w)\big].
\end{align*}
In the third equality we used the mixed-product rule: $(\bm v^T\otimes\bm I_p)^T\bm\phi=(\bm v\otimes\bm I_p)(1\otimes\bm\phi)=\bm v\otimes\bm\phi$. In the fourth, we used that the $i$-th row of $\bm\Phi$ is $(\bm v(A_i)\otimes\bm\phi(\bm X_i))^T$. All of these expectations are finite, because $|w_i\bm\phi(\bm X_i)^T\tilde{\bm\beta}_n(A_i)|\le|w_i|\max_j|\bm\phi(\bm X_i)^T\bm\beta(j/m_n)|$. This completes the proof.
\end{proof}
\subsection{Proof of Proposition~\ref{lem:bj_intervals}}
\begin{proof}
Let $\bm e_j$ denote the $j^{\mathrm{th}}$ canonical basis vector of $\mathbb R^{m_n}$, so that $\bm v^{(j)}=\bm e_j$ by construction.

\emph{Part (a).} For $j=1$, if $a\in B^{(1)}$, i.e. $a< \frac{3}{2m_n}$, then $|a-1/m_n|\leq|a-j/m_n|$ for all $2\leq j\leq m_n$, so $1/m_n$ is a nearest grid point and $\bm v(a)=\bm v^{(1)}$. Next, for $2\leq j\leq m_n-1$, let $a\in B^{(j)}=[\frac{2j-1}{2m_n},\frac{2j+1}{2m_n})$. If $a$ strictly lies between the midpoints $\frac{j-1/2}{m_n}$ and $\frac{j+1/2}{m_n}$, then note that $\left\lvert a - \frac{j}{m_n}\right\rvert<\frac{1}{2m_n}$, while $\left\lvert a - \frac{j-1}{m_n}\right\rvert > \left\lvert \frac{j-1/2}{m_n} - \frac{j-1}{m_n}\right\rvert = \frac{1}{2m_n}$ and $\left\lvert a - \frac{j+1}{m_n}\right\rvert > \left\lvert \frac{j+1/2}{m_n} - \frac{j+1}{m_n}\right\rvert = \frac{1}{2m_n}$, so that $j/m_n$ is a nearest grid point. In case $a = \frac{j-1/2}{m_n}$, then its nearest grid point is a tie between $\frac{j-1}{m_n}$ and $\frac{j}{m_n}$, which the convention of resolving ties to the larger grid point assigns to $j/m_n$. Hence $\bm v(a)=\bm v^{(j)}$ throughout $B^{(j)}$. Finally, if $a\in B^{(m_n)}$ then $a$ is closer to $1$ than to any other grid point (its left endpoint being a tie assigned to $1$), so $\bm v(a)=\bm v^{(m_n)}$. This proves part (a).

\emph{Part (b).} Since $\{B^{(j)}\}$ partition $[0,1]$, part (a) gives, for each $i$,
\[
\bm \nu (f,\bm X_i)=\int_{\mathcal A}\bm v(a)f(da,\bm X_i)=\sum_{j=1}^{m_n}\bm v^{(j)}f(B^{(j)},\bm X_i).
\]
Hence, with $\hat{\bm r}_j=\frac1n\sum_{i=1}^n\bm\phi(\bm X_i)f(B^{(j)},\bm X_i)$, the right-hand side of \eqref{eqn:balancing} satisfies
\[
\sum_{i=1}^n\bm \nu (f,\bm X_i)\otimes\bm\phi(\bm X_i)=\sum_{j=1}^{m_n}\bm v^{(j)}\otimes\Big\{\sum_{i=1}^n\bm\phi(\bm X_i)f(B^{(j)},\bm X_i)\Big\}=\sum_{j=1}^{m_n}\bm v^{(j)}\otimes n\hat{\bm r}_j.
\]
Similarly, since $\bm v(A_i)=\bm v^{(j)}$ for $i\in S_j$ and $\bm\Phi_j^\top\bm w_{S_j}=\sum_{i\in S_j}w_i\bm\phi(\bm X_i)$,
\[
\bm\Phi^\top\bm w=\sum_{i=1}^n w_i\{\bm v(A_i)\otimes\bm\phi(\bm X_i)\}=\sum_{j=1}^{m_n}\bm v^{(j)}\otimes\bm\Phi_j^\top\bm w_{S_j}.
\]
Because $\{\bm v^{(1)},\ldots,\bm v^{(m_n)}\}$ are the canonical basis vectors of $\mathbb R^{m_n}$, the two Kronecker expansions occupy disjoint blocks, so \eqref{eqn:balancing} holds if and only if the blocks agree, i.e. $\bm\Phi_j^\top\bm w_{S_j}=n\hat{\bm r}_j$ for every $j=1,\ldots,m_n$. The proof now follows.
\end{proof}

\subsection{Proof of Proposition~\ref{lem:weights_characterization}}
\begin{proof}
    Fix $j\in\{1,\ldots,m_n\}$. Eliminating the slack $\bm\xi_j=\bm\Phi_j^\top\bm w_{S_j}-n\hat{\bm r}_j$ from the constraint in \eqref{eqn:final_optimization}, the bin-$j$ subproblem reduces to the unconstrained minimization of
    \[
    g(\bm w_{S_j}):=\tfrac{1}{2}\|\bm w_{S_j}\|^2+\tfrac{1}{2n\lambda_{nj}}\|\bm\Phi_j^\top\bm w_{S_j}-n\hat{\bm r}_j\|^2 .
    \]
    Since $\nabla^2 g=\bm I_{n_j}+(n\lambda_{nj})^{-1}\bm\Phi_j\bm\Phi_j^\top\succ\bm 0$, $g$ is strictly convex and $g(\bm w)\to \infty$, as $\|\bm w\|\to \infty$, so its first-order condition $\bm w_{S_j}+(n\lambda_{nj})^{-1}\bm\Phi_j(\bm\Phi_j^\top\bm w_{S_j}-n\hat{\bm r}_j)=\bm 0$ characterizes the unique minimizer. Rearranging gives $(n\lambda_{nj}\bm I_{n_j}+\bm\Phi_j\bm\Phi_j^\top)\hat{\bm w}_{S_j}=n\bm\Phi_j\hat{\bm r}_j$ and hence, by the identity $(\rho\bm I+\bm A\bm A^\top)^{-1}\bm A=\bm A(\bm A^\top\bm A+\rho\bm I)^{-1}$ with $\rho=n\lambda_{nj}$ and $\bm A=\bm\Phi_j$, we get,
    \[
    \hat{\bm w}_{S_j}=n\bm\Phi_j(\bm\Phi_j^\top\bm\Phi_j+n\lambda_{nj}\bm I_p)^{-1}\hat{\bm r}_j
    =\bm\Phi_j(\hat{\bm M}_j+\lambda_{nj}\bm I_p)^{-1}\hat{\bm r}_j=\bm\Phi_j\hat{\bm c}^{(n)}_j,
    \]
    where the second equality uses $\bm\Phi_j^\top\bm\Phi_j+n\lambda_{nj}\bm I_p=n(\hat{\bm M}_j+\lambda_{nj}\bm I_p)$ and the definition of $\hat{\bm c}^{(n)}_j$. Note that, since $\lambda_{nj}>0$, the matrix $\hat{\bm M}_j+\lambda_{nj}\bm I_p$ is invertible for every $j$ irrespective of $n_j$ or the rank of $\bm\Phi_j$; in particular the subproblem is always feasible with a unique solution. Since the rows of $\bm\Phi_j$ are $\bm\phi(\bm X_i)^\top$, $i\in S_j$, the display reads $\hat w_{f,i}=\bm\phi(\bm X_i)^\top\hat{\bm c}^{(n)}_j$ for every $i\in S_j$. As $j$ was arbitrary, the claim follows.
\end{proof}

\subsection{Proof of Theorem~\ref{thm:consistency}}
We first state an auxiliary lemma that will be used in the proof of Theorem \ref{thm:consistency}. Its proof is given after the proof of the theorem.

\begin{lemma}
\label{lem:uniform_first_stage_bounds}
Suppose Assumptions \ref{assm:iid}--\ref{assm:rate_choice_consistency} hold. Define the population analogues of $\hat{\bm M}_j$, $\hat{\bm r}_j$ and $\hat{\bm c}_j^{(n)}$:
\begin{align*}
\bm M_j=\mathbb E\left[\bm\phi(\bm X_1)\bm\phi(\bm X_1)^\top I(A_1\in B^{(j)})\right],\quad
\bm r_j=\mathbb E\left[\bm\phi(\bm X_1)f(B^{(j)},\bm X_1)\right],\quad\textrm{and}\quad\bm c^{(n)}_j=\bm M_j^{-1}\bm r_j.
\end{align*}
Let $\|\bm M\|$ denote the operator norm of any matrix $\bm M$. Then,
\begin{align}
\max_{1\leq j\leq m_n}\|\hat{\bm M}_j-\bm M_j\|=O_{\mathbb P}\left(\sqrt{\frac{\log m_n}{nm_n}}\right),
\label{eqn:Mhat_uniform_bound}
\end{align}
\begin{align}
\max_{1\leq j\leq m_n}\|\hat{\bm r}_j-\bm r_j\|=O_{\mathbb P}\left(\sqrt{\frac{\log m_n}{nm_n}}\right),
\label{eqn:rhat_uniform_bound}
\end{align}
and
\begin{align}
\max_{1\leq j\leq m_n}\|\hat{\bm c}^{(n)}_j-\bm c^{(n)}_j\|=o_{\mathbb P}(1).
\label{eqn:chat_uniform_convergence}
\end{align}
Consequently,
\begin{align}
\max_{1\leq j\leq m_n}\|\hat{\bm c}^{(n)}_j\|=O_{\mathbb P}(1).
\label{eqn:chat_uniform_bounded}
\end{align}
\end{lemma}

\begin{proof}[Proof of Theorem \ref{thm:consistency}]
    Throughout, let $j(a)$ denote the index with $a\in B^{(j(a))}$, and recall from Lemma~\ref{lem:uniform_first_stage_bounds} the population quantities $\bm M_j=\mathbb E[\bm\phi(\bm X_1)\bm\phi(\bm X_1)^\top\mathbb I(A_1\in B^{(j)})]$, $\bm r_j=\mathbb E[\bm\phi(\bm X_1)f(B^{(j)},\bm X_1)]$ and $\bm c^{(n)}_j=\bm M_j^{-1}\bm r_j$, together with the bounds
    \[
    \max_{1\le j\le m_n}\|\hat{\bm c}^{(n)}_j-\bm c^{(n)}_j\|=o_{\mathbb P}(1),\qquad \max_{1\le j\le m_n}\|\hat{\bm c}^{(n)}_j\|=O_{\mathbb P}(1),\qquad \max_{1\le j\le m_n}\|\bm c^{(n)}_j\|=O(1).
    \]
    By Proposition~\ref{lem:weights_characterization}, $T(\hat{\bm w}_f)=\frac1n\sum_{i=1}^nY_i\bm\phi(\bm X_i)^\top\hat{\bm c}^{(n)}_{j(A_i)}$. Writing $\xi_i = Y_i - m(A_i, \bm X_i)$, due to Assumption~\ref{assm:signal_plus_noise_general} decompose $T(\hat{\bm w}_f)=S_n+R_n$ with
    \begin{align}
    \label{eqn:sn_and_rn}
    S_n=\frac1n\sum_{i=1}^n m(A_i,\bm X_i)\bm\phi(\bm X_i)^\top\hat{\bm c}^{(n)}_{j(A_i)},\qquad R_n=\frac1n\sum_{i=1}^n \xi_i\,\bm\phi(\bm X_i)^\top\hat{\bm c}^{(n)}_{j(A_i)}.
    \end{align}

    We will first show that the noise term, $R_n = o_{\mathbb P}(1)$. Let $\mathcal F_n=\sigma(\{(\bm X_i,A_i)\}_{i=1}^n)$. Note that since $\lambda_{nj}$ is $\mathcal F_n$-measurable for all $j$ (Assumption~\ref{assm:rate_choice_consistency}(b)), $\hat{\bm c}^{(n)}_{j(A_i)}$ is $\mathcal F_n$-measurable. Next, since $\mathbb E[\xi_i\mid\mathcal F_n]=0$ and the $\xi_i$ are independent across $i$ given $\mathcal F_n$, we have $\mathbb E[R_n\mid\mathcal F_n]=0$ and, with $\|\bm\phi\|\le C$ (Assumption~\ref{assm:regularity}(a)),
    \[
    \operatorname{Var}(R_n\mid\mathcal F_n)=\frac1{n^2}\sum_{i=1}^n\mathbb E[\xi_i^2\mid A_i,\bm X_i]\{\bm\phi(\bm X_i)^\top\hat{\bm c}^{(n)}_{j(A_i)}\}^2\le\frac{C^2}{n}\Big\{\frac1n\sum_{i=1}^n\mathbb E[\xi_i^2\mid A_i,\bm X_i]\Big\}\max_{1\le j\le m_n}\|\hat{\bm c}^{(n)}_j\|^2.
    \]
    The bracketed average is $O_{\mathbb P}(1)$ by the weak law of large numbers, since $\mathbb E\big[\mathbb E[\xi_1^2\mid A_1,\bm X_1]\big]=\mathbb E[\xi_1^2]=\mathbb E[\sigma^2(\bm X_1)]<\infty$; hence $n\operatorname{Var}(R_n\mid\mathcal F_n)=O_{\mathbb P}(1)$. Therefore, for any $\varepsilon>0$, conditional Chebyshev's inequality gives $\mathbb P(|R_n|>\varepsilon)\le\mathbb E\big[1\wedge\varepsilon^{-2}\operatorname{Var}(R_n\mid\mathcal F_n)\big]\to0$ by bounded convergence theorem, so
    \begin{align}
    R_n=o_{\mathbb P}(1). \label{eqn:noise_negligible_consistency}
    \end{align}

    We next, turn to the signal term $S_n$. First replace $\hat{\bm c}^{(n)}_j$ by $\bm c^{(n)}_j$: since $\frac1n\sum_i|m(A_i,\bm X_i)|=O_{\mathbb P}(1)$ (weak law of large numbers, using $\mathbb E|m(A_1,\bm X_1)|<\infty$) and $\|\bm\phi\|\le C$,
    \[
    \Big|S_n-\frac1n\sum_{i=1}^n m(A_i,\bm X_i)\bm\phi(\bm X_i)^\top\bm c^{(n)}_{j(A_i)}\Big|\le C\Big\{\frac1n\sum_{i=1}^n|m(A_i,\bm X_i)|\Big\}\max_{1\le j\le m_n}\|\hat{\bm c}^{(n)}_j-\bm c^{(n)}_j\|=o_{\mathbb P}(1).
    \]
    Write $\tilde S_n:=\frac1n\sum_i\psi_n(A_i,\bm X_i)$ with $\psi_n(a,\bm x):=m(a,\bm x)\bm\phi(\bm x)^\top\bm c^{(n)}_{j(a)}$. The summands are i.i.d.\ (for fixed $n$), and $|\psi_n(a,\bm x)|\le C\,|m(a,\bm x)|\max_j\|\bm c^{(n)}_j\|$ gives $\mathbb E[\psi_n(A_1,\bm X_1)^2]\le C^2(\max_j\|\bm c_j^{(n)}\|)^2\mathbb E[m(A_1,\bm X_1)^2]=O(1)$; hence $\operatorname{Var}(\tilde S_n)=O(n^{-1})$ and, by Chebyshev's inequality,
    \[
    \tilde S_n=\sum_{j=1}^{m_n}\bm g_j^\top\bm c^{(n)}_j+o_{\mathbb P}(1),\qquad \bm g_j:=\mathbb E[m(A_1,\bm X_1)\bm\phi(\bm X_1)\mathbb I(A_1\in B^{(j)})].
    \]
    It remains to show $\sum_{j=1}^{m_n}\bm g_j^\top\bm c^{(n)}_j\to\mu_f^*$. Define $\bm M(a):=\mathbb E[\bm\phi(\bm X_1)\bm\phi(\bm X_1)^\top e(a,\bm X_1)]$ and $\bm h(a):=\mathbb E[\bm\phi(\bm X_1)m(a,\bm X_1)e(a,\bm X_1)]$, so that $\bm\beta^*(a)=\bm M(a)^{-1}\bm h(a)$. Conditioning on $\bm X_1$ and integrating over $A_1$,
    \[
    \bm M_j=\int_{B^{(j)}}\bm M(a)\,da,\qquad \bm g_j=\int_{B^{(j)}}\bm h(a)\,da=\int_{B^{(j)}}\bm M(a)\bm\beta^*(a)\,da.
    \]
    By strong positivity (Assumption~\ref{assm:pos}) and the full-rank condition (Assumption~\ref{assm:regularity}(d)), $\bm M(a)\succeq\alpha\,\mathbb E[\bm\phi\bm\phi^\top]$, so for $\underline{\kappa}:=\alpha\lambda_{\min}(\mathbb E[\bm\phi\bm\phi^\top])>0$, $\lambda_{\min}(\bm M_j)\ge \underline{\kappa}|B^{(j)}|$, and $\|\bm M(a)\|\le C^2\beta=:\bar{\kappa}$. Using $\bm M_j=\int_{B^{(j)}}\bm M$, for any $a_0\in B^{(j)}$,
    \[
    \bm M_j^{-1}\bm g_j-\bm\beta^*(a_0)=\bm M_j^{-1}\int_{B^{(j)}}\bm M(a)\big(\bm\beta^*(a)-\bm\beta^*(a_0)\big)\,da,
    \]
    hence $\|\bm M_j^{-1}\bm g_j-\bm\beta^*(a_0)\|\le\|\bm M_j^{-1}\|\,\bar{\kappa}\,|B^{(j)}|\sup_{a\in B^{(j)}}\|\bm\beta^*(a)-\bm\beta^*(a_0)\|\le(\bar{\kappa}/\underline{\kappa})\,\omega(|B^{(j)}|)$, where $\omega$ is the modulus of continuity of $\bm\beta^*$. By continuity of $a\mapsto e(a,\bm x)$ and of $a\mapsto m(a,\bm x)$ (Assumption~\ref{assm:continuity}), the boundedness $\|\bm\phi\|\le C$ and $e\le\beta$, and boundedness of $m$, dominated convergence shows $\bm M(\cdot)$ and $\bm h(\cdot)$ are continuous on the compact $\mathcal A$; as $\bm M(\cdot)$ is uniformly invertible, $\bm\beta^*=\bm M^{-1}\bm h$ is continuous, hence uniformly continuous, so $\omega(\delta)\to0$ as $\delta\downarrow0$. Since $\max_j|B^{(j)}|\le 3/(2m_n)\to0$, using $\bm g_j^\top\bm c^{(n)}_j=(\bm M_j^{-1}\bm g_j)^\top\bm r_j$, $\bm r_j=\mathbb E[\bm\phi(\bm X_1)f(B^{(j)},\bm X_1)]$, and $\mu_f^*=\mathbb E[\sum_j\int_{B^{(j)}}\bm\phi(\bm X_1)^\top\bm\beta^*(a)f(da,\bm X_1)]$,
    \begin{align*}
    \Big|\sum_{j=1}^{m_n}\bm g_j^\top\bm c^{(n)}_j-\mu_f^*\Big|
    &=\Big|\mathbb E\Big[\sum_{j=1}^{m_n}\int_{B^{(j)}}\bm\phi(\bm X_1)^\top\big(\bm M_j^{-1}\bm g_j-\bm\beta^*(a)\big)f(da,\bm X_1)\Big]\Big|\\
    &\le C\,\frac{\bar{\kappa}}{\underline{\kappa}}\,\omega\!\Big(\frac{3}{2m_n}\Big)\,\mathbb E\big[|f|(\mathcal A,\bm X_1)\big]\le CC_f\,\frac{\bar{\kappa}}{\underline{\kappa}}\,\omega\!\Big(\frac{3}{2m_n}\Big)\to0,
    \end{align*}
    using Assumption~\ref{assm:regularity}(b). Combining the displays, $S_n\Pto\mu_f^*$, and with \eqref{eqn:noise_negligible_consistency}, $T(\hat{\bm w}_f)\Pto\mu_f^*$.
\end{proof}

\begin{proof}[Proof of Lemma \ref{lem:uniform_first_stage_bounds}]
    As per Assumption~\ref{assm:regularity}, $\|\bm\phi(\bm x)\|\leq C$ for all $\bm x\in\mathcal X$. For the bins $\mathcal A=[0,1]$ is divided into in Proposition \ref{lem:bj_intervals}, we have,
    \begin{align}
    \frac{1}{2m_n}\leq |B^{(j)}|\leq \frac{3}{2m_n},\qquad j=1,\ldots,m_n.
    \label{eqn:bin_length_bounds}
    \end{align}
    We begin with lower bounding the eigenvalues of $\bm M_j$. For any $\bm u\in\mathbb R^p$,
    \begin{align*}
    \bm u^\top\bm M_j\bm u
    &=\mathbb E\left[(\bm u^\top\bm\phi(\bm X_1))^2\mathbb I(A_1\in B^{(j)})\right]\\
    &=\mathbb E\left[(\bm u^\top\bm\phi(\bm X_1))^2\mathbb P(A_1\in B^{(j)}\mid \bm X_1)\right]\\
    &=\mathbb E\left[(\bm u^\top\bm\phi(\bm X_1))^2\int_{B^{(j)}}e(a,\bm X_1)da\right].
    \end{align*}
    By strong positivity in Assumption \ref{assm:pos},
    \begin{align*}
    \int_{B^{(j)}}e(a,\bm X_1)da\geq \alpha |B^{(j)}|.
    \end{align*}
    Therefore,
    \begin{align*}
    \bm u^\top\bm M_j\bm u
    &\geq \alpha |B^{(j)}|\mathbb E\left[(\bm u^\top\bm\phi(\bm X_1))^2\right]\\
    &=\alpha |B^{(j)}|\bm u^\top\mathbb E[\bm\phi(\bm X_1)\bm\phi(\bm X_1)^\top]\bm u\\
    &\geq \alpha |B^{(j)}|\lambda_{\min}\left(\mathbb E[\bm\phi(\bm X_1)\bm\phi(\bm X_1)^\top]\right)\|\bm u\|^2.
    \end{align*}
    Combining this with Equation \eqref{eqn:bin_length_bounds} and the fact that $\lambda_{\min}\left(\mathbb E[\bm\phi(\bm X_1)\bm\phi(\bm X_1)^\top]\right)>0$ from Assumption \ref{assm:regularity}, there exists $c_M>0$ such that
    \begin{align}
    \lambda_{\min}(\bm M_j)\geq \frac{c_M}{m_n},\qquad j=1,\ldots,m_n.
    \label{eqn:M_min_eigen_bound}
    \end{align}
    
    Next, we bound $\bm r_j$. By the local regularity of the counterfactual weights from Assumption \ref{assm:regularity}, and the boundedness of $\bm\phi(\cdot)$,
    \begin{align*}
    \|\bm r_j\| =\left\|\mathbb E\left[\bm\phi(\bm X_1)f(B^{(j)},\bm X_1)\right]\right\| \leq \mathbb E\left[\|\bm\phi(\bm X_1)\||f(B^{(j)},\bm X_1)|\right] \leq C\mathbb E[|f|(B^{(j)},\bm X_1)] \leq CC_f^\ast |B^{(j)}|.
    \end{align*}
    Combining this with Equation \eqref{eqn:bin_length_bounds}, there exists $C_r<\infty$ such that
    \begin{align}
    \max_{1\leq j\leq m_n}\|\bm r_j\|\leq \frac{C_r}{m_n}.
    \label{eqn:r_population_bound}
    \end{align}
    Combining Equations \eqref{eqn:M_min_eigen_bound} and \eqref{eqn:r_population_bound},
    \begin{align}
    \max_{1\leq j\leq m_n}\|\bm c^{(n)}_j\|
    \leq \max_{1\leq j\leq m_n}\|\bm M_j^{-1}\|\max_{1\leq j\leq m_n}\|\bm r_j\|
    \leq \frac{m_n}{c_M}\frac{C_r}{m_n}=O(1).
    \label{eqn:c_population_bound}
    \end{align}
    
    We next prove Equation \eqref{eqn:Mhat_uniform_bound}. Fix $j\in\{1,\ldots,m_n\}$ and coordinates $k,\ell\in\{1,\ldots,p\}$. Define
    \begin{align*}
    Z_{i,j,k\ell}=\phi_k(\bm X_i)\phi_\ell(\bm X_i)\mathbb I(A_i\in B^{(j)}),
    \end{align*}
    where $\phi_k(\cdot)$ denotes the $k^{\rm th}$ coordinate of $\bm \phi(\cdot)$, for any $1\leq k \leq p$. Then, the entry at the $k^{\rm th}$ row and the $\ell^{\rm th}$ column of $\hat{\bm M}_j-\bm M_j$ is given by
    \begin{align*}
    [\hat{\bm M}_j-\bm M_j]_{k\ell}=\frac{1}{n}\sum_{i=1}^n\left\{Z_{i,j,k\ell}-\mathbb E[Z_{i,j,k\ell}]\right\}.
    \end{align*}
    Since $\|\bm\phi(\bm x)\|\leq C$, we have $|Z_{i,j,k\ell}|\leq C^2$. Also,
    \begin{align*}
    \operatorname{Var}(Z_{i,j,k\ell})
    &\leq \mathbb E[Z_{i,j,k\ell}^2] \leq C^4\mathbb P(A_i\in B^{(j)})=C^4\mathbb E\left[\int_{B^{(j)}}e(a,\bm X_i)da\right] \leq C^4\beta |B^{(j)}| \leq \frac{3C^4\beta}{2m_n}.
    \end{align*}
    Next, define the centered random variables
    \begin{align*}
    \widetilde Z_{i,j,k\ell}=Z_{i,j,k\ell}-\mathbb E[Z_{i,j,k\ell}].
    \end{align*}
    Then $\mathbb E[\widetilde Z_{i,j,k\ell}]=0$ and
    \begin{align*}
    |\widetilde Z_{i,j,k\ell}|
    \leq |Z_{i,j,k\ell}|+\mathbb E[|Z_{i,j,k\ell}|]
    \leq 2C^2.
    \end{align*}
    Moreover,
    \begin{align*}
    \frac{1}{n}\sum_{i=1}^n\operatorname{Var}(\widetilde Z_{i,j,k\ell})
    =\operatorname{Var}(Z_{1,j,k\ell})
    \leq \frac{3C^4\beta}{2m_n}.
    \end{align*}
    By Bernstein's inequality for independent mean-zero bounded random variables,
    \begin{align*}
    \mathbb P\left(\left|\frac{1}{n}\sum_{i=1}^n\widetilde Z_{i,j,k\ell}\right|>t\right)
    \leq 2\exp\left[-\frac{nt^2/2}{3C^4\beta/(2m_n)+2C^2t/3}\right].
    \end{align*}
    Hence,
    \begin{align*}
    \mathbb P\left(\left|[\hat{\bm M}_j-\bm M_j]_{k\ell}\right|>t\right)
    \leq 2\exp\left[-\frac{nt^2/2}{3C^4\beta/(2m_n)+2C^2t/3}\right].
    \end{align*}
    Let
    \begin{align*}
    t_n=L\sqrt{\frac{\log m_n}{nm_n}},
    \end{align*}
    where $L>0$ is a constant to be chosen. By the growth conditions on $m_n$ under Assumption~\ref{assm:rate_choice_consistency}, $\frac{m_n\log m_n}{n}\to0$. Thus, $t_n=o(m_n^{-1})$. Hence, for all sufficiently large $n$,
    \begin{align*}
    2C^2t_n/3\leq 3C^4\beta/(2m_n).
    \end{align*}
    Therefore, for all sufficiently large $n$,
    \begin{align*}
    \mathbb P\left(\left|[\hat{\bm M}_j-\bm M_j]_{k\ell}\right|>t_n\right)
    \leq 2\exp\left[-\frac{nt_n^2}{6C^4\beta/m_n}\right]
    =2\exp\left[-\frac{L^2\log m_n}{6C^4\beta}\right].
    \end{align*}
    Applying a union bound over $j=1,\ldots,m_n$ and $k,\ell=1,\ldots,p$,
    \begin{align*}
    \mathbb P\left(\max_{1\leq j\leq m_n}\max_{1\leq k,\ell\leq p}\left|[\hat{\bm M}_j-\bm M_j]_{k\ell}\right|>t_n\right)
    \leq 2p^2m_n\exp\left[-\frac{L^2\log m_n}{6C^4\beta}\right].
    \end{align*}
    Choosing $L$ large enough so that $L^2/(6C^4\beta)>2$, the right-hand side converges to zero. Since $p$ is fixed, there exists a constant $C_p<\infty$ such that, for every $p\times p$ matrix $\bm A$,
    \begin{align*}
    \|\bm A\|\leq C_p\max_{1\leq k,\ell\leq p}|A_{k\ell}|.
    \end{align*}
    Equation \eqref{eqn:Mhat_uniform_bound} follows.
    
    Next, we invoke a similar argument to prove Equation \eqref{eqn:rhat_uniform_bound}. Fix $j\in\{1,\ldots,m_n\}$ and $k\in\{1,\ldots,p\}$. Define
    \begin{align*}
    U_{i,j,k}=\phi_k(\bm X_i)f(B^{(j)},\bm X_i).
    \end{align*}
    Then the $k^{\rm th}$ entry of $\hat{\bm r}_j-\bm r_j$ is given by
    \begin{align*}
    [\hat{\bm r}_j-\bm r_j]_k=\frac{1}{n}\sum_{i=1}^n\left\{U_{i,j,k}-\mathbb E[U_{i,j,k}]\right\}.
    \end{align*}
    By the boundedness of the counterfactual weights from Assumption \ref{assm:regularity}, we get
    \begin{align*}
    |U_{i,j,k}|
    \leq C|f|(B^{(j)},\bm X_i)
    \leq C|f|(\mathcal A,\bm X_i)
    \leq CC_f.
    \end{align*}
    Additionally, using the local regularity of the counterfactual weights from Assumptions \ref{assm:regularity},
    \begin{align*}
    \operatorname{Var}(U_{i,j,k})
    &\leq \mathbb E[U_{i,j,k}^2]\\
    &\leq C^2\mathbb E[f(B^{(j)},\bm X_i)^2]\\
    &\leq C^2\mathbb E\left[|f|^2(B^{(j)},\bm X_i)\right]\\
    &\leq C^2\mathbb E\left[|f|(B^{(j)},\bm X_i)|f|(\mathcal A,\bm X_i)\right]\\
    &\leq C^2C_f\mathbb E[|f|(B^{(j)},\bm X_i)]\\
    &\leq C^2C_fC_f^\ast |B^{(j)}|\\
    &\leq \frac{3C^2C_fC_f^\ast}{2m_n},
    \end{align*}
    where the last inequality follows from Equation~\eqref{eqn:bin_length_bounds}. Define
    \begin{align*}
    \widetilde U_{i,j,k}=U_{i,j,k}-\mathbb E[U_{i,j,k}].
    \end{align*}
    Then $\mathbb E[\widetilde U_{i,j,k}]=0$,
    \begin{align*}
    |\widetilde U_{i,j,k}|\leq 2CC_f,
    \end{align*}
    and
    \begin{align*}
    \frac{1}{n}\sum_{i=1}^n\operatorname{Var}(\widetilde U_{i,j,k})
    =\operatorname{Var}(U_{1,j,k})
    \leq \frac{3C^2C_fC_f^\ast}{2m_n}.
    \end{align*}
    Bernstein's inequality again implies that, for any $t>0$,
    \begin{align*}
    \mathbb P\left(\left|[\hat{\bm r}_j-\bm r_j]_k\right|>t\right)
    \leq 2\exp\left[-\frac{nt^2/2}{3C^2C_fC_f^\ast/(2m_n)+2CC_ft/3}\right].
    \end{align*}
    Using the same $t_n=L\sqrt{\log m_n/(nm_n)}$ as above, and using $t_n=o(m_n^{-1})$, for all sufficiently large $n$,
    \begin{align*}
    2CC_ft_n/3\leq 3C^2C_fC_f^\ast/(2m_n).
    \end{align*}
    Thus, for all sufficiently large $n$,
    \begin{align*}
    \mathbb P\left(\left|[\hat{\bm r}_j-\bm r_j]_k\right|>t_n\right)
    \leq 2\exp\left[-\frac{nt_n^2}{6C^2C_fC_f^\ast/m_n}\right]
    =2\exp\left[-\frac{L^2\log m_n}{6C^2C_fC_f^\ast}\right].
    \end{align*}
    Applying the same union bound argument as with the matrices $\{\hat {\bm M}_j\}$ with $L$ chosen large enough so that $L^2/(6C^2C_fC_f^\ast)>2$, followed by a similar entrywise-to-vector norm argument proves Equation \eqref{eqn:rhat_uniform_bound}.
    
    It remains to prove Equation \eqref{eqn:chat_uniform_convergence}. Let
    \begin{align*}
    \Delta_{M,n}=\max_{1\leq j\leq m_n}\|\hat{\bm M}_j-\bm M_j\|,\qquad
    \Delta_{r,n}=\max_{1\leq j\leq m_n}\|\hat{\bm r}_j-\bm r_j\|.
    \end{align*}
    By Equations \eqref{eqn:Mhat_uniform_bound} and \eqref{eqn:rhat_uniform_bound},
    \begin{align}
    \label{eqn:Delta_rates}
    \Delta_{M,n}=O_{\mathbb P}\left(\sqrt{\frac{\log m_n}{nm_n}}\right),\qquad
    \Delta_{r,n}=O_{\mathbb P}\left(\sqrt{\frac{\log m_n}{nm_n}}\right).
    \end{align}
    Since $m_n\log m_n/n\to0$,
    \begin{align}
    m_n\Delta_{M,n}=o_{\mathbb P}(1),\qquad m_n\Delta_{r,n}=o_{\mathbb P}(1).
    \label{eqn:scaled_first_stage_errors}
    \end{align}
    Next, define the event
    \begin{align*}
    \mathcal E_n=\left\{\Delta_{M,n}\leq \frac{c_M}{2m_n}\right\}.
    \end{align*}
    By Equation \eqref{eqn:scaled_first_stage_errors}, $\mathbb P(\mathcal E_n)\to1$ as $n\to\infty$. On the event $\mathcal E_n$, for all sufficiently large $n$ and uniformly over $j=1,\ldots,m_n$, Weyl's eigenvalue inequality \citep[Theorem 4.3.1]{hornjohnson2013} gives
    \begin{align*}
    \lambda_{\min}(\hat{\bm M}_j+\lambda_{nj}\bm I_p)&\geq  \lambda_{\min}({\bm M}_j+\lambda_{nj}\bm I_p) + \lambda_{\min}\left(\hat{\bm M}_j - {\bm M}_j\right) \\
    &\geq \lambda_{\min}(\bm M_j+\lambda_{nj}\bm I_p)-\|\hat{\bm M}_j-\bm M_j\|\\
    &=\lambda_{\min}(\bm M_j)+\lambda_{nj}-\|\hat{\bm M}_j-\bm M_j\|\\
    &\geq \frac{c_M}{m_n}-\Delta_{M,n}\\
    &\geq \frac{c_M}{2m_n}.
    \end{align*}
    Since $\mathbb P(\mathcal E_n)\to1$, the previous display holds uniformly over $j=1,\ldots,m_n$ with probability tending to one as $n\to\infty$. Therefore, with probability tending to one as $n\to\infty$,
    \begin{align}
    \max_{1\leq j\leq m_n}\|(\hat{\bm M}_j+\lambda_{nj}\bm I_p)^{-1}\|\leq \frac{2m_n}{c_M}.
    \label{eqn:inverse_hatM_bound}
    \end{align}
    Now observe that
    \begin{align*}
    \hat{\bm c}^{(n)}_j-\bm c^{(n)}_j
    &=(\hat{\bm M}_j+\lambda_{nj}\bm I_p)^{-1}\hat{\bm r}_j-\bm M_j^{-1}\bm r_j\\
    &=(\hat{\bm M}_j+\lambda_{nj}\bm I_p)^{-1}
    \left\{\hat{\bm r}_j-(\hat{\bm M}_j+\lambda_{nj}\bm I_p)\bm c^{(n)}_j\right\}\\
    &=(\hat{\bm M}_j+\lambda_{nj}\bm I_p)^{-1}
    \left\{(\hat{\bm r}_j-\bm r_j)-(\hat{\bm M}_j-\bm M_j)\bm c^{(n)}_j-\lambda_{nj}\bm c^{(n)}_j\right\},
    \end{align*}
    where we used $\bm M_j\bm c^{(n)}_j=\bm r_j$. Hence, by Equations \eqref{eqn:c_population_bound} and \eqref{eqn:inverse_hatM_bound}, on an event whose probability tends to one as $n\to\infty$,
    \begin{align*}
    \max_{1\leq j\leq m_n}\|\hat{\bm c}^{(n)}_j-\bm c^{(n)}_j\|
    &\leq \frac{2m_n}{c_M}\left\{\Delta_{r,n}+\Delta_{M,n}\max_{1\leq j\leq m_n}\|\bm c^{(n)}_j\|+\max_{1\leq j\leq m_n}\lambda_{nj}\max_{1\leq j\leq m_n}\|\bm c^{(n)}_j\|\right\}\\
    &\leq O_{\mathbb P}(m_n\Delta_{r,n})+O_{\mathbb P}(m_n\Delta_{M,n})+O_{\mathbb P}\left(m_n\max_{1\leq j\leq m_n}\lambda_{nj}\right)\\
    &= O_{\mathbb P}\left(\sqrt{\frac{m_n \log m_n}{n}}\right) + O_{\mathbb P}\left(m_n\max_{1\leq j\leq m_n}\lambda_{nj}\right) \\
    &=o_{\mathbb P}(1),
    \end{align*}
    where the penultimate step follows from Equation~\eqref{eqn:Delta_rates} and the last equality follows from the fact that $m_n\log m_n/n\to 0$ and the rate condition $m_n\max_{1\leq j\leq m_n}\lambda_{nj} = o_{\mathbb P}(1)$. This proves Equation \eqref{eqn:chat_uniform_convergence}. Finally, Equation \eqref{eqn:chat_uniform_bounded} follows from Equation \eqref{eqn:chat_uniform_convergence} and Equation \eqref{eqn:c_population_bound}. This completes the proof.
\end{proof}

\subsection{Proof of Corollary~\ref{cor:correct_specification}}
\label{sec:proof_cor_correct_specification}
\begin{proof}
    Suppose $m(a,\bm x)=\bm\phi(\bm x)^\top\bm\beta(a)$. Then, for every $a$,
    \[
    \mathbb E[\bm\phi(\bm X_1)m(a,\bm X_1)e(a,\bm X_1)]=\mathbb E[\bm\phi(\bm X_1)\bm\phi(\bm X_1)^\top e(a,\bm X_1)]\bm\beta(a),
    \]
    so, by the definition of $\bm\beta^*$, $\bm\beta^*(a)=\mathbb E[\bm\phi(\bm X_1)\bm\phi(\bm X_1)^\top e(a,\bm X_1)]^{-1}\mathbb E[\bm\phi(\bm X_1)\bm\phi(\bm X_1)^\top e(a,\bm X_1)]\bm\beta(a)=\bm\beta(a)$. Hence
    \[
    \mu_f^*=\mathbb E\Big[\int_{\mathcal A}\bm\phi(\bm X_1)^\top\bm\beta(a)f(da,\bm X_1)\Big]=\mathbb E\Big[\int_{\mathcal A}m(a,\bm X_1)f(da,\bm X_1)\Big],
    \]
    On the other hand, since $f(\cdot,\bm X_1)$ is $\bm X_1$-measurable, Assumption~\ref{assm:unconfound} and the tower property give
    \[
    \mu_f=\mathbb E\Big[\int_{\mathcal A}Y_1(a)f(da,\bm X_1)\Big]=\mathbb E\Big[\int_{\mathcal A}\mathbb E[Y_1(a)\mid\bm X_1]f(da,\bm X_1)\Big]=\mathbb E\Big[\int_{\mathcal A}m(a,\bm X_1)f(da,\bm X_1)\Big],
    \]
    where the conditional expectation and the integral can be interchanged due to Tonelli's theorem and the condition that $\mathbb E\left[\int |Y_i(a)||f|(da, \bm X_i)\right]<\infty$. Therefore $\mu_f=\mu_f^*$.
\end{proof}

\subsection{Proof of Theorem~\ref{thm:clt}}
\label{sec:proof_clt}

Let us define
\begin{align}
    H_n(\bm x):=\sum_{j=1}^{m_n}\bm\beta_j^\top\bm\phi(\bm x)f(B^{(j)},\bm x).\label{eqn:Hn_definition_consistency}
\end{align}
We prove Theorem \ref{thm:clt} using two auxiliary lemmas.

\begin{lemma}
\label{lem:clt_linear_representation}
Under Assumptions \ref{assm:iid}--\ref{assm:rate_choice_normality},
\begin{align}
\sqrt n\left(T(\hat{\bm w}_f)-\mu_f\right)=\frac{1}{\sqrt n}\sum_{i=1}^n\left(H_n(\bm X_i)-\mathbb E[H_n(\bm X_1)]\right)+\frac{1}{\sqrt n}\sum_{i=1}^n\sigma(\bm X_i)\epsilon_i\bm\phi(\bm X_i)^\top\bm c^{(n)}_{j(A_i)}+o_{\mathbb P}(1).
\label{eqn:clt_linear_representation}
\end{align}
\end{lemma}

\begin{lemma}
\label{lem:leading_triangular_array_clt}
Under Assumptions \ref{assm:iid}--\ref{assm:rate_choice_normality},
\begin{align}
\frac{1}{\sqrt n\eta_n}\sum_{i=1}^n\left[H_n(\bm X_i)-\mathbb E[H_n(\bm X_1)]+\sigma(\bm X_i)\epsilon_i\bm\phi(\bm X_i)^\top\bm c^{(n)}_{j(A_i)}\right]\dto N(0,1).
\label{eqn:leading_triangular_array_clt}
\end{align}
\end{lemma}

\begin{proof}[Proof of Theorem \ref{thm:clt}]
By Lemma \ref{lem:clt_linear_representation},
\begin{align*}
\sqrt n\left(T(\hat{\bm w}_f)-\mu_f\right)=\frac{1}{\sqrt n}\sum_{i=1}^n\left[H_n(\bm X_i)-\mathbb E[H_n(\bm X_1)]+\sigma(\bm X_i)\epsilon_i\bm\phi(\bm X_i)^\top\bm c^{(n)}_{j(A_i)}\right]+o_{\mathbb P}(1).
\end{align*}
We will show in the proof of Lemma \ref{lem:leading_triangular_array_clt} that $\liminf_{n\to\infty}\eta_n^2>0$. Hence, $\eta_n$ is bounded away from zero for all sufficiently large $n$. Therefore,
\begin{align*}
\frac{\sqrt n\left(T(\hat{\bm w}_f)-\mu_f\right)}{\eta_n}=\frac{1}{\sqrt n\eta_n}\sum_{i=1}^n\left[H_n(\bm X_i)-\mathbb E[H_n(\bm X_1)]+\sigma(\bm X_i)\epsilon_i\bm\phi(\bm X_i)^\top\bm c^{(n)}_{j(A_i)}\right]+o_{\mathbb P}(1).
\end{align*}
The first term on the right-hand side converges in distribution to $N(0,1)$ by Lemma \ref{lem:leading_triangular_array_clt}. Slutsky's theorem gives
\begin{align*}
\frac{\sqrt n\left(T(\hat{\bm w}_f)-\mu_f\right)}{\eta_n}\dto N(0,1).
\end{align*}
This completes the proof.
\end{proof}

\begin{proof}[Proof of Lemma \ref{lem:clt_linear_representation}]
By Proposition \ref{lem:weights_characterization},
\begin{align*}
T(\hat{\bm w}_f)=\frac{1}{n}\sum_{i=1}^nY_i\bm\phi(\bm X_i)^\top\hat{\bm c}^{(n)}_{j(A_i)}.
\end{align*}
Using Assumption \ref{assm:signal_plus_noise}, write
\begin{align}
\label{eqn:sn_plus_rn}
T(\hat{\bm w}_f)=S_n+R_n,
\end{align}
where
\begin{align*}
S_n=\frac{1}{n}\sum_{i=1}^n\bm\beta(A_i)^\top\bm\phi(\bm X_i)\bm\phi(\bm X_i)^\top\hat{\bm c}^{(n)}_{j(A_i)},\qquad R_n=\frac{1}{n}\sum_{i=1}^n\sigma(\bm X_i)\epsilon_i\bm\phi(\bm X_i)^\top\hat{\bm c}^{(n)}_{j(A_i)}.
\end{align*}

We first analyze the signal component. Let $\bm\beta_j=\bm\beta(j/m_n)$. By the construction of the bins in Proposition \ref{lem:bj_intervals}, if $A_i\in B^{(j)}$, then $|A_i-j/m_n|\leq 1/m_n$. Hence, using Equation \eqref{eqn:modulus of continuity},
\begin{align*}
\sqrt n\left|S_n-\frac{1}{n}\sum_{i=1}^n\bm\beta_{j(A_i)}^\top\bm\phi(\bm X_i)\bm\phi(\bm X_i)^\top\hat{\bm c}^{(n)}_{j(A_i)}\right|&\leq \sqrt n M(1/m_n)C^2\max_{1\leq j\leq m_n}\|\hat{\bm c}^{(n)}_j\|\\
&=o_{\mathbb P}(1),
\end{align*}
where the last equality follows from Assumption \ref{assm:rate_choice_normality}(a) and Equation \eqref{eqn:chat_uniform_bounded}. Now,
\begin{align*}
\frac{1}{n}\sum_{i=1}^n\bm\beta_{j(A_i)}^\top\bm\phi(\bm X_i)\bm\phi(\bm X_i)^\top\hat{\bm c}^{(n)}_{j(A_i)}=\sum_{j=1}^{m_n}\bm\beta_j^\top\hat{\bm M}_j\hat{\bm c}^{(n)}_j.
\end{align*}
Since $(\hat{\bm M}_j+\lambda_{nj}\bm I_p)\hat{\bm c}^{(n)}_j=\hat{\bm r}_j$, we have $\hat{\bm M}_j\hat{\bm c}^{(n)}_j=\hat{\bm r}_j-\lambda_{nj}\hat{\bm c}^{(n)}_j$. Therefore,
\begin{align*}
\sqrt n\left|\sum_{j=1}^{m_n}\bm\beta_j^\top\hat{\bm M}_j\hat{\bm c}^{(n)}_j-\sum_{j=1}^{m_n}\bm\beta_j^\top\hat{\bm r}_j\right|&\leq \sqrt n C_\beta\max_{1\leq j\leq m_n}\|\hat{\bm c}^{(n)}_j\|\sum_{j=1}^{m_n}\lambda_{nj}\\
&\leq C_\beta\max_{1\leq j\leq m_n}\|\hat{\bm c}^{(n)}_j\|\sqrt n m_n\max_{1\leq j\leq m_n}\lambda_{nj}\\
&=o_{\mathbb P}(1),
\end{align*}
where $C_\beta=\sup_{a\in\mathcal A}\|\bm\beta(a)\|<\infty$, and the last equality follows from Assumption \ref{assm:rate_choice_normality}(b) and Equation \eqref{eqn:chat_uniform_bounded}. The definition of $\hat{\bm r}_j$ gives the exact identity,
\begin{align*}
\sum_{j=1}^{m_n}\bm\beta_j^\top\hat{\bm r}_j=\frac{1}{n}\sum_{i=1}^nH_n(\bm X_i).
\end{align*}
Thus,
\begin{align}
\sqrt n\left(S_n-\frac{1}{n}\sum_{i=1}^nH_n(\bm X_i)\right)=o_{\mathbb P}(1).
\label{eqn:signal_linearization_clt}
\end{align}

Finally, since $\mu_f=\mathbb E[\int_{\mathcal A}\bm\phi(\bm X_1)^\top\bm\beta(a)f(da,\bm X_1)]$, $\mathbb E[H_n(\bm X_1)]=\mathbb E[\int_{\mathcal A}\bm\phi(\bm X_1)^\top\bm\beta_{j(a)}f(da,\bm X_1)]$, and $|a-j(a)/m_n|\le 1/m_n$,
    \begin{align}
    |\mathbb E[H_n(\bm X_1)]-\mu_f|\le CC_f\sup_{a\in\mathcal A}\|\bm\beta_{j(a)}-\bm\beta(a)\|\le CC_f\,M(1/m_n).\label{eqn:Hn_bias_consistency}
    \end{align}
Combining Equations \eqref{eqn:signal_linearization_clt} and \eqref{eqn:Hn_bias_consistency}, we obtain
\begin{align}
\sqrt n(S_n-\mu_f)=\frac{1}{\sqrt n}\sum_{i=1}^n\left(H_n(\bm X_i)-\mathbb E[H_n(\bm X_1)]\right)+o_{\mathbb P}(1).
\label{eqn:signal_linear_representation_clt}
\end{align}
We next analyze the noise component. Define
\begin{align*}
R_n^0=\frac{1}{n}\sum_{i=1}^n\sigma(\bm X_i)\epsilon_i\bm\phi(\bm X_i)^\top\bm c^{(n)}_{j(A_i)}.
\end{align*}
We show that $\sqrt n(R_n-R_n^0)=o_{\mathbb P}(1)$. Again, let $\mathcal F_n$ denote the $\sigma$-algebra generated by $\{(\bm X_i,A_i)\}_{i=1}^n$. Conditional on $\mathcal F_n$, the coefficients $\hat{\bm c}^{(n)}_j$ and $\bm c^{(n)}_j$ are fixed. By Assumptions \ref{assm:unconfound} and \ref{assm:signal_plus_noise}, $\mathbb E[\epsilon_i\mid\mathcal F_n] = 0$, $\operatorname{Var}(\epsilon_i\mid\mathcal F_n) = 1$, and the errors are independent. Therefore,
\begin{align*}
\operatorname{Var}\left(\sqrt n(R_n-R_n^0)\mid\mathcal F_n\right)&=\frac{1}{n}\sum_{i=1}^n\sigma^2(\bm X_i)\left[\bm\phi(\bm X_i)^\top\left(\hat{\bm c}^{(n)}_{j(A_i)}-\bm c^{(n)}_{j(A_i)}\right)\right]^2\\
&\leq C^2\left(\frac{1}{n}\sum_{i=1}^n\sigma^2(\bm X_i)\right)\max_{1\leq j\leq m_n}\|\hat{\bm c}^{(n)}_j-\bm c^{(n)}_j\|^2.
\end{align*}
By Assumption \ref{assm:asymp_regularity}(b), $n^{-1}\sum_{i=1}^n\sigma^2(\bm X_i)=O_{\mathbb P}(1)$, and by Equation \eqref{eqn:chat_uniform_convergence}, $\max_{1\leq j\leq m_n}\|\hat{\bm c}^{(n)}_j-\bm c^{(n)}_j\|=o_{\mathbb P}(1)$. Hence,
\begin{align*}
\operatorname{Var}\left(\sqrt n(R_n-R_n^0)\mid\mathcal F_n\right)=o_{\mathbb P}(1).
\end{align*}
By the same conditional Chebyshev's inequality argument used to establish Equation~\eqref{eqn:noise_negligible_consistency}, $\sqrt n(R_n-R_n^0)=o_{\mathbb P}(1)$. Hence,
\begin{align}
\sqrt nR_n=\frac{1}{\sqrt n}\sum_{i=1}^n\sigma(\bm X_i)\epsilon_i\bm\phi(\bm X_i)^\top\bm c^{(n)}_{j(A_i)}+o_{\mathbb P}(1).
\label{eqn:noise_linear_representation_clt}
\end{align}
Combining Equations \eqref{eqn:signal_linear_representation_clt} and \eqref{eqn:noise_linear_representation_clt} proves Equation \eqref{eqn:clt_linear_representation}.
\end{proof}

\begin{proof}[Proof of Lemma \ref{lem:leading_triangular_array_clt}]
Define
\begin{align*}
\psi_{n,i}=\left(H_n(\bm X_i)-\mathbb E[H_n(\bm X_1)]\right)+\sigma(\bm X_i)\epsilon_i\bm\phi(\bm X_i)^\top\bm c^{(n)}_{j(A_i)}.
\end{align*}
For each fixed $n$, $\psi_{n,1},\ldots,\psi_{n,n}$ are independent and identically distributed. By Assumptions \ref{assm:unconfound} and \ref{assm:signal_plus_noise}, $\mathbb E[\epsilon_i\mid\bm X_i,A_i]=0$, so $\mathbb E[\psi_{n,i}]=0$. Moreover, the covariance between the two summands in $\psi_{n,i}$ is zero because
\begin{align*}
\mathbb E\left[\left(H_n(\bm X_i)-\mathbb E[H_n(\bm X_1)]\right)\sigma(\bm X_i)\epsilon_i\bm\phi(\bm X_i)^\top\bm c^{(n)}_{j(A_i)}\right]=0.
\end{align*}
Therefore, by Equation \eqref{eqn:var_exprn},
\begin{align*}
\operatorname{Var}(\psi_{n,i})=\eta_n^2.
\end{align*}
We next show that $\eta_n^2$ is bounded away from zero. Since the first component in Equation \eqref{eqn:var_exprn} is nonnegative,
\begin{align*}
\eta_n^2&\geq \mathbb E\left[\sigma^2(\bm X_1)\left(\bm\phi(\bm X_1)^\top\bm c^{(n)}_{j(A_1)}\right)^2\right]\\
&\geq \sigma_*^2\sum_{j=1}^{m_n}\mathbb E\left[\left(\bm\phi(\bm X_1)^\top\bm c^{(n)}_j\right)^2\mathbb I(A_1\in B^{(j)})\right]\\
&=\sigma_*^2\sum_{j=1}^{m_n}(\bm c^{(n)}_j)^\top\bm M_j\bm c^{(n)}_j\\
&=\sigma_*^2\sum_{j=1}^{m_n}\bm r_j^\top\bm M_j^{-1}\bm r_j.
\end{align*}
On the other hand, by the boundedness of $\bm\phi(\cdot)$, the upper bound on $e(\cdot,\cdot)$ in Assumption \ref{assm:pos}, and Equation \eqref{eqn:bin_length_bounds} (and similar to the proof of Equation~\eqref{eqn:M_min_eigen_bound}), there exists $C_M^+<\infty$ such that
\begin{align*}
\lambda_{\max}(\bm M_j)\leq \frac{C_M^+}{m_n},\qquad j=1,\ldots,m_n.
\end{align*}
Thus,
\begin{align*}
\sum_{j=1}^{m_n}\bm r_j^\top\bm M_j^{-1}\bm r_j\geq \frac{m_n}{C_M^+}\sum_{j=1}^{m_n}\|\bm r_j\|^2\geq \frac{1}{C_M^+}\left(\sum_{j=1}^{m_n}\|\bm r_j\|\right)^2,
\end{align*}
where the last inequality follows from Cauchy--Schwarz. Assumption \ref{assm:asymp_regularity}(c) therefore implies
\begin{align}
\liminf_{n\to\infty}\eta_n^2>0.
\label{eqn:eta_non_degenerate_clt}
\end{align}
To prove the CLT, we verify Lyapunov's condition for the triangular array $\{\psi_{n,i}/\eta_n:1\leq i\leq n\}$. Since $\operatorname{Var}(\psi_{n,i})=\eta_n^2$, Lyapunov's condition requires that, for some $\delta>0$,
\begin{align}
\frac{1}{(n\eta_n^2)^{1+\delta/2}}\sum_{i=1}^n\mathbb E[|\psi_{n,i}|^{2+\delta}]\to0.
\label{eqn:lyapunov_condition}
\end{align}
Because $\psi_{n,1},\ldots,\psi_{n,n}$ are identically distributed for each fixed $n$, the left-hand side of Equation \eqref{eqn:lyapunov_condition} equals
\begin{align*}
\frac{n\mathbb E[|\psi_{n,1}|^{2+\delta}]}{(n\eta_n^2)^{1+\delta/2}}=\frac{\mathbb E[|\psi_{n,1}|^{2+\delta}]}{n^{\delta/2}(\eta_n^2)^{1+\delta/2}}.
\end{align*}
We now show that the numerator is uniformly bounded. Next, note that $H_n$ is uniformly bounded in $n$ under Assumption~\ref{assm:regularity}(a) and (b) and the continuity of $\bm \beta()$. Also, Equation \eqref{eqn:c_population_bound} gives $\max_{1\leq j\leq m_n}\|\bm c^{(n)}_j\|=O(1)$. Since $\bm\phi(\cdot)$ is bounded by Assumption \ref{assm:regularity}(a), $\sigma(\cdot)$ is bounded above by Assumption \ref{assm:asymp_regularity}(b), and Assumption \ref{assm:asymp_regularity}(a) imposes a $(2+\delta)$ moment condition on $\epsilon_i$, there exists a constant $C_\psi<\infty$ such that
\begin{align}
\sup_{n\geq1}\mathbb E[|\psi_{n,1}|^{2+\delta}]\leq C_\psi.
\label{eqn:psi_uniform_moment_bound}
\end{align}
By Equation \eqref{eqn:eta_non_degenerate_clt}, there exist constants $\underline \eta>0$ and $N<\infty$ such that $\eta_n^2\geq \underline \eta$ for all $n\geq N$. Therefore, for all $n\geq N$, Equations \eqref{eqn:psi_uniform_moment_bound} and \eqref{eqn:eta_non_degenerate_clt} imply
\begin{align*}
\frac{1}{(n\eta_n^2)^{1+\delta/2}}\sum_{i=1}^n\mathbb E[|\psi_{n,i}|^{2+\delta}]
&=\frac{\mathbb E[|\psi_{n,1}|^{2+\delta}]}{n^{\delta/2}(\eta_n^2)^{1+\delta/2}}\\
&\leq \frac{C_\psi}{n^{\delta/2}\underline \eta^{1+\delta/2}}\to0.
\end{align*}
Thus, Lyapunov's condition in Equation \eqref{eqn:lyapunov_condition} holds. Therefore, Lyapunov's central limit theorem for triangular arrays implies Equation \eqref{eqn:leading_triangular_array_clt}.
\end{proof}

\subsection{Proof of Proposition~\ref{lem:limiting_variance}}

\begin{proof}
The absolute continuity of $\bm Q$ follows directly from Assumption~\ref{assm:regularity}(c) and the boundedness of $\bm \phi(\bm x)$ due to Assumption~\ref{assm:regularity}(a).

We make another observation that the continuity of $a\mapsto e(a, \bm x)$ implying the continuity of the map $a\mapsto \bm \phi(\bm X_1)\bm \phi(\bm X_1)^Te(a, \bm X_1)$ together with the boundedness of the latter map, implies the continuity of $a\mapsto \bm M(a) = \mathbb E\left[\bm \phi(\bm X_1)\bm \phi(\bm X_1)^Te(a, \bm X_1)\right]$, due to the Dominated Convergence Theorem.

Now, recall from Equation \eqref{eqn:var_exprn} that
\begin{align*}
\eta_n^2=\operatorname{Var}(H_n(\bm X_1))+\mathbb E\left[\sigma^2(\bm X_1)\left(\bm\phi(\bm X_1)^\top\bm c^{(n)}_{j(A_1)}\right)^2\right].
\end{align*}
We analyze the two terms separately.

First, define
\begin{align*}
G_f(\bm X_1)=\int_{\mathcal A}\bm\phi(\bm X_1)^\top\bm\beta(a)f(da,\bm X_1).
\end{align*}
By Equation \eqref{eqn:Hn_definition_consistency},
\begin{align*}
H_n(\bm X_1)=\sum_{j=1}^{m_n}\bm\beta(j/m_n)^\top\bm\phi(\bm X_1)f(B^{(j)},\bm X_1)=\int_{\mathcal A}\bm\phi(\bm X_1)^\top\bm\beta(j(a)/m_n)f(da,\bm X_1),
\end{align*}
where, recall that $a\in B^{(j(a))}$. Therefore, by Assumption \ref{assm:regularity}(a), Assumption \ref{assm:regularity}(b), and the definition of the modulus of continuity in Equation \eqref{eqn:modulus of continuity},
\begin{align*}
|H_n(\bm X_1)-G_f(\bm X_1)|&\leq \int_{\mathcal A}\left|\bm\phi(\bm X_1)^\top\left(\bm\beta(j(a)/m_n)-\bm\beta(a)\right)\right||f|(da,\bm X_1)\\
&\leq C M(1/m_n)|f|(\mathcal A,\bm X_1)\\
&\leq CC_fM(1/m_n).
\end{align*}
Since $M(1/m_n)\to0$, it follows that $H_n(\bm X_1)\to G_f(\bm X_1)$ in $L^2$. Hence,
\begin{align}
\operatorname{Var}(H_n(\bm X_1))\to \operatorname{Var}(G_f(\bm X_1)).
\label{eqn:Hn_variance_limit}
\end{align}
We next identify $G_f(\bm X_1)$ with the conditional mean in the statement of the proposition. By Assumption \ref{assm:signal_plus_noise},
\begin{align*}
Y_1(a)=\bm\phi(\bm X_1)^\top\bm\beta(a)+\sigma(\bm X_1)\epsilon_1(a).
\end{align*}
Now note that $f(\cdot,\bm X_1)$ is a finite signed measure, $|f|(\mathcal A,\bm X_1)$ is uniformly bounded by Assumption \ref{assm:regularity}(b), $\bm\phi(\cdot)$ and $\bm\beta(\cdot)$ are bounded, $\sigma(\cdot)$ is bounded above by Assumption \ref{assm:asymp_regularity}(b), and Assumption \ref{assm:asymp_regularity}(a) gives that $\sup_{a\in \mathcal A}\mathbb E\left[|\epsilon_1(a)|\right]<\infty$. Thus, we get that
\begin{align*}
    \int_{\mathcal A}\mathbb E\left[|Y_1(a)|\mid \bm X_1\right]|f|(da, \bm X_1)\leq \int_{\mathcal A}\left[CC_{\beta} + {\sigma^*}\sup_{a\in \mathcal A}\mathbb E\left(|\epsilon_1(a)|\right)\right]|f|(da, \bm X_1)<\infty,
\end{align*}
where $C_{\beta}$ is a bound on $\|\bm \beta(\cdot)\|$. Thus, by Tonelli's theorem,
\begin{align*}
\mathbb E\left[\int_{\mathcal A}Y_1(a)f(da,\bm X_1)\middle| \bm X_1\right]=\int_{\mathcal A}\mathbb E[Y_1(a)\mid \bm X_1]f(da,\bm X_1) =\int_{\mathcal A}\bm\phi(\bm X_1)^\top\bm\beta(a)f(da,\bm X_1) =G_f(\bm X_1),
\end{align*}
where the second equality uses $\mathbb E[\epsilon_1(a)\mid \bm X_1]=0$ from Assumption \ref{assm:signal_plus_noise}. Therefore,
\begin{align}
\operatorname{Var}(H_n(\bm X_1))\to \operatorname{Var}\left(\mathbb E\left[\int_{\mathcal A}Y_1(a)f(da,\bm X_1)\middle| \bm X_1\right]\right).
\label{eqn:signal_variance_limit}
\end{align}

We next analyze the second term in Equation \eqref{eqn:var_exprn}. For each bin, write
\begin{align*}
\bm r_j=\mathbb E[\bm\phi(\bm X_1)f(B^{(j)},\bm X_1)]=\bm Q(B^{(j)})=\int_{B^{(j)}}\bm q(a)da
\end{align*}
and
\begin{align*}
\bm M_j=\mathbb E[\bm\phi(\bm X_1)\bm\phi(\bm X_1)^\top\mathbb I(A_1\in B^{(j)})]=\int_{B^{(j)}}\bm M(a)da.
\end{align*}
Let $a_j=j/m_n$ and define
\begin{align*}
\bar{\bm M}_j=\frac{1}{|B^{(j)}|}\int_{B^{(j)}}\bm M(a)da,\qquad \bar{\bm q}_j=\frac{1}{|B^{(j)}|}\int_{B^{(j)}}\bm q(a)da.
\end{align*}
This gives $\bm c_j^{(n)}=\bm M_j^{-1}\bm r_j=\bar{\bm M}_j^{-1}\bar{\bm q}_j$. We now show that these bin averages of $\bm M(a)$ and $\bm q(a)$ converge uniformly to their values at the grid points. Since $a\mapsto\bm M(a)$ and $a\mapsto\bm q(a)$ are continuous on the compact set $\mathcal A$, they are uniformly continuous. Define the moduli of continuity
\begin{align*}
\omega_M(\delta)=\sup_{|a-b|\leq\delta}\|\bm M(a)-\bm M(b)\|,\qquad \omega_q(\delta)=\sup_{|a-b|\leq\delta}\|\bm q(a)-\bm q(b)\|.
\end{align*}
Then $\omega_M(\delta)\to0$ and $\omega_q(\delta)\to0$ as $\delta\downarrow0$. Moreover, by the construction of the bins in Proposition \ref{lem:bj_intervals}, for every $a\in B^{(j)}$, $|a-a_j|\leq 1/m_n$. Therefore,
\begin{align*}
\|\bar{\bm M}_j-\bm M(a_j)\|&=\left\|\frac{1}{|B^{(j)}|}\int_{B^{(j)}}\left(\bm M(a)-\bm M(a_j)\right)da\right\| \leq \frac{1}{|B^{(j)}|}\int_{B^{(j)}}\|\bm M(a)-\bm M(a_j)\|da \leq \omega_M(1/m_n),
\end{align*}
and similarly,
\begin{align*}
\|\bar{\bm q}_j-\bm q(a_j)\|\leq \omega_q(1/m_n).
\end{align*}
Taking maxima over $j=1,\ldots,m_n$ gives
\begin{align}
\max_{1\leq j\leq m_n}\|\bar{\bm M}_j-\bm M(a_j)\|\to0,\qquad \max_{1\leq j\leq m_n}\|\bar{\bm q}_j-\bm q(a_j)\|\to0.
\label{eqn:bin_average_uniform_convergence}
\end{align}

Moreover, by strong positivity in Assumption \ref{assm:pos} and the full-rank condition in Assumption \ref{assm:regularity}(d), there exists $\underline \kappa>0$ such that $\lambda_{\min}(\bm M(a))\geq \underline \kappa,\forall a\in\mathcal A$. This implies that $\lambda_{\min}(\bar{\bm M}_j)\geq \underline \kappa,\forall j=1,\ldots,m_n$. 

Thus, $\|\bm M(a_j)^{-1}\|\leq \underline \kappa^{-1}$ and $\|\bar{\bm M}_j^{-1}\|\leq \underline \kappa^{-1}$. Using the identity $A^{-1}-B^{-1}=A^{-1}(B-A)B^{-1}$ with $A=\bar{\bm M}_j$ and $B=\bm M(a_j)$,
\begin{align*}
\|\bar{\bm M}_j^{-1}-\bm M(a_j)^{-1}\|&\leq \|\bar{\bm M}_j^{-1}\|\|\bar{\bm M}_j-\bm M(a_j)\|\|\bm M(a_j)^{-1}\|\\
&\leq \underline \kappa^{-2}\|\bar{\bm M}_j-\bm M(a_j)\|.
\end{align*}
Combining this with Equation \eqref{eqn:bin_average_uniform_convergence},
\begin{align}
\max_{1\leq j\leq m_n}\|\bar{\bm M}_j^{-1}-\bm M(a_j)^{-1}\|\to0.
\label{eqn:bin_inverse_uniform_convergence}
\end{align}
Define $\bm c(a)=\bm M(a)^{-1}\bm q(a)$. Since $\bm q(\cdot)$ is continuous on a compact set, there exists $C_q<\infty$ such that $\sup_{a\in\mathcal A}\|\bm q(a)\|\leq C_q$. Hence,
\begin{align*}
\|\bm c_j^{(n)}-\bm c(a_j)\|&=\|\bar{\bm M}_j^{-1}\bar{\bm q}_j-\bm M(a_j)^{-1}\bm q(a_j)\|\\
&\leq \|\bar{\bm M}_j^{-1}\|\|\bar{\bm q}_j-\bm q(a_j)\|+\|\bar{\bm M}_j^{-1}-\bm M(a_j)^{-1}\|\|\bm q(a_j)\|\\
&\leq \underline \kappa^{-1}\|\bar{\bm q}_j-\bm q(a_j)\|+C_q\|\bar{\bm M}_j^{-1}-\bm M(a_j)^{-1}\|.
\end{align*}
Taking maxima and using Equations \eqref{eqn:bin_average_uniform_convergence} and \eqref{eqn:bin_inverse_uniform_convergence}, we obtain
\begin{align}
\max_{1\leq j\leq m_n}\|\bm c_j^{(n)}-\bm c(a_j)\|\to0.
\label{eqn:c_population_uniform_limit}
\end{align}

Now, the second term in Equation \eqref{eqn:var_exprn} can be written as
\begin{align*}
\mathbb E\left[\sigma^2(\bm X_1)\left(\bm\phi(\bm X_1)^\top\bm c^{(n)}_{j(A_1)}\right)^2\right]=\sum_{j=1}^{m_n}\int_{B^{(j)}}\mathbb E\left[\sigma^2(\bm X_1)e(a,\bm X_1)\left(\bm\phi(\bm X_1)^\top\bm c_j^{(n)}\right)^2\right]da.
\end{align*}
We first replace $\bm c_j^{(n)}$ with $\bm c(a_j)$. Let $\Delta_n=\max_{1\leq j\leq m_n}\|\bm c_j^{(n)}-\bm c(a_j)\|$. By Equation \eqref{eqn:c_population_uniform_limit}, $\Delta_n\to0$. Also, Equation \eqref{eqn:c_population_bound}, the uniform boundedness of $\bm q(\cdot)$, and the uniform lower eigenvalue bound on $\bm M(\cdot)$ imply that $\max_j\|\bm c_j^{(n)}\|+\sup_{a\in\mathcal A}\|\bm c(a)\|=O(1)$. Therefore, using $|u^2-v^2|\leq |u-v|(|u|+|v|)$,
\begin{align*}
&\sum_{j=1}^{m_n}\int_{B^{(j)}}\left|\mathbb E\left[\sigma^2(\bm X_1)e(a,\bm X_1)\left[\left(\bm\phi(\bm X_1)^\top\bm c_j^{(n)}\right)^2-\left(\bm\phi(\bm X_1)^\top\bm c(a_j)\right)^2\right]\right]\right|da\\
&\quad\leq \sum_{j=1}^{m_n}\int_{B^{(j)}}\mathbb E\left[\sigma^2(\bm X_1)e(a,\bm X_1)\left|\bm\phi(\bm X_1)^\top\left(\bm c_j^{(n)}-\bm c(a_j)\right)\right|\left|\bm\phi(\bm X_1)^\top\left(\bm c_j^{(n)}+\bm c(a_j)\right)\right|\right]da\\
&\quad\leq C'\Delta_n\sum_{j=1}^{m_n}|B^{(j)}|\\
&\quad=C'\Delta_n\to0,
\end{align*}
for some constant $C'<\infty$, where we used the boundedness of $\bm\phi(\cdot)$, $\sigma(\cdot)$, and $e(\cdot,\cdot)$. Thus,
\begin{align}
\sum_{j=1}^{m_n}\int_{B^{(j)}}\left|\mathbb E\left[\sigma^2(\bm X_1)e(a,\bm X_1)\left[\left(\bm\phi(\bm X_1)^\top\bm c_j^{(n)}\right)^2-\left(\bm\phi(\bm X_1)^\top\bm c(a_j)\right)^2\right]\right]\right|da\to0.
\label{eqn:noise_replace_cj_by_ca}
\end{align}

Next, we replace $\bm c(a_j)$ by $\bm c(a)$ inside each bin. Since $\bm M(\cdot)$ and $\bm q(\cdot)$ are continuous, and $\lambda_{\min}(\bm M(a))$ is uniformly bounded away from zero, the map $a\mapsto\bm c(a)=\bm M(a)^{-1}\bm q(a)$ is continuous on the compact set $\mathcal A$, hence uniformly continuous. Let
\begin{align*}
\omega_c(\delta)=\sup_{|a-b|\leq\delta}\|\bm c(a)-\bm c(b)\|.
\end{align*}
Then $\omega_c(\delta)\to0$ as $\delta\downarrow0$. For every $a\in B^{(j)}$, $|a-a_j|\leq 1/m_n$, so $\|\bm c(a_j)-\bm c(a)\|\leq \omega_c(1/m_n)$. Using the same inequality $|u^2-v^2|\leq |u-v|(|u|+|v|)$ and the same boundedness arguments as above, for some $C''<\infty$,
\begin{align*}
&\sum_{j=1}^{m_n}\int_{B^{(j)}}\left|\mathbb E\left[\sigma^2(\bm X_1)e(a,\bm X_1)\left[\left(\bm\phi(\bm X_1)^\top\bm c(a_j)\right)^2-\left(\bm\phi(\bm X_1)^\top\bm c(a)\right)^2\right]\right]\right|da\\
&\quad\leq C''\omega_c(1/m_n)\sum_{j=1}^{m_n}|B^{(j)}|\\
&\quad=C''\omega_c(1/m_n)\to0.
\end{align*}
Consequently,
\begin{align}
\sum_{j=1}^{m_n}\int_{B^{(j)}}\mathbb E\left[\sigma^2(\bm X_1)e(a,\bm X_1)\left(\bm\phi(\bm X_1)^\top\bm c(a_j)\right)^2\right]da-\int_{\mathcal A}\mathbb E\left[\sigma^2(\bm X_1)e(a,\bm X_1)\left(\bm\phi(\bm X_1)^\top\bm c(a)\right)^2\right]da\to0.
\label{eqn:noise_replace_caj_by_ca}
\end{align}
Combining Equations \eqref{eqn:noise_replace_cj_by_ca} and \eqref{eqn:noise_replace_caj_by_ca}, we obtain
\begin{align}
\mathbb E\left[\sigma^2(\bm X_1)\left(\bm\phi(\bm X_1)^\top\bm c^{(n)}_{j(A_1)}\right)^2\right]\to \int_{\mathcal A}\mathbb E\left[\sigma^2(\bm X_1)e(a,\bm X_1)\left(\bm\phi(\bm X_1)^\top\bm M(a)^{-1}\bm q(a)\right)^2\right]da.
\label{eqn:noise_variance_limit}
\end{align}
Combining Equations \eqref{eqn:signal_variance_limit} and \eqref{eqn:noise_variance_limit} yields
\begin{align*}
\eta_n^2\to \operatorname{Var}\left(\mathbb E\left[\int_{\mathcal A}Y_1(a)f(da,\bm X_1)\middle| \bm X_1\right]\right)+\int_{\mathcal A}\mathbb E\left[\sigma^2(\bm X_1)e(a,\bm X_1)\left(\bm\phi(\bm X_1)^\top\bm M(a)^{-1}\bm q(a)\right)^2\right]da.
\end{align*}
This completes the proof.
\end{proof}




\subsection{Proof of Theorem~\ref{thm:bal_eff_gain}}
\begin{proof}
Throughout, $\sigma^2(\bm x)=\sigma^2$ and $f(da,\bm x)=f(a,\bm x)da$. Let $\epsilon_1:=\epsilon_1(A_1)$. By Assumptions~\ref{assm:unconfound} and \ref{assm:signal_plus_noise_general}, $\mathbb E[\epsilon_1\mid A_1,\bm X_1]=0$ and $\mathbb E[\epsilon_1^2\mid A_1,\bm X_1]=1$. Moreover, since $A_1\mid\bm X_1$ has density $e(\cdot,\bm X_1)$, for any integrable $g$, $\mathbb E[g(A_1,\bm X_1)/e(A_1,\bm X_1)\mid\bm X_1]=\int_{\mathcal A}g(a,\bm X_1)da$.

Let us first consider the asymptotic variance of the IPW estimator. Note that $T_{\rm IPW}$ is an average of the i.i.d.\ terms $\zeta_i:=Y_if(A_i,\bm X_i)/e(A_i,\bm X_i)$ (Assumption~\ref{assm:iid}). Since $Y_1=m(A_1,\bm X_1)+\sigma\epsilon_1$, we can write
\[
\zeta_1=m(A_1,\bm X_1)\frac{f(A_1,\bm X_1)}{e(A_1,\bm X_1)}+\sigma\epsilon_1\frac{f(A_1,\bm X_1)}{e(A_1,\bm X_1)}.
\]
The second term has mean zero conditional on $(A_1,\bm X_1)$, so it is uncorrelated with the first, and its variance is $\sigma^2\mathbb E[f^2(A_1,\bm X_1)/e^2(A_1,\bm X_1)]=\sigma^2\int_{\mathcal A}\mathbb E[f^2(a,\bm X_1)/e(a,\bm X_1)]da$. This gives the stated expression for $\sigma^2_{\rm IPW}$.

We now turn to the asymptotic variance of the balancing estimator. By Proposition~\ref{lem:limiting_variance}, under homoskedasticity,
\[
\sigma^2_{\rm bal}=\Var\left(\mathbb E\left[\int_{\mathcal A}Y_1(a)f(da,\bm X_1)\,\middle|\,\bm X_1\right]\right)+\sigma^2\int_{\mathcal A}\mathbb E\left[e(a,\bm X_1)\left\{\bm\phi(\bm X_1)^\top\bm M(a)^{-1}\bm q(a)\right\}^2\right]da,
\]
where $\bm M(a)=\mathbb E[\bm\phi(\bm X_1)\bm\phi(\bm X_1)^\top e(a,\bm X_1)]$ is invertible, since $\bm M(a)\succeq\alpha\,\mathbb E[\bm\phi(\bm X_1)\bm\phi(\bm X_1)^\top]$ by Assumptions~\ref{assm:pos} and~\ref{assm:regularity}(d). For the first component, interchanging the integral and the conditional expectation as in the proof of Proposition~\ref{lem:limiting_variance}, we obtain
\[
\mathbb E\left[\int_{\mathcal A}Y_1(a)f(da,\bm X_1)\,\middle|\,\bm X_1\right]=\int_{\mathcal A}m(a,\bm X_1)f(a,\bm X_1)da=\mathbb E\left[m(A_1,\bm X_1)\frac{f(A_1,\bm X_1)}{e(A_1,\bm X_1)}\,\middle|\,\bm X_1\right],
\]
which yields the stated first component of $\sigma^2_{\rm bal}$.

For the second component, when $f$ is a density, we have $\bm q(a)=\mathbb E\left[\bm\phi(\bm X_1)f(a,\bm X_1)\right]$ (see Proposition~\ref{lem:limiting_variance}). Then, under homoskedasticity, the second component of $\sigma^2_{\rm bal}$ equals
\[
\begin{aligned}
& \sigma^2\int_{\mathcal A}\mathbb E\left[e(a,\bm X_1)\left\{\bm\phi(\bm X_1)^\top\bm M(a)^{-1}\mathbb E[\bm\phi(\bm X_1)f(a,\bm X_1)]\right\}^2\right]da\\
= \ & \sigma^2\int_{\mathcal A}\mathbb E\left[\left\{\bm\Pi_a\left(\frac{f(a,\bm X_1)}{\sqrt{e(a,\bm X_1)}}\right)\right\}^2\right]da,
\end{aligned}
\]
where $\bm\Pi_a$ is the projection operator defined in the statement of the theorem. Indeed, writing $\bm u_a:=\bm\phi(\bm X_1)\sqrt{e(a,\bm X_1)}$ and $\ell_a:=f(a,\bm X_1)/\sqrt{e(a,\bm X_1)}$, we have $\mathbb E[\bm u_a\bm u_a^\top]=\bm M(a)$ and $\mathbb E[\bm u_a\ell_a]=\bm q(a)$, so that $\bm\Pi_a(\ell_a)=\bm u_a^\top\bm M(a)^{-1}\bm q(a)=\sqrt{e(a,\bm X_1)}\,\bm\phi(\bm X_1)^\top\bm M(a)^{-1}\bm q(a)$.

It remains to compare the two variances component by component. For the first components, the law of total variance gives
\[
\Var\left(\mathbb E\left[m(A_1,\bm X_1)\frac{f(A_1,\bm X_1)}{e(A_1,\bm X_1)}\,\middle|\,\bm X_1\right]\right)\leq\Var\left(m(A_1,\bm X_1)\frac{f(A_1,\bm X_1)}{e(A_1,\bm X_1)}\right).
\]
For the second components, fix $a$. The variable $\ell_a$ is square integrable, since $|f|\leq C_f$ and $e\geq\alpha$ (Assumptions~\ref{assm:regularity}(b) and~\ref{assm:pos}). Moreover, $\mathbb E[\bm u_a\{\ell_a-\bm\Pi_a(\ell_a)\}]=\bm q(a)-\bm M(a)\bm M(a)^{-1}\bm q(a)=\bm 0$, so $\bm\Pi_a$ is the $\mathcal L^2$ orthogonal projection onto the span of the entries of $\bm u_a$. Hence,
\[
\mathbb E\left[\ell_a^2\right]=\mathbb E\left[\{\bm\Pi_a(\ell_a)\}^2\right]+\mathbb E\left[\{\ell_a-\bm\Pi_a(\ell_a)\}^2\right]\geq\mathbb E\left[\{\bm\Pi_a(\ell_a)\}^2\right].
\]
Since $\mathbb E[\ell_a^2]=\mathbb E[f^2(a,\bm X_1)/e(a,\bm X_1)]$, integrating over $a\in\mathcal A$ and multiplying by $\sigma^2$ shows that the second component of $\sigma^2_{\rm bal}$ is at most that of $\sigma^2_{\rm IPW}$. Combining the two comparisons yields $\sigma^2_{\rm bal}\leq\sigma^2_{\rm IPW}$.
\end{proof}
\subsection{Proof of Theorem~\ref{thm:variance_estimation}}
\label{sec:proof_variance_estimation}

We first give the full statement of
Theorem~\ref{thm:variance_estimation}.

\begin{theorem*}[Full statement of Theorem~\ref{thm:variance_estimation}]
    Suppose Assumption~\ref{assm:variance_estimation} holds in addition to the assumptions of
    Theorem~\ref{thm:clt}. For each $i=1,\dots,n$, define the estimated influence-function value
    \begin{align}
    \label{eqn:if_estimator}
        \hat\psi_i=\hat w_{f,i}\left(Y_i-\hat m_{\rm bal}(A_i,\bm X_i)\right)
        +\int_{\mathcal A}\hat m_{\rm bal}(a,\bm X_i)f(da,\bm X_i)-T(\hat{\bm w}_f),
    \end{align}
    and let $\bar{\hat\psi}=n^{-1}\sum_{i=1}^n\hat\psi_i$. Then the variance estimator
    \begin{align}
    \label{eqn:var_estimator}
        \hat\eta_n^2=\frac{1}{n}\sum_{i=1}^n\left(\hat\psi_i-\bar{\hat\psi}\right)^2
    \end{align}
    satisfies $\hat\eta_n^2-\eta_n^2=o_{\mathbb P}(1)$. Consequently, for any $\alpha\in(0,1)$, an asymptotic
    $100(1-\alpha)\%$ confidence interval for $\mu_f$ is
    \begin{align*}
        C_f^{(1-\alpha)}=\left(T(\hat{\bm w}_f)\pm z_{1-\alpha/2}\frac{\hat\eta_n}{\sqrt n}\right).
    \end{align*}
\end{theorem*}
\begin{proof}[Proof of Theorem \ref{thm:variance_estimation}]
The proof uses two lemmas stated below.

\begin{lemma}
\label{lem:variance_outcome_side_uniform_bounds}
Under Assumptions \ref{assm:iid}--\ref{assm:variance_estimation},
\begin{align*}
\max_{1\leq j\leq m_n}\left\|\hat{\bm q}_j-\mathbb E[\hat{\bm q}_j]\right\|=o_{\mathbb P}(m_n^{-1}),
\qquad
\max_{1\leq j\leq m_n}\left\|\hat{\bm\beta}^{(n)}_j-\bm\beta_j\right\|=o_{\mathbb P}(1),
\end{align*}
and consequently $\max_{1\leq j\leq m_n}\|\hat{\bm\beta}^{(n)}_j\|=O_{\mathbb P}(1)$. Here, recall the notations $\hat{\bm q}_j$ and $\hat{\bm \beta}_j^{(n)}$ from Equation~\ref{eqn:or_model}.
\end{lemma}

\begin{lemma}
\label{lem:estimated_influence_l2_consistency}
Let
\begin{align*}
\psi_{n,i}=H_n(\bm X_i)-\mathbb E[H_n(\bm X_1)]
+\sigma(\bm X_i)\epsilon_i\bm\phi(\bm X_i)^\top\bm c^{(n)}_{j(A_i)},
\end{align*}
with $H_n$ as in Equation~\eqref{eqn:Hn_definition_consistency}. Then, under
Assumptions \ref{assm:iid}--\ref{assm:variance_estimation}, there is a sequence $\Delta_n=o_{\mathbb P}(1)$,
not depending on $i$, such that
\begin{align}
\left|\hat\psi_i-\psi_{n,i}\right|\leq \Delta_n\left(1+|Y_i|\right),\qquad i=1,\dots,n.
\label{eqn:if_uniform_approximation}
\end{align}
Consequently,
\begin{align}
\frac{1}{n}\sum_{i=1}^n\left(\hat\psi_i-\psi_{n,i}\right)^2=o_{\mathbb P}(1).
\label{eqn:if_l2_consistency_variance}
\end{align}
\end{lemma}

For $\bm u=(u_1,\dots,u_n)^\top\in\mathbb R^n$ write $\|\bm u\|_n:=(n^{-1}\sum_{i=1}^nu_i^2)^{1/2}$, and set
$\hat{\bm\psi}=(\hat\psi_1,\dots,\hat\psi_n)^\top$ and $\bm\psi_n=(\psi_{n,1},\dots,\psi_{n,n})^\top$. 

We first begin with showing that $\eta_n^2 = O(1)$. The proof of Lemma~\ref{lem:leading_triangular_array_clt} shows that, for
each fixed $n$, the variables $\psi_{n,1},\dots,\psi_{n,n}$ are i.i.d.\ with $\mathbb E[\psi_{n,1}]=0$ and
$\mathbb E[\psi_{n,1}^2]=\operatorname{Var}(\psi_{n,1})=\eta_n^2$, and that
$\sup_{n\geq1}\mathbb E[|\psi_{n,1}|^{2+\delta}]\leq C_\psi<\infty$ by
Equation~\eqref{eqn:psi_uniform_moment_bound}, where $\delta$ is the exponent of
Assumption~\ref{assm:asymp_regularity}(a). Now, $\eta_n^2 = \mathbb E\left[ \psi_{n,1}^2\right] =  \mathbb E\left[ \left(\psi_{n,1}^{2+\delta}\right)^{\frac{2}{2+\delta}}\right]\leq \left(\mathbb E[\psi_{n,1}^2]\right)^{\frac{2}{2+\delta}} \leq C_\psi^{2/(2+\delta)}$, where the second to last inequality follows from Jensen's inequality; so
$\eta_n^2=O(1)$.

Next, the proof of Theorem~\ref{thm:variance_estimation} follows the following two steps: (1) We show that $\|\hat{\bm \psi}\|^2 = \eta_n^2 + o_{\mathbb P}(1)$, followed by that  (2) $\bar{\hat{\psi}}:=\frac{1}{n}\sum_{i=1}^n \psi_{i,n}$.

To establish (1), let $\kappa=\min\{\delta/2,1\}$ and $U_{n,i}=\psi_{n,i}^2-\eta_n^2$. Then $\mathbb E[U_{n,i}]=0$ and, since
$2+2\kappa\leq2+\delta$ and $\eta_n^2=O(1)$, we have $\sup_{n\geq1}\mathbb E[|U_{n,1}|^{1+\kappa}]<\infty$.
As $1+\kappa\in[1,2]$, the von Bahr--Esseen \citep{vonbahresseen1965} inequality gives
\begin{align*}
\mathbb E\left[\left|\frac1n\sum_{i=1}^nU_{n,i}\right|^{1+\kappa}\right]
\leq\frac{2}{n^{1+\kappa}}\sum_{i=1}^n\mathbb E\left[|U_{n,i}|^{1+\kappa}\right]\lesssim n^{-\kappa}\to0,
\end{align*}
and therefore, $\frac{1}{n}\sum_{i=1}^n U_{n,i}\stackrel{\mathcal L^{1+k}}{\to} 0\implies \frac{1}{n}\sum_{i=1}^n U_{n,i}\Pto 0$, and hence,
\begin{align}
\|\bm\psi_n\|^2=\frac1n\sum_{i=1}^n\psi_{n,i}^2=\eta_n^2+o_{\mathbb P}(1)=O_{\mathbb P}(1).
\label{eqn:oracle_if_second_moment_lln}
\end{align}

By Equation~\eqref{eqn:if_l2_consistency_variance},
$\|\hat{\bm\psi}-\bm\psi_n\|=o_{\mathbb P}(1)$. Hence the triangle inequality together
with Equation~\eqref{eqn:oracle_if_second_moment_lln} gives $\|\hat{\bm\psi}\|_n=O_{\mathbb P}(1)$ and
\begin{align*}
\left|\,\|\hat{\bm\psi}\|^2-\|\bm\psi_n\|^2\,\right|
=\left|\,\|\hat{\bm\psi}\|-\|\bm\psi_n\|\,\right|
\left(\|\hat{\bm\psi}\|+\|\bm\psi_n\|\right)
\leq\|\hat{\bm\psi}-\bm\psi_n\|\cdot O_{\mathbb P}(1)=o_{\mathbb P}(1),
\end{align*}
so that
\begin{align}
\frac1n\sum_{i=1}^n\hat\psi_i^2=\eta_n^2+o_{\mathbb P}(1).
\label{eqn:hatpsi_second_moment}
\end{align}

Turning to (2), note that averaging Equation~\eqref{eqn:if_estimator} over $i$ and using
Equation~\eqref{eqn:or_model}, we get,
\begin{align*}
\bar{\hat\psi}
=T(\hat{\bm w}_f)-\frac1n\sum_{i=1}^n\hat w_{f,i}\,\hat m_{\rm bal}(A_i,\bm X_i)
=T(\hat{\bm w}_f)-\sum_{j=1}^{m_n}\left(\hat{\bm c}^{(n)}_j\right)^\top\hat{\bm M}_j\hat{\bm\beta}^{(n)}_j,
\end{align*}
the last equality by grouping the observations according to the bin containing $A_i$ and using the
definition of $\hat{\bm M}_j$. Since $\hat{\bm M}_j$ is symmetric with
$\hat{\bm M}_j\hat{\bm c}^{(n)}_j=\hat{\bm r}_j-\lambda_{nj}\hat{\bm c}^{(n)}_j$, and since
$\sum_{j=1}^{m_n}\hat{\bm r}_j^\top\hat{\bm\beta}^{(n)}_j=T(\hat{\bm w}_f)$ by
Equation~\eqref{eqn:or_model}, the two copies of $T(\hat{\bm w}_f)$ cancel and
\begin{align}
\bar{\hat\psi}=\sum_{j=1}^{m_n}\lambda_{nj}\left(\hat{\bm c}^{(n)}_j\right)^\top\hat{\bm\beta}^{(n)}_j.
\label{eqn:psi_bar_ridge_slack}
\end{align}
Therefore, by Equation~\eqref{eqn:chat_uniform_bounded},
Lemma~\ref{lem:variance_outcome_side_uniform_bounds} and Assumption~\ref{assm:rate_choice_consistency}(b),
\begin{align*}
\left|\bar{\hat\psi}\right|
\leq\left(m_n\max_{1\leq j\leq m_n}\lambda_{nj}\right)
\max_{1\leq j\leq m_n}\left\|\hat{\bm c}^{(n)}_j\right\|
\max_{1\leq j\leq m_n}\left\|\hat{\bm\beta}^{(n)}_j\right\|=o_{\mathbb P}(1).
\end{align*}
Since $\hat\eta_n^2=n^{-1}\sum_{i=1}^n\hat\psi_i^2-\bar{\hat\psi}^2$,
Equation~\eqref{eqn:hatpsi_second_moment} yields $\hat\eta_n^2-\eta_n^2=o_{\mathbb P}(1)$.

Finally, by Equation~\eqref{eqn:eta_non_degenerate_clt},
$\liminf_{n\to\infty}\eta_n^2>0$, so $\hat\eta_n/\eta_n\Pto1$. Theorem~\ref{thm:clt} and Slutsky's theorem
then give
\begin{align*}
\frac{\sqrt n\left(T(\hat{\bm w}_f)-\mu_f\right)}{\hat\eta_n}\dto N(0,1),
\end{align*}
and the stated confidence interval follows immediately.
\end{proof}

\begin{proof}[Proof of Lemma \ref{lem:variance_outcome_side_uniform_bounds}]
Let $r=\zeta+\delta'$, with $\zeta>2$ and $\delta'>0$ from Assumption~\ref{assm:variance_estimation}(a).

\textbf{Proof of first display of the lemma.} Let us first establish the bound for $\hat{\bm q}_j$. Since $p$ is fixed it suffices to argue coordinatewise. Fix
$k\in\{1,\dots,p\}$ and set $Z_{i,j,k}=Y_i\phi_k(\bm X_i)\mathbb I(A_i\in B^{(j)})$ and
$\widetilde Z_{i,j,k}=Z_{i,j,k}-\mathbb E[Z_{i,j,k}]$, so that
$[\hat{\bm q}_j-\mathbb E[\hat{\bm q}_j]]_k=n^{-1}\sum_{i=1}^n\widetilde Z_{i,j,k}$. By
Assumptions~\ref{assm:unconfound} and \ref{assm:signal_plus_noise_general}, the error process
$\{\epsilon_i(a)\}_{a\in\mathcal A}$ is independent of $A_i$ given $\bm X_i$, so
$\mathbb E[|\epsilon_i|^s\mid A_i,\bm X_i]\leq\sup_{a\in\mathcal A}\mathbb E[|\epsilon_1(a)|^s]<\infty$ for
$s\in\{2,r\}$, the case $s=r$ by Assumption~\ref{assm:variance_estimation}(a). Combining this with
$Y_i=\bm\phi(\bm X_i)^\top\bm\beta(A_i)+\sigma(\bm X_i)\epsilon_i$
(Assumptions~\ref{assm:signal_plus_noise_general} and \ref{assm:signal_plus_noise}), the inequality
$|u+v|^s\leq2^{s-1}(|u|^s+|v|^s)$, and the boundedness of $\bm\phi(\cdot)$, $\bm\beta(\cdot)$ and
$\sigma(\cdot)$ (Assumptions~\ref{assm:regularity}(a) and \ref{assm:asymp_regularity}(b)),
\begin{align*}
\mathbb E\left[|Y_i|^s\,\mathbb I(A_i\in B^{(j)})\right]\lesssim\mathbb P\left(A_i\in B^{(j)}\right),
\qquad s\in\{2,r\}.
\end{align*}
By Assumption~\ref{assm:pos} and Equation~\eqref{eqn:bin_length_bounds},
$\mathbb P(A_i\in B^{(j)})=\mathbb E[\int_{B^{(j)}}e(a,\bm X_i)da]\leq\beta|B^{(j)}|\lesssim m_n^{-1}$. Since
$|\phi_k(\bm x)|\leq C$, it follows that, uniformly in $i,j,k$,
\begin{align}
\mathbb E\left[|\widetilde Z_{i,j,k}|^r\right]=O(m_n^{-1}),
\qquad
\mathbb E\left[\widetilde Z_{i,j,k}^2\right]=O(m_n^{-1}).
\label{eqn:Z_moment_bounds_variance}
\end{align}
By Rosenthal's inequality \citep{rosenthal1970} and Equation~\eqref{eqn:Z_moment_bounds_variance},
\begin{align*}
\mathbb E\left[\left|\frac1n\sum_{i=1}^n\widetilde Z_{i,j,k}\right|^r\right]
\lesssim\frac{1}{n^r}\left(\sum_{i=1}^n\mathbb E\left[\widetilde Z_{i,j,k}^2\right]\right)^{r/2}
+\frac{1}{n^r}\sum_{i=1}^n\mathbb E\left[|\widetilde Z_{i,j,k}|^r\right]
=O\left(\frac{1}{(nm_n)^{r/2}}+\frac{1}{n^{r-1}m_n}\right),
\end{align*}
so that, for any $\varepsilon>0$, Markov's inequality and a union bound over $j=1,\dots,m_n$ give
\begin{align*}
\mathbb P\left(\max_{1\leq j\leq m_n}
\left|\left[\hat{\bm q}_j-\mathbb E[\hat{\bm q}_j]\right]_k\right|>\frac{\varepsilon}{m_n}\right)
\leq\frac{m_n^{r+1}}{\varepsilon^r}\max_{1\leq j\leq m_n}
\mathbb E\left[\left|\left[\hat{\bm q}_j-\mathbb E[\hat{\bm q}_j]\right]_k\right|^r\right]
=O\left(\varepsilon^{-r}\left\{\frac{m_n^{1+r/2}}{n^{r/2}}+\frac{m_n^r}{n^{r-1}}\right\}\right).
\end{align*}
Since $r>\zeta>2$, we have $1+2/r<1+2/\zeta$ and $r/(r-1)=1+1/(r-1)<1+2/\zeta$. Hence, by
Assumption~\ref{assm:variance_estimation}(b),
\begin{align*}
\frac{m_n^{1+r/2}}{n^{r/2}}=\left(\frac{m_n^{1+2/r}}{n}\right)^{r/2}
\leq\left(\frac{m_n^{1+2/\zeta}}{n}\right)^{r/2}\to0,
\qquad
\frac{m_n^r}{n^{r-1}}=\left(\frac{m_n^{r/(r-1)}}{n}\right)^{r-1}
\leq\left(\frac{m_n^{1+2/\zeta}}{n}\right)^{r-1}\to0.
\end{align*}
As $p$ is fixed, this proves the first display of the lemma.

\textbf{Proof of second display of the lemma.} We next prove the bound for $\hat{\bm\beta}^{(n)}_j$. Since
$Y_1=\bm\phi(\bm X_1)^\top\bm\beta(A_1)+\sigma(\bm X_1)\epsilon_1$ with
$\mathbb E[\epsilon_1\mid A_1,\bm X_1]=0$, $\mathbb E[\hat{\bm q}_j]
=\mathbb E\left[\bm\phi(\bm X_1)\bm\phi(\bm X_1)^\top\bm\beta(A_1)\mathbb I(A_1\in B^{(j)})\right]$, so that, 

$$\mathbb E[\hat{\bm q}_j]-\bm M_j\bm\beta_j
=\mathbb E\left[\bm\phi(\bm X_1)\bm\phi(\bm X_1)^\top
\left\{\bm\beta(A_1)-\bm\beta_j\right\}\mathbb I(A_1\in B^{(j)})\right].$$

By Proposition~\ref{lem:bj_intervals}, $|A_1-j/m_n|\leq1/m_n$ whenever $A_1\in B^{(j)}$; hence
Assumptions~\ref{assm:regularity}(a) and~\ref{assm:pos}, Equation~\eqref{eqn:bin_length_bounds} and the definition of
the modulus of continuity in Equation~\eqref{eqn:modulus of continuity} give
\begin{align}
\max_{1\leq j\leq m_n}\left\|\mathbb E[\hat{\bm q}_j]-\bm M_j\bm\beta_j\right\|
 = O\left(M(1/m_n)\max_{1\leq j\leq m_n}\left|B^{(j)}\right|\right)
\lesssim\frac{M(1/m_n)}{m_n}.
\label{eqn:q_population_beta_bias}
\end{align}
Next, from the definition of $\hat{\bm\beta}^{(n)}_j$ and $\bm M_j\bm c^{(n)}_j=\bm r_j$,
\begin{align*}
\hat{\bm\beta}^{(n)}_j-\bm\beta_j
=\left(\hat{\bm M}_j+\lambda_{nj}\bm I_p\right)^{-1}
\left\{\hat{\bm q}_j-\left(\hat{\bm M}_j+\lambda_{nj}\bm I_p\right)\bm\beta_j\right\}
\end{align*}
and, adding and subtracting $\mathbb E[\hat{\bm q}_j]$ and $\bm M_j\bm\beta_j$,
\begin{align*}
\hat{\bm\beta}^{(n)}_j-\bm\beta_j
=\left(\hat{\bm M}_j+\lambda_{nj}\bm I_p\right)^{-1}
\left\{\left(\hat{\bm q}_j-\mathbb E[\hat{\bm q}_j]\right)
+\left(\mathbb E[\hat{\bm q}_j]-\bm M_j\bm\beta_j\right)
-\left(\hat{\bm M}_j-\bm M_j\right)\bm\beta_j-\lambda_{nj}\bm\beta_j\right\}.
\end{align*}
By Equation~\eqref{eqn:inverse_hatM_bound},
$\max_{1\leq j\leq m_n}\|(\hat{\bm M}_j+\lambda_{nj}\bm I_p)^{-1}\|\leq2m_n/c_M$ on an event whose
probability tends to one; so on that event,
\begin{align*}
\max_{1\leq j\leq m_n}\left\|\hat{\bm\beta}^{(n)}_j-\bm\beta_j\right\|
\leq\frac{2m_n}{c_M}\Bigg\{&\max_{1\leq j\leq m_n}\left\|\hat{\bm q}_j-\mathbb E[\hat{\bm q}_j]\right\|
+\max_{1\leq j\leq m_n}\left\|\mathbb E[\hat{\bm q}_j]-\bm M_j\bm\beta_j\right\|\\
&+C_\beta\max_{1\leq j\leq m_n}\left\|\hat{\bm M}_j-\bm M_j\right\|
+C_\beta\max_{1\leq j\leq m_n}\lambda_{nj}\Bigg\}.
\end{align*}
Multiplied by $m_n$, the four terms in braces are, respectively, $o_{\mathbb P}(1)$ by proof of first display of the lemma;
$O(M(1/m_n))=o(1)$ by Equation~\eqref{eqn:q_population_beta_bias} and
Assumption~\ref{assm:rate_choice_normality}(a); $O_{\mathbb P}(\sqrt{m_n\log m_n/n})=o_{\mathbb P}(1)$ by
Equation~\eqref{eqn:Mhat_uniform_bound} and Assumption~\ref{assm:rate_choice_consistency}(a); and $o_{\mathbb P}(1)$ by
Assumption~\ref{assm:rate_choice_consistency}(b). This proves the second display, and the triangle inequality then gives
$\max_{1\leq j\leq m_n}\|\hat{\bm\beta}^{(n)}_j\|\leq C_\beta+o_{\mathbb P}(1)=O_{\mathbb P}(1)$.
\end{proof}

\begin{proof}[Proof of Lemma \ref{lem:estimated_influence_l2_consistency}]
Write $\Delta_{c,n}=\max_{1\leq j\leq m_n}\|\hat{\bm c}^{(n)}_j-\bm c^{(n)}_j\|$ and
$\Delta_{\beta,n}=\max_{1\leq j\leq m_n}\|\hat{\bm\beta}^{(n)}_j-\bm\beta_j\|$, both $o_{\mathbb P}(1)$ by
Equation~\eqref{eqn:chat_uniform_convergence} and Lemma~\ref{lem:variance_outcome_side_uniform_bounds}. By
Equation~\eqref{eqn:Hn_definition_consistency} and Assumptions~\ref{assm:regularity}(a) and (b), the
function $H_n$ is uniformly bounded: for every $\bm x$,
\begin{align}
|H_n(\bm x)|\leq CC_\beta\sum_{j=1}^{m_n}|f|\left(B^{(j)},\bm x\right)
=CC_\beta|f|(\mathcal A,\bm x)\leq CC_\beta C_f.
\label{eqn:Hn_uniform_bound_variance}
\end{align}
Decompose
\begin{align*}
\hat\psi_i-\psi_{n,i}
=&\underbrace{\hat w_{f,i}\left(Y_i-\hat m_{\rm bal}(A_i,\bm X_i)\right)
-\sigma(\bm X_i)\epsilon_i\bm\phi(\bm X_i)^\top\bm c^{(n)}_{j(A_i)}}_{(\mathrm i)_i}\\
+&\underbrace{\int_{\mathcal A}\hat m_{\rm bal}(a,\bm X_i)f(da,\bm X_i) -H_n(\bm X_i)}_{(\mathrm{ii})_i}
-\underbrace{\left\{T(\hat{\bm w}_f)-\mathbb E[H_n(\bm X_1)]\right\}}_{(\mathrm{iii})}.
\end{align*}

By the definitions of $\hat m_{\rm bal}$ and $H_n$ and
Proposition~\ref{lem:bj_intervals},
$(\mathrm{ii})_i=\sum_{j=1}^{m_n}(\hat{\bm\beta}^{(n)}_j-\bm\beta_j)^\top\bm\phi(\bm X_i)f(B^{(j)},\bm X_i)$,
so that, arguing as in Equation~\eqref{eqn:Hn_uniform_bound_variance}, for every $i$,
\begin{align}
\left|(\mathrm{ii})_i\right|
\leq C\Delta_{\beta,n}\,|f|(\mathcal A,\bm X_i)\leq CC_f\Delta_{\beta,n}.
\label{eqn:term_two_bound_variance}
\end{align}

\emph{The term $(\mathrm{iii})$.} By Equation~\eqref{eqn:or_model},
$T(\hat{\bm w}_f)=n^{-1}\sum_{i=1}^n\int_{\mathcal A}\hat m_{\rm bal}(a,\bm X_i)f(da,\bm X_i)$, hence,
\begin{align*}
(\mathrm{iii})=\frac1n\sum_{i=1}^n(\mathrm{ii})_i
+\left\{\frac1n\sum_{i=1}^nH_n(\bm X_i)-\mathbb E[H_n(\bm X_1)]\right\},
\end{align*}
so that, by Equation~\eqref{eqn:term_two_bound_variance},
\begin{align}
\left|(\mathrm{iii})\right|\leq CC_f\Delta_{\beta,n}+R_n,
\qquad\textrm{where}\qquad
R_n:=\left|\frac1n\sum_{i=1}^nH_n(\bm X_i)-\mathbb E[H_n(\bm X_1)]\right|=o_{\mathbb P}(1)
\label{eqn:term_three_bound_variance}
\end{align}
by the weak law of large numbers, since the $H_n(\bm X_i)$ are i.i.d.\ and uniformly bounded by
Equation~\eqref{eqn:Hn_uniform_bound_variance}.

By Assumptions~\ref{assm:signal_plus_noise_general} and
\ref{assm:signal_plus_noise},
\begin{align*}
Y_i-\hat m_{\rm bal}(A_i,\bm X_i)
=\sigma(\bm X_i)\epsilon_i
+\bm\phi(\bm X_i)^\top\left\{\bm\beta(A_i)-\hat{\bm\beta}^{(n)}_{j(A_i)}\right\},
\end{align*}
so, using $\hat w_{f,i}=\bm\phi(\bm X_i)^\top\hat{\bm c}^{(n)}_{j(A_i)}$,
\begin{align*}
(\mathrm i)_i=\sigma(\bm X_i)\epsilon_i\,\bm\phi(\bm X_i)^\top
\left\{\hat{\bm c}^{(n)}_{j(A_i)}-\bm c^{(n)}_{j(A_i)}\right\}
+\hat w_{f,i}\,\bm\phi(\bm X_i)^\top
\left\{\bm\beta(A_i)-\hat{\bm\beta}^{(n)}_{j(A_i)}\right\}.
\end{align*}
Now $|\sigma(\bm X_i)\epsilon_i|=|Y_i-\bm\phi(\bm X_i)^\top\bm\beta(A_i)|\leq|Y_i|+CC_\beta$, while
$|\hat w_{f,i}|\leq C\max_{1\leq j\leq m_n}\|\hat{\bm c}^{(n)}_j\|=O_{\mathbb P}(1)$ by
Equation~\eqref{eqn:chat_uniform_bounded}, and, by Proposition~\ref{lem:bj_intervals} and
Equation~\eqref{eqn:modulus of continuity},
$\|\bm\beta(A_i)-\hat{\bm\beta}^{(n)}_{j(A_i)}\|\leq M(1/m_n)+\Delta_{\beta,n}$. Hence, for every $i$,
\begin{align*}
\left|(\mathrm i)_i\right|
\leq C\left(|Y_i|+CC_\beta\right)\Delta_{c,n}
+C^2\max_{1\leq j\leq m_n}\left\|\hat{\bm c}^{(n)}_j\right\|
\left\{M(1/m_n)+\Delta_{\beta,n}\right\}.
\end{align*}

Collecting the bound on $(\mathrm i)_i$ with Equations~\eqref{eqn:term_two_bound_variance} and
\eqref{eqn:term_three_bound_variance}, and using $|Y_i|+CC_\beta\leq(1+CC_\beta)(1+|Y_i|)$ for the one term
that involves $Y_i$ and $1\leq1+|Y_i|$ to absorb those that do not, we obtain
Equation~\eqref{eqn:if_uniform_approximation} with
\begin{align*}
\Delta_n:=C(1+CC_\beta)\Delta_{c,n}
+\left\{C^2\max_{1\leq j\leq m_n}\left\|\hat{\bm c}^{(n)}_j\right\|+2CC_f\right\}\Delta_{\beta,n}
+C^2\max_{1\leq j\leq m_n}\left\|\hat{\bm c}^{(n)}_j\right\|M(1/m_n)+R_n.
\end{align*}
This does not depend on $i$, and it is $o_{\mathbb P}(1)$: indeed $\Delta_{c,n}=o_{\mathbb P}(1)$ and
$\Delta_{\beta,n}=o_{\mathbb P}(1)$, while
$\max_{1\leq j\leq m_n}\|\hat{\bm c}^{(n)}_j\|=O_{\mathbb P}(1)$ by
Equation~\eqref{eqn:chat_uniform_bounded}, $M(1/m_n)\to0$ by
Assumption~\ref{assm:rate_choice_normality}(a), and $R_n=o_{\mathbb P}(1)$ by
Equation~\eqref{eqn:term_three_bound_variance}. Finally,
$Y_1=\bm\phi(\bm X_1)^\top\bm\beta(A_1)+\sigma(\bm X_1)\epsilon_1$ together with
$\mathbb E[\epsilon_1^2]=1$, gives $\mathbb E[Y_1^2]\leq2C^2C_\beta^2+2{\sigma^*}^2<\infty$, so
$n^{-1}\sum_{i=1}^nY_i^2=O_{\mathbb P}(1)$ by the weak law of large numbers. Therefore, using
$(1+|Y_i|)^2\leq2(1+Y_i^2)$,
\begin{align*}
\frac1n\sum_{i=1}^n\left(\hat\psi_i-\psi_{n,i}\right)^2
\leq2\Delta_n^2\left(1+\frac1n\sum_{i=1}^nY_i^2\right)=o_{\mathbb P}(1),
\end{align*}
which is Equation~\eqref{eqn:if_l2_consistency_variance}.
\end{proof}

\subsection{Proof of Equation~\eqref{eqn:or_model}}
\label{sec:or_model_proof}
The equation follow from the following lines of implication:
\begin{align*}
T(\hat{\bm w}_f)
&= \frac{1}{n}\sum_{i=1}^n Y_i\,\bm\phi(\bm X_i)^\top\hat{\bm c}^{(n)}_{j(A_i)}\\[2pt]
&= \frac{1}{n}\sum_{i=1}^n\sum_{1\le j\le m_n} Y_i\,\mathbb{I}(A_i\in B^{(j)})\;\bm\phi(\bm X_i)^\top(\hat{\bm M}_j+\lambda_{nj}\bm I_p)^{-1}\hat{\bm r}_j\\[2pt]
&= \sum_{1\le j\le m_n}\Big(\underbrace{\frac{1}{n}\sum_{i=1}^n Y_i\,\bm\phi(\bm X_i)\,\mathbb{I}(A_i\in B^{(j)})}_{=\;\hat{\bm q}_j}\Big)^{\!\top}(\hat{\bm M}_j+\lambda_{nj}\bm I_p)^{-1}\hat{\bm r}_j\\[2pt]
&= \sum_{1\le j\le m_n}\hat{\bm q}_j^{\top}(\hat{\bm M}_j+\lambda_{nj}\bm I_p)^{-1}\hat{\bm r}_j\\[2pt]
&= \sum_{1\le j\le m_n}\hat{\bm r}_j^{\top}(\hat{\bm M}_j+\lambda_{nj}\bm I_p)^{-1}\hat{\bm q}_j \\[2pt]
&= \sum_{1\le j\le m_n}\Big(\frac{1}{n}\sum_{i=1}^n \bm\phi(\bm X_i)\,f(B^{(j)},\bm X_i)\Big)^{\!\top}(\hat{\bm M}_j+\lambda_{nj}\bm I_p)^{-1}\hat{\bm q}_j\\[2pt]
&= \frac{1}{n}\sum_{i=1}^n\bm\phi(\bm X_i)^\top\sum_{1\le j\le m_n} f(B^{(j)},\bm X_i)\,(\hat{\bm M}_j+\lambda_{nj}\bm I_p)^{-1}\hat{\bm q}_j\\[2pt]
&= \frac{1}{n}\sum_{i=1}^n\bm\phi(\bm X_i)^\top\sum_{1\le j\le m_n}\int_{B^{(j)}}\hat{\bm\beta}^{(n)}(a)\,f(a,\bm X_i)\,da\\[2pt]
&= \frac{1}{n}\sum_{i=1}^n\int_{\mathcal A}\bm\phi(\bm X_i)^\top\,\hat{\bm\beta}^{(n)}(a)\,f(a,\bm X_i)\,da.
\end{align*}

\subsection{Proof of Theorem~\ref{thm:aipw_clt}}
\label{sec:proof_aipw_clt}
\begin{proof}
    Write $\varphi_0(A_i,\bm X_i):=\varphi(A_i,\bm X_i;m,e)$ and, for $i\in I_k$,
    $\hat\varphi_i:=\varphi(A_i,\bm X_i;\hat m_{(k)},\hat e_{(k)})$, so that
    $\hat T^{\hat m_{\rm bal},\hat e}_{\rm AIPW}=\frac1n\sum_{k=1}^K\sum_{i\in I_k}\hat\varphi_i$.

    By Assumption~\ref{assm:unconfound} and $Y_1=Y_1(A_1)$ we have
    $\mathbb E[Y_1-m(A_1,\bm X_1)\mid A_1,\bm X_1]=0$, so the augmentation term of $\varphi_0$ has mean zero
    and $\mathbb E[\varphi_0(A_1,\bm X_1)]=\mathbb E\big[\int_{\mathcal A}m(a,\bm X_1)f(a,\bm X_1)\,da\big] = \mu_f$. Next, write
    \begin{align}
    \label{eqn:aipw_decomposition}
    \sqrt n\big(\hat T^{\hat m_{\rm bal},\hat e}_{\rm AIPW}-\mu_f\big)
    =\underbrace{\frac1{\sqrt n}\sum_{i=1}^n\big(\varphi_0(A_i,\bm X_i)-\mu_f\big)}_{=:G_n}
    +\sum_{k=1}^K R_{n,k},
    \quad
    R_{n,k}:=\frac1{\sqrt n}\sum_{i\in I_k}\big(\hat\varphi_i-\varphi_0(A_i,\bm X_i)\big).
    \end{align}
    Fix $k$ and condition on $\mathcal D_k:=\{(Y_i,\bm X_i,A_i):i\in I_k^c\}$, which determines
    $\hat m_{(k)}$ and $\hat e_{(k)}$. Since the folds are drawn independently of the data, given
    $\mathcal D_k$ the observations $\{(Y_i,A_i,\bm X_i):i\in I_k\}$ are i.i.d.\ from the population law and
    independent of the nuisance estimators. Split $R_{n,k}=R_{n,k}^{\rm var}+R_{n,k}^{\rm bias}$, where
    \begin{align*}
    R_{n,k}^{\rm bias}=\frac{|I_k|}{\sqrt n}\,
    \mathbb E\big[\hat\varphi_i-\varphi_0(A_i,\bm X_i)\mid\mathcal D_k\big],
    \qquad
    R_{n,k}^{\rm var}=R_{n,k}-R_{n,k}^{\rm bias}.
    \end{align*}

    \noindent We will first show that $R_{n,k}^{\rm var} = o_{\mathbb P}(1)$. Decompose for $i\in I_k$,
    \begin{align}
    \label{eqn:aipw_var_split}
    \hat\varphi_i-\varphi_0(A_i,\bm X_i)
    =\underbrace{\int_{\mathcal A}f(a,\bm X_i)\big(\hat m_{(k)}-m\big)(a,\bm X_i)\,da}_{=:U_i}
    +\underbrace{f(A_i,\bm X_i)\left[\frac{Y_i-\hat m_{(k)}(A_i,\bm X_i)}{\hat e_{(k)}(A_i,\bm X_i)}
    -\frac{Y_i-m(A_i,\bm X_i)}{e(A_i,\bm X_i)}\right]}_{=:V_i}.
    \end{align}
    For $U_i$, Assumption~\ref{assm:regularity}(b), the Cauchy--Schwarz inequality on $\mathcal A$ (which has unit Lebesgue measure
    by Assumption~\ref{assm:bdd}), and $e\geq\alpha$ due to Assumption~\ref{assm:pos} gives
    \begin{align*}
    \mathbb E\big[U_i^2\mid\mathcal D_k\big]
    \leq& C_f^2\,\mathbb E\left[\int_{\mathcal A}\big(\hat m_{(k)}-m\big)^2(a,\bm X_i)\,da\middle|\mathcal D_k\right]
    \leq \frac{ C_f^2}{\alpha}\,
    \mathbb E\left[\int_{\mathcal A}\big(\hat m_{(k)}-m\big)^2(a,\bm X_i)e(a,\bm X_i)\,da\middle|\mathcal D_k\right]\\
    =&\frac{ C_f^2}{\alpha}\big\|\hat m_{(k)}-m\big\|^2.
    \end{align*}
    Next, for $V_i$, note that,
    \begin{align*}
    \frac{Y_i-\hat m_{(k)}}{\hat e_{(k)}}-\frac{Y_i-m}{e}
    =\frac{\big(Y_i-m\big)\big(e-\hat e_{(k)}\big)+\big(m-\hat m_{(k)}\big)e}{\hat e_{(k)}\,e},
    \end{align*}
    all functions being evaluated at $(A_i,\bm X_i)$, so that by the condition $\hat e_{(k)}\geq c_e$ and Assumptions~\ref{assm:regularity}(b) and~\ref{assm:pos},
    \begin{align*}
    |V_i|\leq \frac{C_f}{c_e\alpha}\big|Y_i-m(A_i,\bm X_i)\big|\,
    \big|\big(e-\hat e_{(k)}\big)(A_i,\bm X_i)\big|
    +\frac{C_f\beta}{c_e\alpha}\big|\big(m-\hat m_{(k)}\big)(A_i,\bm X_i)\big|.
    \end{align*}
    Since $\hat e_{(k)}$ is $\mathcal D_k$-measurable and $(A_i, \bm X_i)\notin \mathcal D_k$, conditioning on
    $(A_i,\bm X_i)$ in the first term and Assumption~\ref{assm:asymp_regularity}(b) gives
    \begin{align*}
    \mathbb E\left[\big(Y_i-m(A_i,\bm X_i)\big)^2\big(e-\hat e_{(k)}\big)^2(A_i,\bm X_i)\;\middle|\;\mathcal D_k\right]
    =&\mathbb E\left[\operatorname{Var}\big(Y_i\mid A_i,\bm X_i\big)
    \big(e-\hat e_{(k)}\big)^2(A_i,\bm X_i)\;\middle|\;\mathcal D_k\right]\\
    \leq&{\sigma^*}^2\big\|\hat e_{(k)}-e\big\|^2,
    \end{align*}
    while the second term contributes $\mathbb E[(m-\hat m_{(k)})^2(A_1,\bm X_1)]=\|\hat m_{(k)}-m\|^2$.
    Combining the two displays with $(u+v)^2\leq2u^2+2v^2$,
    \begin{align}
    \label{eqn:aipw_var_bound}
    \mathbb E\left[\big(\hat\varphi_i-\varphi_0(A_i,\bm X_i)\big)^2\;\middle|\;\mathcal D_k\right]
    \lesssim \big\|\hat m_{(k)}-m\big\|^2+\big\|\hat e_{(k)}-e\big\|^2=o_{\mathbb P}(1).
    \end{align}
    Now, given $\mathcal D_k$, the summands of $R_{n,k}^{\rm var}$ are i.i.d.\ with mean zero, so, using
    $|I_k|\leq n$ and Equation~\eqref{eqn:aipw_var_bound},
    \begin{align*}
    \operatorname{Var}\big(R_{n,k}^{\rm var}\mid\mathcal D_k\big)
    =\frac{|I_k|}{n}\operatorname{Var}\big(\hat\varphi_i-\varphi_0(A_i,\bm X_i)\mid\mathcal D_k\big)
    \leq \mathbb E\left[\big(\hat\varphi_i-\varphi_0(A_i,\bm X_i)\big)^2\;\middle|\;\mathcal D_k\right]
    =o_{\mathbb P}(1).
    \end{align*}
    Therefore, for any $\varepsilon>0$, conditional Chebyshev's inequality gives
    $\mathbb P(|R_{n,k}^{\rm var}|>\varepsilon)\leq\mathbb E[W_n]$ with
    $W_n:=1\wedge\varepsilon^{-2}\operatorname{Var}(R_{n,k}^{\rm var}\mid\mathcal D_k)$. As $0\leq W_n\leq1$
    and $W_n\Pto0$, we have $\mathbb E[W_n]\leq\delta+\mathbb P(W_n>\delta)$ for every $\delta>0$, implying
    $\limsup_n\mathbb E[W_n]\leq\delta$ and so $\mathbb E[W_n]\to0$. Consequently
    \begin{align}
    \label{eqn:aipw_var_negligible}
    R_{n,k}^{\rm var}=o_{\mathbb P}(1).
    \end{align}

    \noindent Next, we will show that $R_{n,k}^{\rm bias}=o_{\mathbb P}(1)$. Let $(Y,A,\bm X)$ be an observation independent of $\mathcal D_k$, with the nuisance
    estimators held fixed. Conditioning on $\bm X$, integrating over $A\sim e(\cdot,\bm X)$ and using
    $\mathbb E[Y\mid A=a,\bm X]=m(a,\bm X)$ in the augmentation term,
    \begin{align*}
    \mathbb E\big[\hat\varphi_i\mid\mathcal D_k\big]
    =\mathbb E\left[\int_{\mathcal A}f(a,\bm X)\left\{\hat m_{(k)}(a,\bm X)
    +\frac{e(a,\bm X)}{\hat e_{(k)}(a,\bm X)}\big(m(a,\bm X)-\hat m_{(k)}(a,\bm X)\big)\right\}da
    \;\middle|\;\mathcal D_k\right].
    \end{align*}
    Subtracting $\mathbb E[\varphi_0(A_i,\bm X_i)]=\mathbb E[\int_{\mathcal A}f(a,\bm X)m(a,\bm X)da]$ and
    collecting the terms in $\hat m_{(k)}-m$,
    \begin{align}
    \label{eqn:aipw_mixed_bias}
    \mathbb E\big[\hat\varphi_i-\varphi_0(A_i,\bm X_i)\mid\mathcal D_k\big]
    =\mathbb E\left[\int_{\mathcal A}f(a,\bm X)\,
    \frac{\big(\hat m_{(k)}-m\big)(a,\bm X)\big(\hat e_{(k)}-e\big)(a,\bm X)}{\hat e_{(k)}(a,\bm X)}\,da
    \;\middle|\;\mathcal D_k\right].
    \end{align} 
    Invoking Assumption~\ref{assm:regularity}(b), the condition $\hat e_{(k)}\geq c_e$, almost surely, and the Cauchy--Schwarz inequality on the product of
    $\mathbb P_{\bm X}$ and Lebesgue measure on $\mathcal A$, and $e\geq\alpha$ applied to each factor,
    \begin{align*}
    \left|\mathbb E\big[\hat\varphi_i-\varphi_0(A_i,\bm X_i)\mid\mathcal D_k\big]\right|
    \leq &\frac{C_f}{c_e}
    \left\{\mathbb E\left[\int_{\mathcal A}\big(\hat m_{(k)}-m\big)^2 da\middle| \mathcal D_k\right]\right\}^{1/2}
    \left\{\mathbb E\left[\int_{\mathcal A}\big(\hat e_{(k)}-e\big)^2 da\middle| \mathcal D_k\right]\right\}^{1/2}\\
    \leq &\frac{C_f}{c_e\alpha}
    \left\{\mathbb E\left[\int_{\mathcal A}\big(\hat m_{(k)}-m\big)^2 e(a, \bm X_1)da\middle| \mathcal D_k\right]\right\}^{1/2}\\
    \times & \left\{\mathbb E\left[\int_{\mathcal A}\big(\hat e_{(k)}-e\big)^2e(a, \bm X_1) da\middle| \mathcal D_k\right]\right\}^{1/2}\\
    \leq &\frac{C_f}{c_e\alpha}\big\|\hat m_{(k)}-m\big\|\,\big\|\hat e_{(k)}-e\big\|,
    \end{align*}
    which is $o_{\mathbb P}(n^{-1/2})$ by the rate condition of Theorem~\ref{thm:aipw_clt}. Hence,
    \begin{align}
    \label{eqn:aipw_bias_negligible}
    R_{n,k}^{\rm bias}=\frac{|I_k|}{\sqrt n}\,o_{\mathbb P}(n^{-1/2})
    =\frac{|I_k|}{n}\,o_{\mathbb P}(1)=o_{\mathbb P}(1).
    \end{align}

    Finally, summing Equations~\eqref{eqn:aipw_var_negligible} and
    \eqref{eqn:aipw_bias_negligible} over the finitely many folds, Equation~\eqref{eqn:aipw_decomposition}
    gives $\sqrt n(\hat T^{\hat m_{\rm bal},\hat e}_{\rm AIPW}-\mu_f)=G_n+o_{\mathbb P}(1)$. By
    Assumption~\ref{assm:iid} the variables $\varphi_0(A_1,\bm X_1),\dots,\varphi_0(A_n,\bm X_n)$ are
    i.i.d.\ with mean $\mu_f$ and variance
    $\sigma^2_{\rm eff}=\operatorname{Var}(\varphi(A_1,\bm X_1;m,e))
    =\mathbb E\varphi^2(A_1,\bm X_1;m,e)-\mu_f^2<\infty$ by assumption, so an application of CLT gives $G_n\dto\mathcal N(0,\sigma^2_{\rm eff})$. Slutsky's theorem yields the claim.
\end{proof}

\subsection{Proof of Theorem~\ref{thm:eff_dominance}}
\begin{proof}
    Under Assumption~\ref{assm:signal_plus_noise} with $\sigma^2(\bm x) = \sigma^2$, set $g(\bm X_1):=\int_{\mathcal A}m(a,\bm X_1)f(a,\bm X_1)da=\mathbb E[\int_{\mathcal A}Y_1(a)f(da,\bm X_1)\mid\bm X_1]$ (where the intergral and expectation can be interchanged, similar to how we did in Proposition~\ref{lem:limiting_variance} using the Tonelli's Theorem); since $Y_1-m(A_1,\bm X_1)=\sigma\epsilon_1$,
    \[
    \varphi(A_1,\bm X_1;m,e)=g(\bm X_1)+\sigma\,\frac{f(A_1,\bm X_1)}{e(A_1,\bm X_1)}\,\epsilon_1.
    \]
    As $\mathbb E[\epsilon_1\mid A_1,\bm X_1]=0$, the two terms are uncorrelated, $\mathbb E[\varphi]=\mathbb E[g(\bm X_1)]=\mu_f$, and $\mathbb E\left[\left(\sigma\,\frac{f(A_1,\bm X_1)}{e(A_1,\bm X_1)}\,\epsilon_1\right)^2\right]=\sigma^2\mathbb E[\tfrac{f^2(A_1,\bm X_1)}{e^2(A_1,\bm X_1)}]=\sigma^2\int_{\mathcal A}\mathbb E[\tfrac{f^2(a,\bm X_1)}{e(a,\bm X_1)}]da$. Therefore
    \[
    \sigma^2_{\rm eff}=\operatorname{Var}(\varphi(A_1,\bm X_1;m,e))=\operatorname{Var}\Big(\mathbb E\big[\textstyle\int_{\mathcal A}Y_1(a)f(da,\bm X_1)\mid\bm X_1\big]\Big)+\sigma^2\int_{\mathcal A}\mathbb E\Big[\frac{f^2(a,\bm X_1)}{e(a,\bm X_1)}\Big]da,
    \]
    which is the stated expression.

    For the comparison, by Proposition~\ref{lem:limiting_variance} under homoskedasticity,
    \[
    \sigma^2_{\rm bal}=\operatorname{Var}\Big(\mathbb E\big[\textstyle\int_{\mathcal A}Y_1(a)f(da,\bm X_1)\mid\bm X_1\big]\Big)+\sigma^2\int_{\mathcal A}\mathbb E\big[e(a,\bm X_1)\{\bm\phi(\bm X_1)^\top\bm M(a)^{-1}\bm q(a)\}^2\big]da,
    \]
    with $\bm q(a)=\mathbb E[\bm\phi(\bm X_1)f(a,\bm X_1)]$ and $\bm M(a)=\mathbb E[\bm\phi(\bm X_1)\bm\phi(\bm X_1)^\top e(a,\bm X_1)]$. The first terms of $\sigma^2_{\rm eff}$ and $\sigma^2_{\rm bal}$ coincide, so it suffices to show, for each $a$,
    \[
    \mathbb E\big[e(a,\bm X_1)\{\bm\phi(\bm X_1)^\top\bm M(a)^{-1}\bm q(a)\}^2\big]\le\mathbb E\Big[\frac{f^2(a,\bm X_1)}{e(a,\bm X_1)}\Big].
    \]
    Fix $a$ and put $\bm u:=\bm\phi(\bm X_1)\sqrt{e(a,\bm X_1)}$ and $\ell:=f(a,\bm X_1)/\sqrt{e(a,\bm X_1)}$. Then $\mathbb E[\bm u\bm u^\top]=\bm M(a)$ and $\mathbb E[\bm u\ell]=\mathbb E[\bm\phi(\bm X_1)f(a,\bm X_1)]=\bm q(a)$, so the $\mathcal L^2$ projection of $\ell$ onto $\operatorname{span}(\bm u)$ is $\hat\ell=\bm u^\top\bm M(a)^{-1}\bm q(a)$, with $\mathbb E[\hat\ell^2]=\bm q(a)^\top\bm M(a)^{-1}\bm q(a)=\mathbb E[e(a,\bm X_1)\{\bm\phi(\bm X_1)^\top\bm M(a)^{-1}\bm q(a)\}^2]$. As a projection does not increase the second moment, $\mathbb E[\hat\ell^2]\le\mathbb E[\ell^2]=\mathbb E[f^2(a,\bm X_1)/e(a,\bm X_1)]$. Integrating over $a$ and multiplying by $\sigma^2$ shows the second term of $\sigma^2_{\rm bal}$ is at most that of $\sigma^2_{\rm eff}$; hence $\sigma^2_{\rm bal}\le\sigma^2_{\rm eff}$.
\end{proof}

\subsection{More on item (c) of Assumption~\ref{assm:asymp_regularity}}
\label{sec:more_on_item_e}
Here we show that the signed measure $\bm Q$, defined in Section~\ref{sec:asymp_normality}, must either satisfy the condition of Assumption~\ref{assm:asymp_regularity}(c) or be a zero measure.
\begin{lemma}
    Assumption~\ref{assm:asymp_regularity}(c) does not hold if and only if $\bm Q$ is the zero measure.
\end{lemma}
\begin{proof}
    To begin with, assume that Assumption~\ref{assm:asymp_regularity}(c) does not hold. Without loss of generality, assume that the measure $\bm Q$ is one-dimensional; if it is multi-dimensional, one can just replicate this proof for each coordinate of $\bm Q$. Hence, from here on, we will use $Q$ instead of the boldfaced notation, $\bm Q$. Now consider any closed interval $I\subseteq [0,1]$. We will first show that $Q((a,b)) = 0$. Now, note that we know,
    \[
        \underset{n\to \infty}{\lim\inf}~\sum_{j=1}^{m_n}\left\|Q(B^{(j)})\right\| = 0.
    \]
    Thus, for any $\epsilon>0$, one can fine some $n_{\epsilon}\in \mathbb N$ such that,
    \[
        \sum_{j=1}^{m_{n_\epsilon}}\left\|Q(B^{(j)})\right\| < \epsilon.
    \]
    Now because $B^{(1)},\cdots, B^{(m_{n_\epsilon})}$ forms a partition of $I$, there exists a finite collection $\{j_1,\cdots, j_k\} \subseteq \{1,\cdots, m_n\}$ such that $I\subseteq \cup_{j=1}^k B^{(j)}$. Hence,
    \[
        \|Q(I)\| \leq \sum_{j=1}^k \|Q(B^{(j)}) \| \leq \sum_{j=1}^{m_n} \|Q(B^{(j)}) \| <\epsilon.
    \]
    Since, $\epsilon>0$ is arbitrary, $Q(I) = 0$. Now, define the collection, $$\mathcal L = \{B:\textrm{$B$ is a measurable subset of $[0,1]$ and }Q(B) = 0\}.$$ Then note that $\mathcal L$ is a $\lambda$-system as $[0,1]\in \mathcal L$, because $[0,1]$ is an interval. If $B\in \mathcal L$, then $Q([0,1]\setminus B) = Q([0,1]) - Q(B) = 0$, so that $[0,1]\setminus B\in \mathcal L$ and finally, using the continuity from below of a finite signed measure, if $B_1\subseteq B_2\subseteq\cdots \in \mathcal L$ then $Q(\cup_{i\geq 1}B_i) = 0$, so that $\cup_{i\geq 1}B_i\in \mathcal L$. Hence, $\mathcal L$ is a $\lambda$-system. Next, note that from our derivation above, $\mathcal L$ contains the set of all closed intervals, which is a $\pi$-system. Hence, by Dynkin's $\pi$-$\lambda$ theorem, $\mathcal L$ contains the $\sigma$-algebra generated by the set of open sub-intervals of $[0,1]$ which is precisely the Borel $\sigma$-algebra of $[0,1]$. In other words, for any Borel set $B\subseteq [0,1]$, $Q(B) = 0$ and hence, $Q$ is the zero measure.

    Conversely, assume that $\bm Q$ is indeed the zero measure. Then it is trivial to show that Assumption~\ref{assm:asymp_regularity} is not satisfied. This completes the proof.
\end{proof}

\subsection{Proof of Lemma~\ref{lem:ridge_imbalance_equiv}}
\label{sec:proof_ridge_imbalance_equiv}
\begin{proof}
Fix any bin $j$ and let $\lambda_{nj}>0$. By Proposition~\ref{lem:weights_characterization}, the ridge solution is
\[
\hat{\bm w}^{\rm ridge}_{S_j}(\lambda_{nj})
=\bm\Phi_j(\hat{\bm M}_j+\lambda_{nj}\bm I_p)^{-1}\hat{\bm r}_j.
\]
Since $\bm\Phi_j^T\bm\Phi_j=n\hat{\bm M}_j$, we have
\[
\bm\Phi_j^T\hat{\bm w}^{\rm ridge}_{S_j}(\lambda_{nj})-n\hat{\bm r}_j
=-n\lambda_{nj}(\hat{\bm M}_j+\lambda_{nj}\bm I_p)^{-1}\hat{\bm r}_j.
\]
Thus, the ridge solution satisfies the constraint in~\eqref{eqn:with_imbalance} with equality for $\delta_j=\delta_j(\lambda_{nj})$, establishing feasibility.

Now, let $\bm w_{S_j}$ be any feasible solution to~\eqref{eqn:with_imbalance}. By the optimality of $\hat{\bm w}^{\rm ridge}_{S_j}(\lambda_{nj})$ in~\eqref{eqn:final_optimization},
\begin{align*}
\frac12\|\hat{\bm w}^{\rm ridge}_{S_j}(\lambda_{nj})\|^2
+\frac{\delta_j^2}{2n\lambda_{nj}}
&\leq
\frac12\|\bm w_{S_j}\|^2
+\frac{\|\bm\Phi_j^T\bm w_{S_j}-n\hat{\bm r}_j\|^2}{2n\lambda_{nj}}\\
&\leq
\frac12\|\bm w_{S_j}\|^2
+\frac{\delta_j^2}{2n\lambda_{nj}}.
\end{align*}
Hence, $\|\hat{\bm w}^{\rm ridge}_{S_j}(\lambda_{nj})\|^2\leq\|\bm w_{S_j}\|^2$. Since the constrained objective is strictly convex and its feasible set is convex, its minimizer is unique. Therefore,
\[
\hat{\bm w}^{\rm ridge}_{S_j}(\lambda_{nj})
=\hat{\bm w}^{\rm imbal}_{S_j}(\delta_j(\lambda_{nj})).
\]

Next, we characterize the feasible values of $\delta_j$. Since $\hat{\bm M}_j=\bm\Phi_j^T\bm\Phi_j/n$, let
\[
\hat{\bm M}_j=\sum_{k=1}^{\rho_j}\tau_{jk}\bm h_k\bm h_k^T
\]
be its eigendecomposition, where $\rho_j$ denotes its rank, $\{\tau_{jk}\}_{k=1}^{\rho_j}$ are its positive eigenvalues, and $\{\bm h_k\}_{k=1}^{p}$ form an orthonormal basis, with the first $\rho_j$ corresponding to positive eigenvalues. Then $\{\bm h_k\}_{k=\rho_j+1}^{p}$ span the null space of $\bm\Phi_j$. Consequently, for any $\bm w_{S_j}$,
\[
\|\bm\Phi_j^T\bm w_{S_j}-n\hat{\bm r}_j\|^2
\geq n^2\sum_{k=\rho_j+1}^{p}(\bm h_k^T\hat{\bm r}_j)^2
=n^2\|\hat{\bm r}_j^\perp\|^2.
\]
Equality is attained by a least-squares solution to $\bm\Phi_j^T\bm w_{S_j}=n\hat{\bm r}_j$. Thus, \eqref{eqn:with_imbalance} is feasible if and only if
\[
\delta_j\geq\delta_{\min}:=n\|\hat{\bm r}_j^\perp\|.
\]

Finally, using the same eigendecomposition, we obtain
\begin{align*}
\delta_j^2(\lambda_{nj})
&=n^2\lambda_{nj}^2
\left\|(\hat{\bm M}_j+\lambda_{nj}\bm I_p)^{-1}
\hat{\bm r}_j\right\|^2\\
&=n^2\sum_{k=1}^{\rho_j}
\frac{\lambda_{nj}^2(\bm h_k^T\hat{\bm r}_j)^2}
{(\tau_{jk}+\lambda_{nj})^2}
+n^2\sum_{k=\rho_j+1}^{p}(\bm h_k^T\hat{\bm r}_j)^2.
\end{align*}
This expression is continuous and nondecreasing in $\lambda_{nj}>0$, and is strictly increasing whenever $\delta_{\min}<n\|\hat{\bm r}_j\|$. Moreover,
\[
\lim_{\lambda_{nj}\to0^+}\delta_j(\lambda_{nj})
=\delta_{\min},
\qquad
\lim_{\lambda_{nj}\to\infty}\delta_j(\lambda_{nj})
=n\|\hat{\bm r}_j\|.
\]
Therefore, whenever $\delta_{\min}<\delta_j<n\|\hat{\bm r}_j\|$, the intermediate value theorem guarantees a unique $\lambda_{nj}>0$ satisfying $\delta_j=\delta_j(\lambda_{nj})$. By the first part of the proof, the corresponding ridge and constrained solutions coincide.

For the boundary cases, the ridge solution admits the representation
\[
\hat{\bm w}^{\rm ridge}_{S_j}(\lambda_{nj})
=\bm\Phi_j\sum_{k=1}^{\rho_j}
\frac{\bm h_k^T\hat{\bm r}_j}{\tau_{jk}+\lambda_{nj}}\bm h_k.
\]
Hence, as $\lambda_{nj}\to0^+$, it converges to
\[
\bm\Phi_j\sum_{k=1}^{\rho_j}
\frac{\bm h_k^T\hat{\bm r}_j}{\tau_{jk}}\bm h_k,
\]
which is the minimum-norm solution attaining the smallest possible imbalance $\delta_{\min}$. Thus,
\[
\hat{\bm w}^{\rm imbal}_{S_j}(\delta_{\min})
=\lim_{\lambda_{nj}\to0^+}
\hat{\bm w}^{\rm ridge}_{S_j}(\lambda_{nj}).
\]
On the other hand, as $\lambda_{nj}\to\infty$, the ridge solution converges to $\bm0$. Since $\bm0$ is feasible for \eqref{eqn:with_imbalance} whenever $\delta_j\geq n\|\hat{\bm r}_j\|$, it is also the unique constrained minimizer in this case. Therefore,
\[
\hat{\bm w}^{\rm imbal}_{S_j}(\delta_j)
=\bm0
=\lim_{\lambda_{nj}\to\infty}
\hat{\bm w}^{\rm ridge}_{S_j}(\lambda_{nj}),
\qquad \delta_j\geq n\|\hat{\bm r}_j\|.
\]
This also covers the degenerate case $\delta_{\min}=n\|\hat{\bm r}_j\|$, in which all feasible constrained solutions and all ridge solutions equal $\bm0$. This completes the proof.
\end{proof}

\section{Further simulations and details}
\label{sec:further_simul}

\subsection{Implementation details of the simulations}
\label{sec:implementation_details}
In this section, we provide with the additional implementation details of the B-OPE procedure from Section~\ref{sec:expt_density}.

When implementing the B-OPE method, the counterfactual action is well-specified when $f$ is a Dirac measure on a counterfactual policy, and in particular is given by $\pi(\bm x)$ for a covariate $\bm x$. When $f$ is a density, the counterfactual action is given by a draw from the distribution represented by $f(\cdot;\bm x)$. Finally, as mentioned in \cite[Section 5.1]{sondhietal2020}, for the choice of the classifier, we used gradient boosted trees with default XGBoost parameters, except for the number of boosting iterations, which we set to be equal to $\lceil\sqrt{20\times n}\rceil$, for a sample size $n$.


Next, we present analogous results as Section~\ref{sec:expt_density} corresponding to $d=5$ and $d=10$, and are summarized below. 

\subsection{Additional simulations}
\label{sec:further_simul_dens}
\begin{figure}[!ht]
    \centering
    \includegraphics{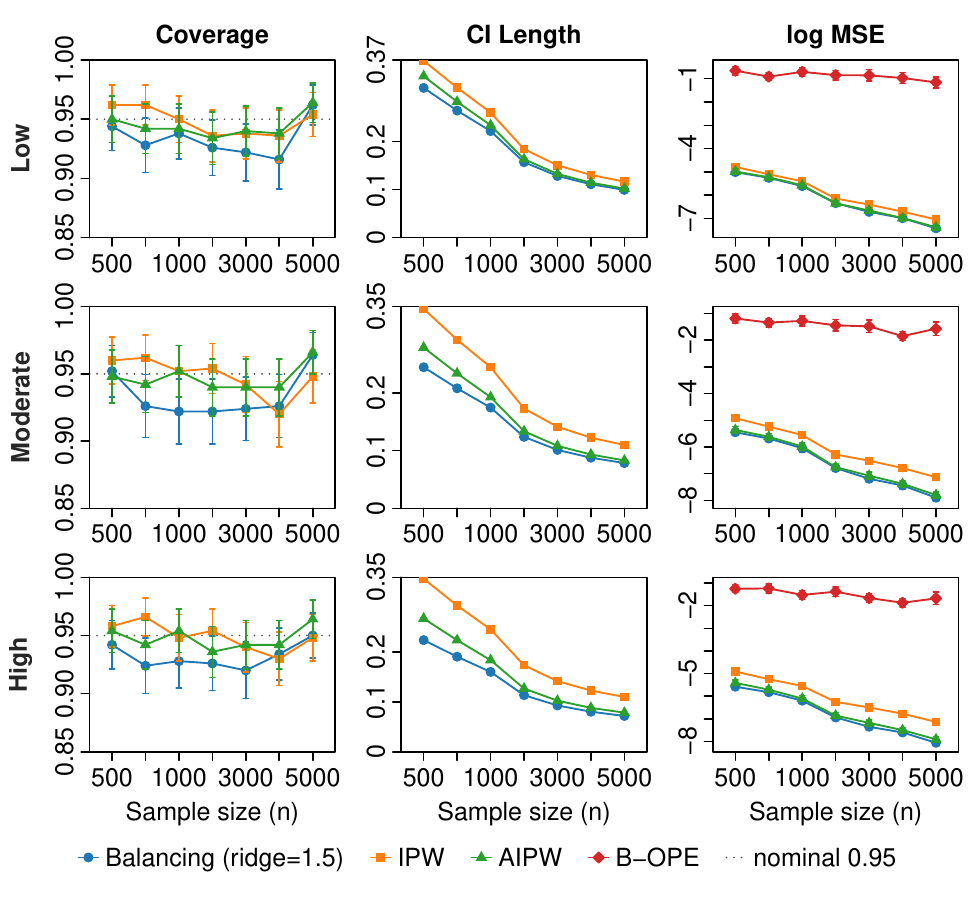}
    \caption{Same setting as Figure~\ref{fig:main_density} except for $d=5$ and for `low', `moderate' and `high' settings, $\alpha_{\rm amp}$ take values 5.1, 2.6 and 1.9 respectively.}
    \label{fig:5_density}
\end{figure}

\begin{figure}[!ht]
    \centering
    \includegraphics{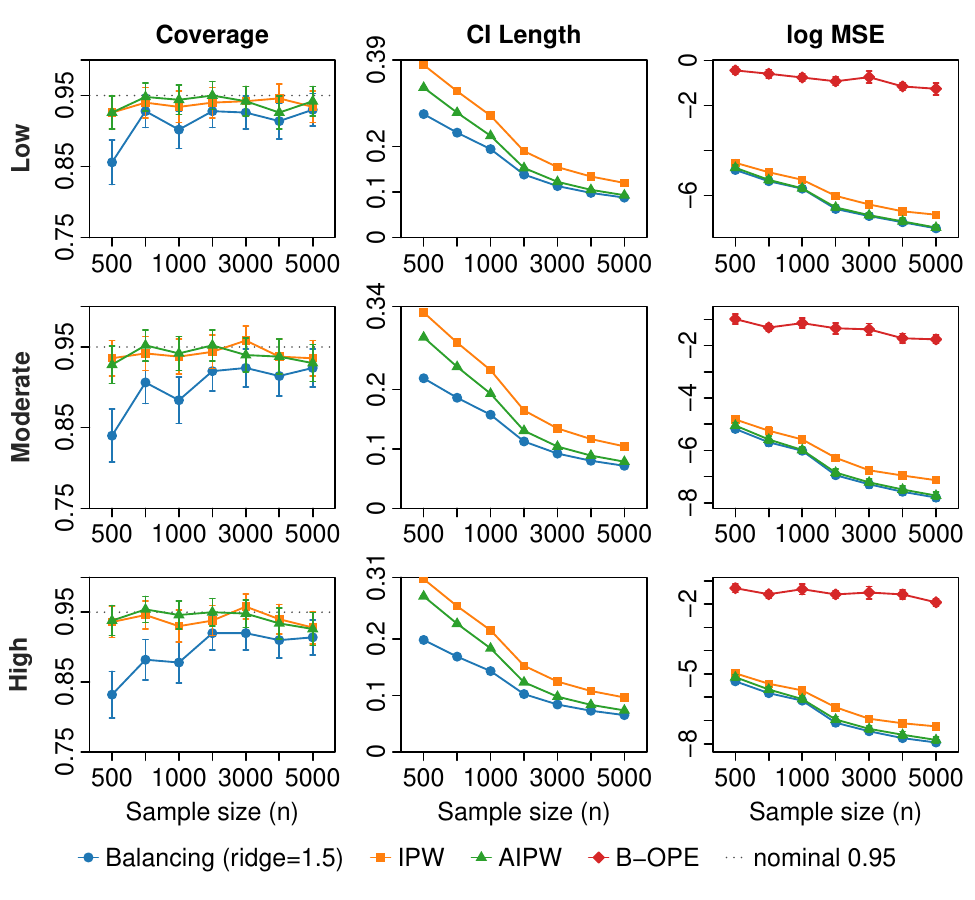}
    \caption{Same setting as Figure~\ref{fig:main_density} except for $d=10$ and for `low', `moderate' and `high' settings, $\alpha_{\rm amp}$ take values 3.3, 1.7 and 1.1 respectively.}
    \label{fig:10_density}
\end{figure}

Analogs of Figure~\ref{fig:main_density} corresponding to $d=5$ and $d=10$ are summarized in Figures~\ref{fig:5_density} and \ref{fig:10_density} respectively. Similar conclusions about the dominance of our balancing estimator extends to both of these cases as well. As Figure~\ref{fig:10_density} prominently points out, we see that with a large $d$, it takes a larger sample size for our balancing-CI to attain nominal coverage.

\subsection{Choice of $\lambda_{nj}$}
\label{sec:lambda_choice}
\begin{figure}[h]
    \centering
    \includegraphics{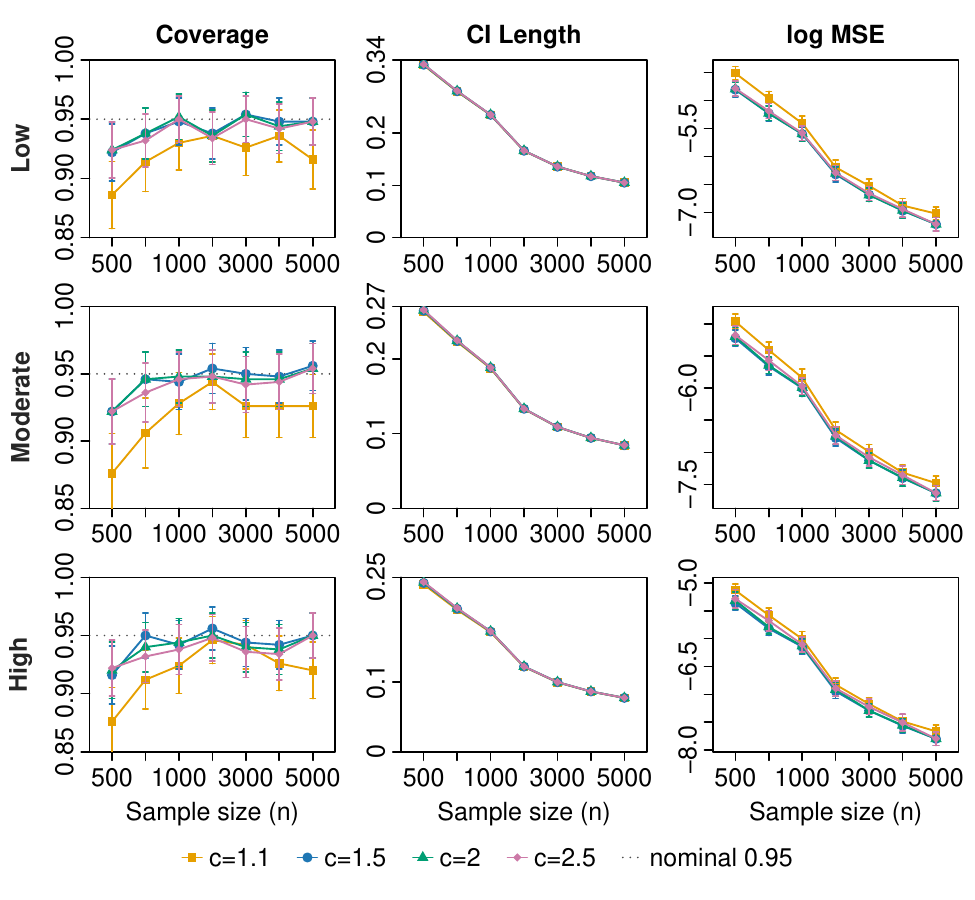}
    \caption{Empirical performance of the balancing estimator for different values of $c$ under the same setting as Figure~\ref{fig:main_density}, i.e., with $d=3$. The error bars represent plus or minus two standard errors.}
    \label{fig:3_ridge}
\end{figure}

\begin{figure}[h]
    \centering
    \includegraphics{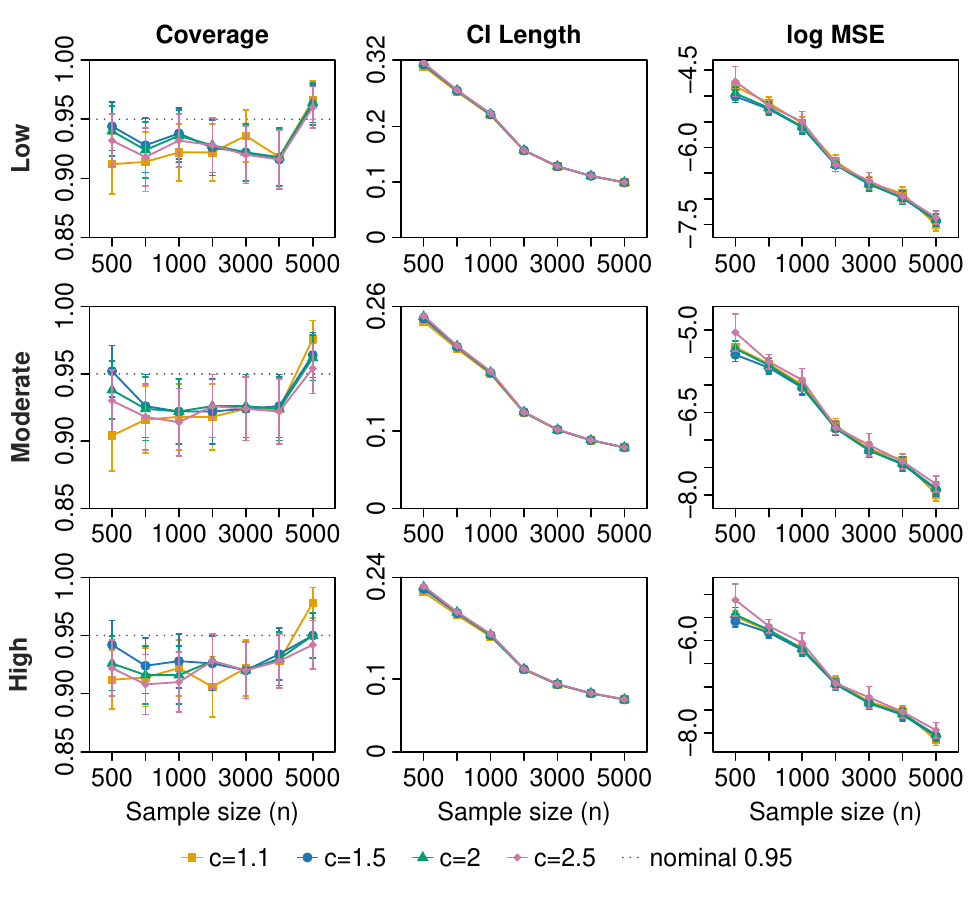}
    \caption{Empirical performance of the balancing estimator for different values of $c$ under the same setting as Figure~\ref{fig:5_density}, i.e., with $d=5$. The error bars represent plus or minus two standard errors.}
    \label{fig:5_ridge}
\end{figure}

\begin{figure}[h]
    \centering
    \includegraphics{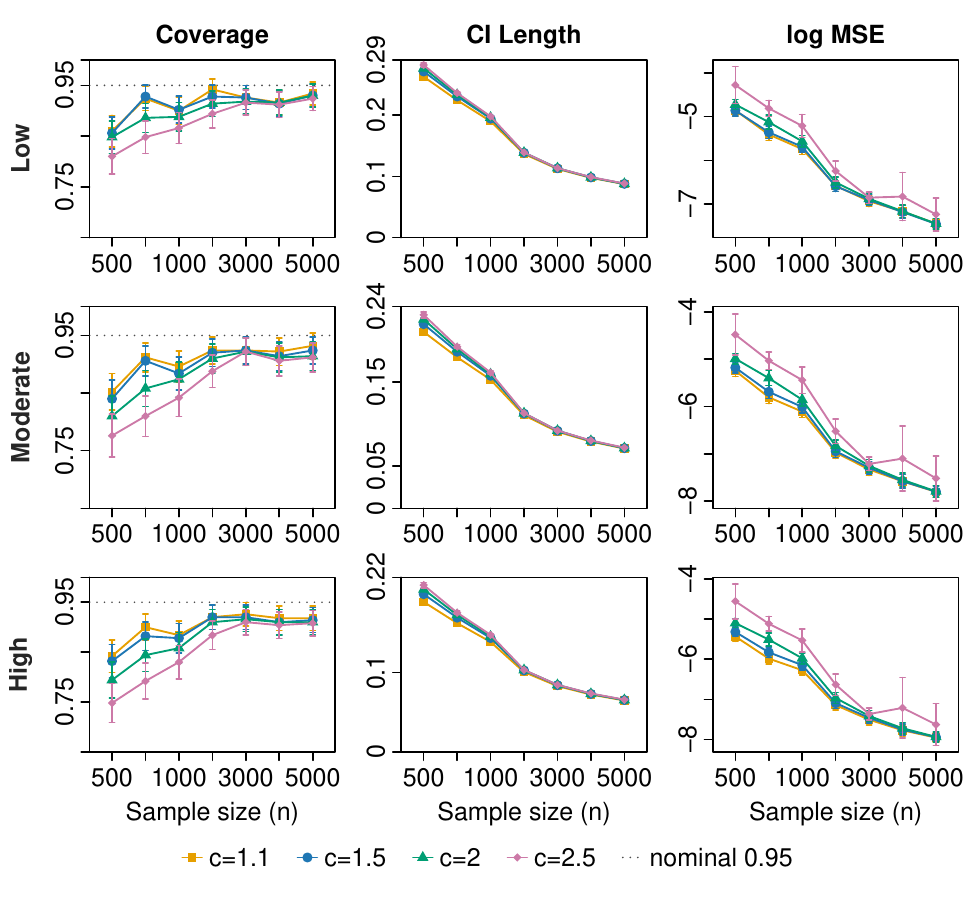}
    \caption{Empirical performance of the balancing estimator for different values of $c$ under the same setting as Figure~\ref{fig:10_density}, i.e., with $d=10$. The error bars represent plus or minus two standard errors.}
    \label{fig:10_ridge}
\end{figure}

In Figures~\ref{fig:3_ridge}, \ref{fig:5_ridge} and \ref{fig:10_ridge}, we present analogs of Figures~\ref{fig:main_density}, \ref{fig:5_density} and  \ref{fig:10_density} where we only compare the performance of the balancing estimator with values of $c$ ranging in the set $\{1.1, 1.5, 2,2.5\}$ respectively (based on the notations introduced in Section~\ref{sec:choice_m_lambda}. These figures show that across the various choices of $d$, the choice of $c=1.5$ produces empirical coverage that is consistently closer to the nominal level of 0.95 compared to the other possible choices of $c$. In terms of the length of the resulting confidence interval and the MSE of the resulting estimator, the different choices of $c$ do not seem to be affecting the performance of the balancing estimator significantly. Hence, $c=1.5$ is our recommended choice.

\subsection{Choice of $m_n$}
\label{sec:mn_choice}
\begin{figure}[h]
    \centering
    \includegraphics{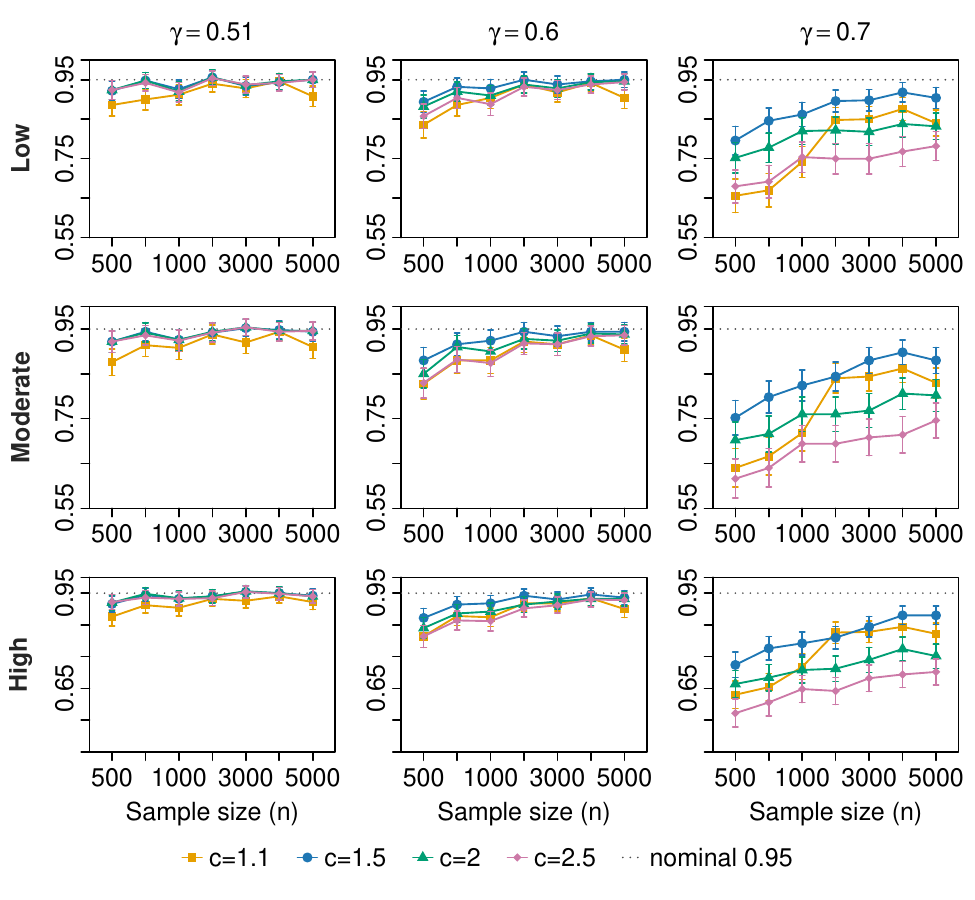}
    \caption{Coverage of the balancing estimator based confidence intervals for different values of $\gamma$ and $c$ for the setting with $d=3$. This is the same setting as Figure~\ref{fig:main_density}.}
    \label{fig:3_sweep}
\end{figure}

\begin{figure}[h]
    \centering
    \includegraphics{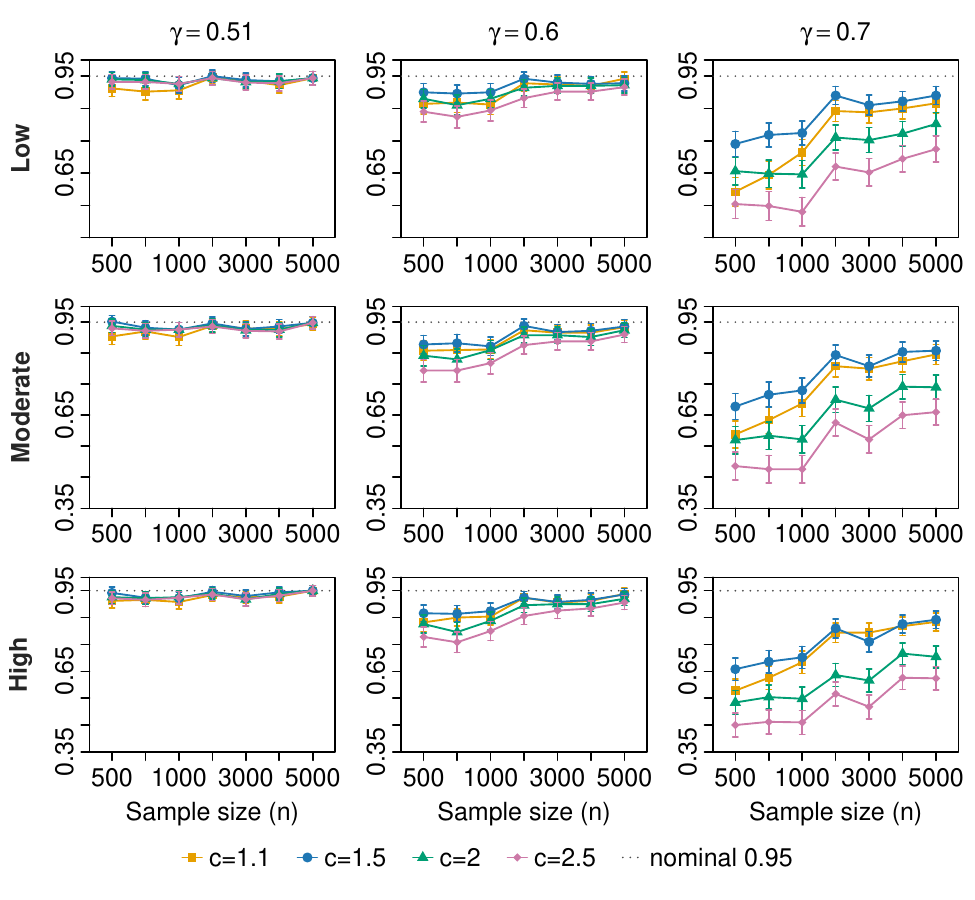}
    \caption{Coverage of the balancing estimator based confidence intervals for different values of $\gamma$ and $c$ for the setting with $d=5$. This is the same setting as Figure~\ref{fig:5_density}.}
    \label{fig:5_sweep}
\end{figure}

\begin{figure}[h]
    \centering
    \includegraphics{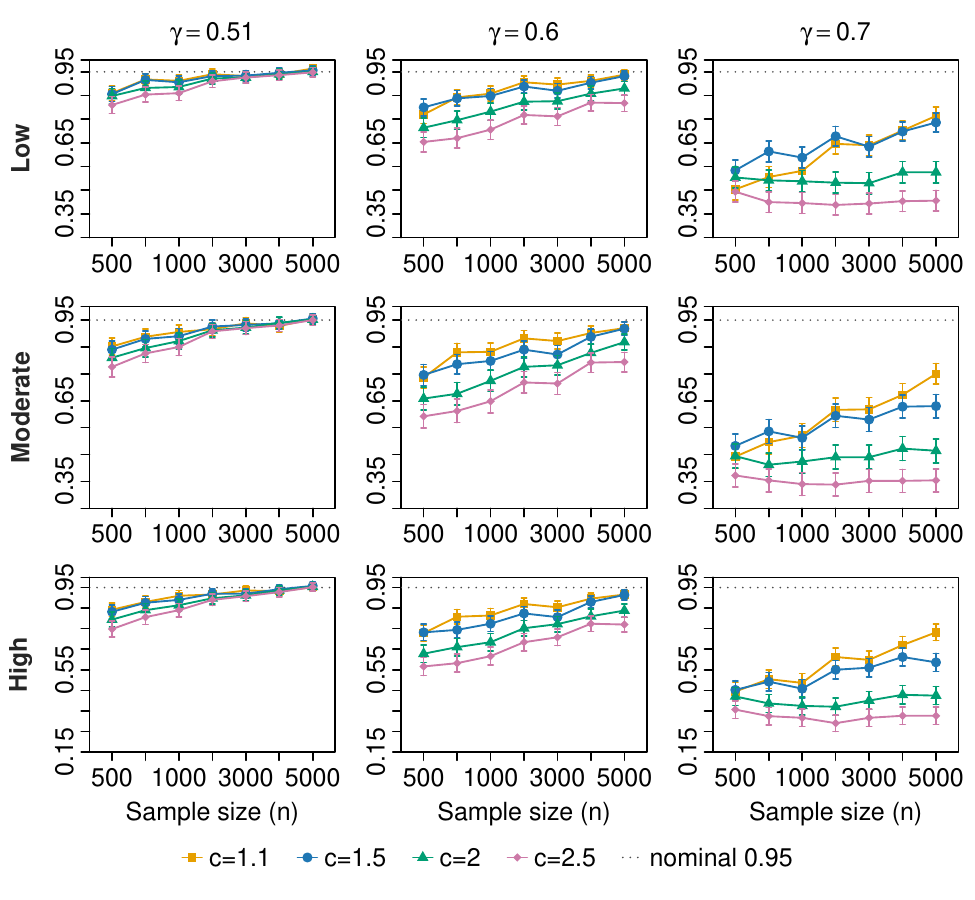}
    \caption{Coverage of the balancing estimator based confidence intervals for different values of $\gamma$ and $c$ for the setting with $d=10$. This is the same setting as Figure~\ref{fig:10_density}.}
    \label{fig:10_sweep}
\end{figure}

In Figures~\ref{fig:3_sweep}, \ref{fig:5_sweep}, and \ref{fig:10_sweep}, we report the empirical coverage of the confidence intervals based on our balancing estimator for various choices of $\gamma$ and $c$, under the simulation settings of Figures~\ref{fig:main_density}, \ref{fig:5_density}, and \ref{fig:10_density}, respectively. We find that across all three settings, our recommended choice of $\gamma = 0.51$ and $c = 1.5$ achieves nominal coverage faster than any other combination considered. This supports our argument in Section~\ref{sec:choice_m_lambda} that although a larger $\gamma$ improves the approximation of $\bm \beta$ by $\tilde{\bm \beta}_n$, it also exacerbates finite-sample bias by increasing $m_n$, and hence the number of bins to balance. Our simulations suggest that the latter effect dominates the former, making nominal coverage harder to attain for larger values of $\gamma$.

\end{document}